\documentclass[aps,prd,twocolumn,preprintnumbers,showpacs,superscriptaddress,nofootinbib,floatfix,10pt]{revtex4-2}
\pdfoutput=1
\usepackage{textcomp}
\usepackage[english]{babel}
\usepackage{amsmath,amssymb,amsbsy,booktabs}
\usepackage{bm}
\usepackage{color}
\usepackage{array}
\usepackage{amstext}
\usepackage{graphicx}
\usepackage{amsfonts}
\usepackage{dcolumn}
\usepackage{rotating}
\usepackage{epstopdf}
\usepackage{esint}
\usepackage{url}
\usepackage[colorlinks=true,
linkcolor=blue,
filecolor=blue,
anchorcolor=blue,
urlcolor=blue,
citecolor=blue
]{hyperref}

\usepackage{array}
\usepackage{booktabs}
\usepackage{multirow}
\newcommand{\tabincell}[2]{\begin{tabular}{@{}#1@{}}#2\end{tabular}}

\usepackage[utf8]{inputenc}

\usepackage{xcolor,colortbl}

\usepackage[T1]{fontenc} % Output font encoding for international characters

\usepackage{cancel}
\usepackage{hhline}
\usepackage{subfigure} 
\usepackage{enumerate}

\allowdisplaybreaks

\begin{document}

\title {\Large \bf \boldmath Searching and identifying leptoquarks through low-energy\\[0.2cm] polarized scattering processes $e^-p\to e^-\Lambda_c$}

\author{Li-Fen Lai}
\email{lailifen@mails.ccnu.edu.cn}
\affiliation{Institute of Particle Physics and Key Laboratory of Quark and Lepton Physics~(MOE),\\
	Central China Normal University, Wuhan, Hubei 430079, China}

\author{Xin-Qiang Li}
\email{xqli@mail.ccnu.edu.cn}
\affiliation{Institute of Particle Physics and Key Laboratory of Quark and Lepton Physics~(MOE),\\
	Central China Normal University, Wuhan, Hubei 430079, China}

\author{Xin-Shuai Yan}
\email{xinshuai@mail.ccnu.edu.cn}
\affiliation{Institute of Particle Physics and Key Laboratory of Quark and Lepton Physics~(MOE),\\
	Central China Normal University, Wuhan, Hubei 430079, China}

\author{Ya-Dong Yang}
\email{yangyd@mail.ccnu.edu.cn}
\affiliation{Institute of Particle Physics and Key Laboratory of Quark and Lepton Physics~(MOE),\\
	Central China Normal University, Wuhan, Hubei 430079, China}
\affiliation{School of Physics and Microelectronics, Zhengzhou University, Zhengzhou, Henan 450001, China}
	
\begin{abstract}
We investigate the potential for searching and identifying the leptoquark (LQ) effects in the charm sector through the low-energy polarized scattering processes $\vec{e}^{\,-}p\to e^-\Lambda_c$, $e^-\vec{p}\to e^-\Lambda_c$, and $\vec{e}^{\,-}\vec{p}\to e^-\Lambda_c$. Considering only the longitudinally polarized processes, we show that the different LQ models can be disentangled from each other by measuring the four spin asymmetries, $A_{L}^e$, $A_{L}^p$, $A_{L3}^{ep}$, and $A_{L6}^{ep}$, constructed in terms of the polarized cross sections. Although it is challenging to accomplish the same goal with transversely polarized processes, we find that investing them in future experiments is especially beneficial, since they can directly probe into the imaginary part of the Wilson coefficients in the general low-energy effective Lagrangian. With our properly designed experimental setups, it is also demonstrated that promising event rates can be expected for all these processes and, even in the worst-case scenario---no LQ signals are observed at all, they can still provide a competitive potential for constraining the new physics, compared with those from the conventional charmed-hadron weak decays and the high-$p_T$ dilepton invariant mass tails at high-energy colliders.
\end{abstract}


\maketitle

\section{Introduction} 
\label{sec:intro}

Many extensions of the Standard Model (SM), such as the grand unified theories~\cite{Pati:1974yy,Georgi:1974sy,Georgi:1974yf,Fritzsch:1974nn,Georgi:1974my,Senjanovic:1982ex,Witten:1985xc,Frampton:1989fu,Murayama:1991ah,Dorsner:2005fq}, predict the existence of a particular type of bosons named leptoquarks (LQs). These hypothetical particles can convert a quark into a lepton and vice versa and, due to such a distinctive character, have a very rich phenomenology in precision experiments and at particle colliders~\cite{Dorsner:2016wpm,Buchmuller:1986zs}. Particularly, several anomalies observed recently in the semileptonic charged- and neutral-current $B$-meson decays~\cite{Bifani:2018zmi,Bernlochner:2021vlv,Albrecht:2021tul,London:2021lfn} as well as in the muon anomalous magnetic moment~\cite{Aoyama:2020ynm} have attracted extensive studies of the LQ interactions, due to their abilities to address the anomalies simultaneously (see, e.g., Refs.~\cite{Alonso:2015sja,Bauer:2015knc,Barbieri:2015yvd,Das:2016vkr,Becirevic:2016yqi,Sahoo:2016pet,Popov:2016fzr,Chen:2017hir,Crivellin:2017zlb,Alok:2017jaf,Buttazzo:2017ixm,Bordone:2018nbg,Becirevic:2018afm,Kumar:2018kmr,Angelescu:2018tyl,Cornella:2019hct,Crivellin:2019dwb,Altmannshofer:2020axr,Saad:2020ihm,Gherardi:2020qhc,Babu:2020hun,Angelescu:2021lln,Cornella:2021sby,Marzocca:2021azj,Julio:2022ton,Crivellin:2022mff,Julio:2022bue}). Often these analyses focus on the LQ couplings to the heavy quarks, but growing interest in the LQ interactions involving the light quarks has also been ignited~(see, e.g., Refs.~\cite{Shanker:1981mj,Shanker:1982nd,Leurer:1993em,Davidson:1993qk,Leurer:1993qx,Carpentier:2010ue,Bobeth:2017ecx,Dorsner:2019vgp,Mandal:2019gff,Su:2019tjn,Crivellin:2020lzu,Crivellin:2021egp,Marzocca:2021miv}).   

Among the various processes used to probe the LQ interactions, the flavor-changing neutral-current (FCNC) ones in the charm sector are the ideal searching ground, due to their absence at the tree level and strong suppression by the Glashow-Iliopoulos-Maiani (GIM)~\cite{Glashow:1970gm} mechanism at the loop level in the SM. The known FCNC processes in the charm sector consist of the rare weak decays of the charmed hadrons~\cite{Burdman:2001tf,Azizi:2010zzb,Paul:2011ar,Cappiello:2012vg,Fajfer:2015mia,Sirvanli:2016wnr,Meinel:2017ggx,Faustov:2018dkn,Gisbert:2020vjx,Faisel:2020php,Fajfer:2021woc,Bause:2020xzj,deBoer:2015boa,DeBoer:2018pdx,Bause:2019vpr,Golz:2021imq,Golz:2022alh,Lees:2011hb,Aaij:2015qmj,BaBar:2020faa,LHCb:2020car}, the high-$p_T$ dilepton invariant mass tails of the processes $pp\to \ell\ell^{(\prime)}$ at high-energy colliders~\cite{Angelescu:2020uug,Fuentes-Martin:2020lea}, as well as the low-energy scattering processes $e^-p\to e^-(\mu^-)\Lambda_c$ we proposed recently~\cite{Lai:2021sww}. Interestingly enough, we found that, based on a set of a high-intensity electron beam and a liquid hydrogen target---both have been used to search for sub-GeV dark vector bosons~\cite{Abrahamyan:2011gv,Essig:2010xa,Essig:2013lka,Allison:2014tpu}---and taking into account the most stringent constraints on the corresponding LQ interactions from other processes, very promising event rates can be expected for both the scattering processes in some specific LQ models~\cite{Lai:2021sww}. Motivated by such a promising prospect of the LQ searches at the low-energy scattering processes (as well as the tenacious hunting for the LQs at the Large Hadron Collider (LHC)~\cite{Diaz:2017lit,Schmaltz:2018nls,Bandyopadhyay:2020wfv,CMS:2012lqv,Aad:2015caa,Sirunyan:2018ruf,CMS:2020wzx,Zyla:2020zbs} and other future facilities~\cite{Bandyopadhyay:2021pld,Qian:2021ihf}), we will address in this paper another interesting question: can the different LQs be systematically disentangled from each other through the low-energy scattering processes? 

A partial answer has been provided in Ref.~\cite{Lai:2021sww}, where we have demonstrated that a combined analysis of the experimental signals of the low-energy scattering processes and the semileptonic $D$-meson decays can distinguish certain scalar LQs from the vector ones. Unfortunately, only the LQs that would not induce tree-level proton decays have been considered there, and the vector LQs are found to be unable to be disentangled from each other~\cite{Lai:2021sww}. Such a deficiency, as will be shown in this work, can be amended by considering the low-energy polarized scattering processes $e^-p\to e^-\Lambda_c$. To be specific, we will consider the FCNC production of $\Lambda_c$ baryons through the low-energy $ep$ scattering processes with only the electron beam polarized ($\vec{e}^{\,-} p \to e^- \Lambda_c$), with only the proton target polarized ($e^-\vec{p}\to e^- \Lambda_c$), and with both of them polarized ($\vec{e}^{\,-}\vec{p}\to e^- \Lambda_c$). Performing a simple analysis of the four spin asymmetries, $A_{L}^e$, $A_{L}^p$, $A_{L3}^{ep} $, and $A_{L6}^{ep}$, constructed in terms of the polarized cross sections, we will show that the different LQs mediating these processes at tree level can be distinguished from each other.

It is also interesting to note that a comprehensive study that worked in a different regime and with a much stricter condition already exists~\cite{Taxil:1999pf}. In particular, focusing on the polarized $ep$ deep inelastic scatterings mediated by the LQs that couple only to the first-generation fermions, the authors of Ref.~\cite{Taxil:1999pf} have shown that the commonly discussed LQ models~\cite{Buchmuller:1986zs} can be disentangled from each other through the precise measurements of some observables, such as the polarized cross sections and spin asymmetries, provided that the high-intensity polarized lepton (both electron and positron) and proton beams are available. By contrast, our proposal will involve much less observables (with the desired measurement precision less demanding), include the case where the LQ couples to quarks belonging to more than a single generation in the mass eigenstate, and is easily adapted to other kinds of new physics (NP) models beyond the SM. 

Of course, all the proposals above cannot be carried out if no LQ signals are observed at all. In this paper, we will show that, thanks to recent advances in the technologies of polarized electron beams and proton targets---both have been resourcefully exploited for studying the nucleon structure (see, e.g., Refs.~\cite{COMPASS:2007rjf,HAPPEX:2011xlw,Arrington:2021alx} and references therein)---promising event rates can be expected for the polarized $ep$ scattering processes, if they are measured with properly designed experimental setups, together with the constraints from the rare charmed-hadron weak decays and the high-$p_T$ dilepton invariant mass tails as input. On the other hand, even in the worst-case scenario---no LQ signals are observed at all, the polarized $ep$ scattering processes can still yield competitive constraints with respect to those obtained from the rare charmed-hadron weak decays and the high-$p_T$ dilepton invariant mass tails. In particular, direct access to the chiral structure of the lepton current in the effective four-fermion operators offers these polarized scattering processes a unique advantage in constraining individually the effective Wilson coefficients (WCs) of the LQ models, which is, otherwise, not available from other conventional FCNC processes in the charm sector. 

The paper is organized as follows. In Sec.~\ref{sec:models}, we start with a 
brief introduction of our theoretical framework, including the most general effective Lagrangian (also the LQ models), the polarized cross sections, and the various spin asymmetries for the polarized scattering processes $e^-p\to e^- \Lambda_c$. In such a framework, we first consider in Sec.~\ref{sec:LPscatter} the longitudinally polarized scattering processes and then in Sec.~\ref{sec:TPscatter} the transversely polarized ones, focusing mainly on their possible applications---the identification of LQ models in particular. With the properly designed experimental setups and the currently existing constraints, we evaluate in Sec.~\ref{sec:Prospect} the prospect for discovering the LQ effects in these low-energy polarized scattering processes. Our conclusions are finally made in Sec.~\ref{sec:con}. For convenience, the helicity-based definitions of the $\Lambda_c\to p$ from factors are given in Appendix~\ref{appendix:form factor}, and the polarized cross sections, together with their relations to the experimentally measurable quantities, are discussed in Appendix~\ref{app:expquantity}, while explicit expressions of the amplitudes squared for both the longitudinally and transversely polarized scattering processes are given in Appendices~\ref{app:LpAmplitude} and \ref{app:TpAmplitude}, respectively.  

\section{Theoretical framework}
\label{sec:models}

\subsection{Effective Lagrangian}

The most general effective Lagrangian responsible for the polarized scattering processes $\ell p\to \ell \Lambda_c$ (or $\ell u \to \ell c$ at the partonic level) can be written as
\begin{align}
\mathcal{L}_{\text{eff}}=\mathcal{L}^{\text{SM}}_{\text{eff}}+\mathcal{L}^{\text{LQ}}_{\text{eff}},
\end{align}
where $\mathcal{L}^{\text{SM}}_{\text{eff}}$ and $\mathcal{L}^{\text{LQ}}_{\text{eff}}$ represent the SM and the LQ contribution, respectively. The SM long-distance effects do not contribute to the scattering processes~\cite{Lai:2021sww}, while the SM short-distance contributions are strongly GIM suppressed~\cite{Burdman:2001tf,Azizi:2010zzb,Paul:2011ar,Cappiello:2012vg,Fajfer:2015mia,Sirvanli:2016wnr,Meinel:2017ggx,Faustov:2018dkn,Gisbert:2020vjx,Faisel:2020php,Fajfer:2021woc,Bause:2020xzj,deBoer:2015boa,DeBoer:2018pdx,Bause:2019vpr,Golz:2021imq,Golz:2022alh}; here, we can safely neglect the contribution from $\mathcal{L}^{\text{SM}}_{\text{eff}}$.

The most general low-energy effective Lagrangian $\mathcal{L}^{\text{LQ}}_{\text{eff}}$ induced by tree-level exchanges of LQs is given by~\cite{Mandal:2019gff} 
\begin{align}
\mathcal{L}^{\text{LQ}}_{\text{eff}}=&\sum_{i,j,m,n}\Big\{ [g^{LL}_{V}]^{ij,mn}(j^{L}_{V})^{ij}  (J^{L}_{V})^{mn}\nonumber \\[-0.15cm]
&\hspace{0.8cm} + [g^{LR}_{V}]^{ij,mn}(j^{L}_{V})^{ij}  (J^{R}_{V})^{mn} \nonumber \\[0.15cm]
&\hspace{0.8cm}+ [g^{RL}_{V}]^{ij,mn}(j^{R}_{V})^{ij}  (J^{L}_{V})^{mn}\nonumber \\[0.15cm]
&\hspace{0.8cm}+ [g^{RR}_{V}]^{ij,mn}(j^{R}_{V})^{ij}  (J^{R}_{V})^{mn} \nonumber \\[0.15cm]
&\hspace{0.8cm}+ [g^{L}_{T}]^{ij,mn}(j^{L}_{T})^{ij}  (J^{L}_{T})^{mn}\nonumber \\[0.15cm]
&\hspace{0.8cm}+ [g^{R}_{T}]^{ij,mn}(j^{R}_{T})^{ij}  (J^{R}_{T})^{mn} \nonumber \\[0.15cm]
&\hspace{0.8cm}+ [g^{L}_{S}]^{ij,mn}(j^{L}_{S})^{ij}  (J^{L}_{S})^{mn}\nonumber \\[0.15cm]
&\hspace{0.8cm}+ [g^{R}_{S}]^{ij,mn}(j^{R}_{S})^{ij}  (J^{R}_{S})^{mn}\Big\}, \label{eq:Lag_LQ}
\end{align}
with 
\begin{align}
(j^{R,L}_{S})^{ij}&\!=\!\bar{\ell}^iP_{R,L}\ell^j\,, \ && (J^{R,L}_{S})^{ij}\!=\!\bar{q}^iP_{R,L} q^j\,, 
\nonumber \\[0.12cm]
(j^{R,L}_{V})^{ij}&\!=\!\bar{\ell}^i\gamma_{\mu}P_{R,L} \ell^j\,,  \  &&(J^{R,L}_{V})^{ij}\!=\! \bar{q}^i\gamma^{\mu}P_{R,L} q^j\,, \nonumber \\[0.12cm]
(j^{R,L}_{T})^{ij}&\!=\!\bar{\ell}^i\sigma_{\mu \nu} P_{R,L} \ell^j\,,  \ &&(J^{R,L}_{T})^{ij}\!=\!\bar{q}^i\sigma^{\mu \nu} P_{R,L} q^j\,, \label{eq:operators}
\end{align}
where $P_{R,L}\!=\!(1\pm \gamma_5)/2$, and $i, j$ and $m, n$ represent the flavor indices of leptons and quarks, respectively. The effective WCs $g$, with their explicit expressions given in terms of the masses and the couplings of the LQs to the SM fermions in a specific model, are obtained by integrating out the heavy LQs, together with proper chiral Fierz transformations (see, e.g., Ref.~\cite{Nishi:2004st}) of the resulting four-fermion operators to the ones given by Eq.~\eqref{eq:Lag_LQ}.

There are seven LQs (three scalar and four vector ones) that can mediate the polarized scattering processes at tree level. Following the same notation as used commonly in literature~\cite{Buchmuller:1986zs,Dorsner:2016wpm}, we present in Table~\ref{tab:LQ} their interactions with the SM fermions, where the left-handed lepton (quark) doublets are denoted by $L_L^i=(\nu_L^i,\ell_L^i)^T$ ($Q_L^i=(u_L^i,d_L^i)^T$), while the right-handed up-(down-)type quark and lepton singlets by $u_R^i$ ($d_R^i$) and $e_R^i$, respectively. Note that, for simplicity, their Hermitian conjugation is not shown explicitly. 

\begin{table}[t]
	\renewcommand*{\arraystretch}{1.6}
	\tabcolsep=0.32cm
	\centering
	\begin{tabular}{ccc}
		\hline \hline
		Model &  $\subset\mathcal{L}_{\text{LQ}}$    &    SM representation   \\
		\hline
	     $S_1$ &\tabincell{c}{$(\lambda^S_{1})_{ij}\bar{Q}_L^{Ci}i\tau_2L_{L}^jS_1$\\ $(\lambda^{\prime S}_{1})_{ij}\bar{u}_{R}^{Ci}e_{R}^{j}S_1$} & $(\bar{3},1,1/3)$        \\ %\cline{2-3}                                                                                                            
		$R_2$ &  \tabincell{c}{$-(\lambda^{R}_2)_{ij}\bar{u}^{i}_{R}i\tau_2L_{L}^{j}R_2$\\ $(\lambda^{\prime R}_2)_{ij}\bar Q_{L}^{i}e_{R}^{j}R_2$} & $(3,2,7/6)$  \\ %\cline{2-3}
	    $S_3$&  $(\lambda^S_{3})_{ij}\bar{Q}^{Ci}_{L}i\tau_2\vec\tau L_{L}^{j}\cdot\vec{S}_3$ & $ (\bar{3},3,1/3)$  \\
		$U_3$  & $(\lambda^U_3)_{ij}\bar{Q}_{L}^{i}\gamma_\mu\vec\tau L_{L}^{j}\cdot\vec{U}_3^\mu$&$ (3,3,2/3)$     \\
		%\cline{2-3}
		$\widetilde{U}_1$ &  $(\lambda^{\widetilde{U}}_1)_{ij}\bar{u}_{R}^{i}\gamma_\mu e_{R}^{j}\widetilde{U}_1^\mu$ & $ (3,1,5/3)$                        \\
		%\cline{2-3}
		$V_2$  &   $(\lambda^V_2)_{ij}\bar Q_{L}^{Ci}\gamma_\mu i\tau_2 e_{R}^{j}V_2^\mu$ & $ (\bar{3},2,5/6)$                           \\
		%\cline{2-3}
		$\widetilde{V}_2$ &  $(\lambda^{\widetilde{V}}_2)_{ij}\bar u_{R}^{Ci}\gamma_\mu i\tau_2 L_{L}^{j}\widetilde{V}_2^\mu$ & $ (\bar 3,2,-1/6)$             \\
		\hline \hline
	\end{tabular}
	\caption{Scalar and vector LQ interactions with the SM fermions, together with their representations under the SM gauge group $\text{SU(3)}_C\otimes \text{SU}(2)_{L}\otimes \text{U}(1)_{Y}$, where $\tau_a$, with $a=1,2,3$, denote the Pauli matrices, and the hypercharge $Y$ is given by $Q_{\text{em}}=T_3+Y$.}
	\label{tab:LQ}
\end{table}

Following the aforementioned procedure, we obtain the effective WCs for each LQ model in Table~\ref{tab:LQ} as 
follows:
\begin{align}\label{eq:LQWilson}
[g^{LL(LR)}_{V}]^{ij,mn}&=k^{LL(LR)}_{V}\frac{(\lambda_I)_{in}\left(\lambda_I\right)^*_{jm}}{M^2}\,, \nonumber \\[0.12cm] 
[g^{RR(RL)}_{V}]^{ij,mn}&=k^{RR(RL)}_{V}\frac{(\lambda_J)_{in}\left(\lambda_J\right)^*_{jm}}{M^2}\,, \nonumber \\[0.12cm] 
[g^{L}_{S}]^{ij,mn}&=k^{L}_{S}\frac{(\lambda_I)_{in}\left(\lambda_J\right)^*_{jm}}{M^2}\,, 
\nonumber \\[0.12cm] 
[g^{R}_{S}]^{ij,mn}&=k^{R}_{S}\frac{(\lambda_J)_{in}\left(\lambda_I\right)^*_{jm}}{M^2}\,, \nonumber \\[0.12cm] 
[g^{L}_{T}]^{ij,mn}&=k^{L}_{T}\frac{(\lambda_I)_{in}\left(\lambda_J\right)^*_{jm}}{M^2}\,, \nonumber \\[0.12cm] 
[g^{R}_{T}]^{ij,mn}&=k^{R}_{T}\frac{(\lambda_J)_{in}\left(\lambda_I\right)^*_{jm}}{M^2}\,, 
\end{align}
where, as done in Ref.~\cite{Lai:2021sww}, we have coincided the mass-eigenstate basis of the left-handed up-type quarks and charged leptons with their flavor basis, and uniformly denoted the LQ masses by $M$, but bearing in mind that they can differ from each other in general. 
The coefficients $k$ can be directly read out from Table~\ref{tab:coeff}, and so are the handy relations,
\begin{align}
[g^{L,R}_{S}]^{ij,mn}=\mp 4 [g^{L,R}_{T}]^{ij,mn}\,, \label{eq:g_denfi_1}
\end{align}
where the $-$ and $+$ signs apply to the LQ models $S_1$ and $R_2$, respectively. It is important to note that the WCs in Eq.~\eqref{eq:LQWilson} are all given at the matching scale $\mu=M$. To connect the LQ coupling constants $\lambda_I$ and $\lambda_J$ to the low-energy polarized scattering processes $e^-p\to e^- \Lambda_c$, they must be evolved to the corresponding low-energy scale through the renormalization group (RG) equation. Moreover, since large mixings of the tensor operators into the scalar ones can arise due to QED and electroweak (EW) one-loop effects~\cite{Gonzalez-Alonso:2017iyc,Aebischer:2017gaw}, both the QCD and EW/QED effects must be taken into account. Taking $M=1$~TeV as the benchmark for the LQ mass,\footnote{Such a choice is motivated by the direct searches for the LQs at LHC, which have already pushed the lower bounds to such an energy scale~\cite{CMS:2012lqv,Aad:2015caa,Sirunyan:2018ruf,CMS:2020wzx,Zyla:2020zbs}. Note that the lower bounds for the vector LQ masses have been pushed roughly up to $1.8$~TeV~\cite{CMS:2020wzx,Zyla:2020zbs}. Nonetheless, we here choose $1$~TeV for both scalar and vector LQs for a simple demonstration.} and performing the RG running from the benchmark scale down to the characteristic scale $\mu=2$~GeV (we refer to Ref.~\cite{Lai:2021sww} for more details), we eventually obtain
\begin{align}\label{eq:RG_high}
g^{\chi}_{S}(2\,\text{GeV})&\approx 2.0\, g^{\chi}_S(1\,\text{TeV})-0.5\, g^{\chi}_T(1\,\text{TeV})\,, \nonumber \\[0.18cm]
g^{\chi}_T(2\,\text{GeV})&\approx 0.8\, g^{\chi}_T(1\,\text{TeV})\,, \nonumber \\[0.15cm]
g_V(2\,\text{GeV})&\simeq\, g_V(1\,\text{TeV})\,, 
\end{align} 
where $\chi=L, R$, and the RG running effects of the vector operators have been neglected, since these operators do not get renormalized under QCD while their RG running effects under EW/QED are only at percent level. With the results in Eq.~\eqref{eq:RG_high}, the scalar-tensor WC relations, $g^{\chi}_{S}(1\,\text{TeV})=\mp4g^{\chi}_T(1\,\text{TeV})$, in  Eq.~\eqref{eq:g_denfi_1} are modified as 
\begin{align}\label{eq:RG_R2}
g^{\chi}_{S}(2\,\text{GeV})\approx \mp\, 9.4\, g^{\chi}_T(2\,\text{GeV})\,, 
\end{align}
at the scale $\mu=2$~GeV for the LQ models $S_1$ and $R_2$, respectively. Note that, for convenience, we will denote the WCs $g(2\,\text{GeV})$ simply by $g$ hereafter.

\begin{table}[t]
	\renewcommand*{\arraystretch}{1.6}
	\setlength\tabcolsep{4pt}
	\centering
	\begin{tabular}{ccccccccccc}
		\hline\hline
		LQ        &  $\lambda_I$     &  $\lambda_J$      &  $k^{LL}_{V}$       &  $k^{RR}_{V}$    &  $k^{LR}_{V}$      &  $k^{RL}_{V}$   &  $k_{S}^L$   &  $k_{S}^R$  &  $k_T^L$       &  $k_{T}^R$  \\
		\hline
		$S_1$         &  $\lambda_{1}^{S}$  &  $\lambda_{1}^{\prime S}$  &   $\frac12$  &  $\frac12$   &  $0$  &  $0$  &  $-\frac12$  &  $-\frac12$  &  $\frac18$  &  $\frac18$  \\
		%\hline
		$S_3$         &  $\lambda_{3}^{S}$  & $\backslash$  &  $\frac12$  &  $0$   &  $0$         &  $0$         &  $0$         &  $0$         &  $0$         &  $0$  \\
		%\hline
		$R_2$         &  $\lambda_{2}^{R}$  &  $\lambda_{2}^{\prime R}$   &  $0$  &  $0$  &  $-\frac12$  &  $-\frac12$   &  $\frac12$  &  $\frac12$  &  $\frac18$  &  $\frac18$  \\
		%\hline
		$ U_3$  &  $\lambda_{3}^{U}$  &  $\backslash$  &  $2$         &  $0$         &  $0$   &  $0$   &  $0$         &  $0$         &  $0$         &  $0$  \\
		%\hline
		$\widetilde{U}_1$         &  $\backslash$  &  $\lambda_{1}^{\widetilde{U}}$   &  $0$   &  $1$   &  $0$         &  $0$         &  $0$         &  $0$         &  $0$         &  $0$  \\
		%\hline
		$ V_2$  & $\lambda_{2}^{V}$  &  $\backslash$ &   $0$         &  $0$         &  $0$   &  $-1$  &  $0$         &  $0$         &  $0$         &  $0$  \\
		%\hline
		$\widetilde V_2$         &  $\backslash$  &  $\lambda_{2}^{\widetilde V}$  & $0$         &  $0$  &  $-1$         &  $0$         &  $0$         &  $0$         &  $0$         &  $0$  \\
		\hline\hline
	\end{tabular}
	\caption{Coefficient matrix for the effective WCs in Eq.~\eqref{eq:LQWilson} for the seven LQs in Table~\ref{tab:LQ}. The entries with ``$\backslash$'' mean that $\lambda_{I,J}$ does not appear for the LQ models in the first column.}
	\label{tab:coeff}
\end{table}

\subsection{Cross section, kinematics, and beam energy}
\label{subsec:Cross section}

The differential cross section of the polarized scattering process $\vec{e}^-(k)+\vec{p}(P)\to e^-(k')+\Lambda_c(P')$, with $P=(m_p, 0)$, $P'=(E_{\Lambda_c}, \pmb{p}')$, $k=(E, \pmb{k})$, and $k'=(E', \pmb{k}')$, is given by 
\begin{align}\label{eq:diff_cross}
d\sigma=&\frac{1}{4[(P\cdot k)^2-m^2_em^2_p]^{1/2}}\frac{d^3\pmb{k}'}{(2\pi)^3}\frac{1}{2E'}\frac{d^3\pmb{p}'}{(2\pi)^3}\frac{1}{2E_{\Lambda_c}}\nonumber \\[0.15cm]
&\times |\mathcal{M}|^2(2\pi)^4\delta^4(P+k-P'-k'),
\end{align}
where the amplitude $\mathcal{M}$ can be written as 
\begin{align}
\mathcal{M}\!=\!\sum g_{\alpha}\langle e^{-} (k^\prime, r^\prime)| j_{\alpha}|\vec{e}^{-}(k,r)\rangle \langle \Lambda_c(P^\prime, s^\prime)|J_{\alpha}|\vec{p}(P, s)\rangle, \label{eq:amplitude}
\end{align} 
with $r$ and $s$ ($r^\prime$ and $s^\prime$) denoting the spins of initial (final) electron and baryon, respectively.
And the amplitude squared $|\mathcal{M}|^2$ is obtained by summing up the initial- and final-state spins; more details are 
elaborated in Appendices~\ref{app:expquantity}, \ref{app:LpAmplitude}, and \ref{app:TpAmplitude}.

The hadronic matrix elements $\langle\Lambda_c(P^\prime,s^\prime)|J_{\alpha}|\vec{p}(P,s)\rangle$ in Eq.~\eqref{eq:amplitude} 
are given by the complex conjugate of $\langle \vec{p}(P,s)|J^{\dagger}_{\alpha}|\Lambda_c(P^\prime,s^\prime)\rangle$, which are further parametrized by the $\Lambda_c \to p$ transition form factors~\cite{Feldmann:2011xf,Meinel:2017ggx,Das:2018sms}. However, since a scattering process generally occupies a different kinematic region from that of a decay, to extend the form factors that are commonly convenient for studying the $\Lambda_c$ weak decays to the scattering process, their parametrization must be analytic in the proper $q^2$ region. Interestingly, there already exists such a parametrization scheme, which was initially proposed to parametrize the $B\to \pi$ vector form factor~\cite{Bourrely:2008za}, and has been recently utilized in the lattice QCD calculation of the $\Lambda_c\to N$ (nucleon) form factors~\cite{Meinel:2017ggx}. We thus adopt the latest lattice QCD results~\cite{Meinel:2017ggx}, given that they provide also an error estimation; for details, we refer the readers to Appendix~\ref{appendix:form factor}. It might be also interesting to note that the same approach has been adopted to explore the quasielastic weak production of $\Lambda_c$ hadron induced by $\bar{\nu}$ scattering off nuclei~\cite{Sobczyk:2019uej}. 

The spinor of the polarized electron or proton is given by $(1+\gamma_5\cancel{\xi})u(k,s)/2$, 
where $\xi^{\mu}$ denotes the spin (or polarization) four-vector.
For the longitudinally polarized electron beam, the polarization four-vector $\xi^{\mu}_e$ in the proton target rest frame (lab frame) is given by~\cite{Barone:2003fy}
\begin{align}
\xi^{\mu}_e=\pm\Big(\dfrac{|\pmb{k}|}{m_e},\dfrac{k^0\pmb{k}}{m_e |\pmb{k}|}\Big)\,,
\end{align}
where the $+$ ($-$) sign corresponds to the case when the electron beam is right-handed (left-handed) polarized. The spin four-vector $\xi_p^{\mu}$ for a longitudinally polarized proton target in the lab frame is given by $\xi_p^{\mu}=\pm(0,0,0,1)$, while $\xi_p^{\mu}=(0,\cos{\beta},\sin{\beta},0)$ for a transversely polarized proton target, 
where $\beta$ is the azimuthal angle between the proton spin directon and the x axis~\cite{Barone:2003fy}.

\begin{figure}[t]
	\centering
	\includegraphics[width=0.48\textwidth]{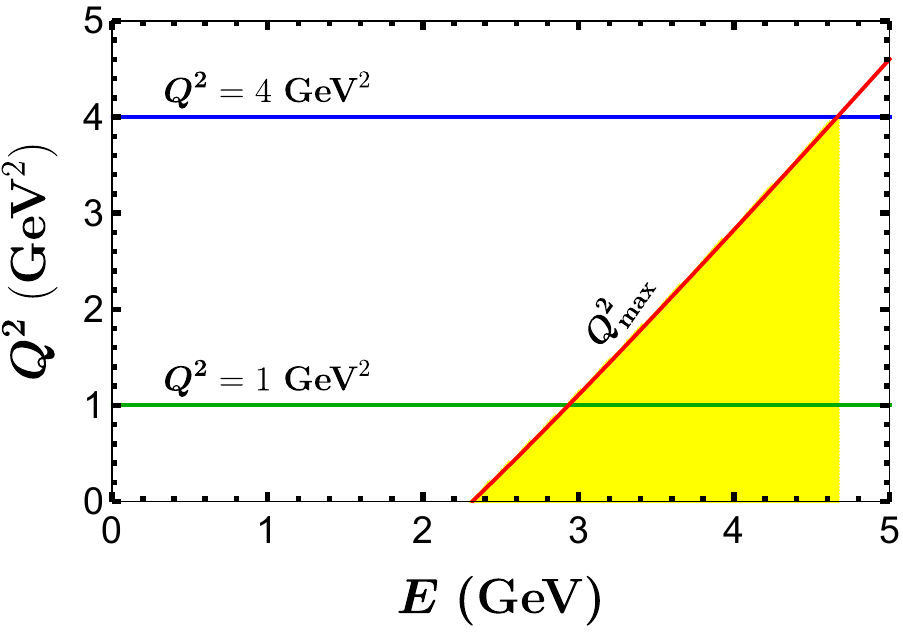}
	\caption{Criteria for selecting the electron beam energy $E$, where the red line denotes the $E-Q^2_{\max}$ relation given by Eq.~\eqref{eq:LFC_Q2_range}, the blue line represents the condition $Q^2\leq 4\,\text{GeV}^2$ required by our theoretical framework, while the green line corresponds to our benchmark scenario with $Q^2=1\,\text{GeV}^2$. The yellow region indicates the eligible $E$ with its corresponding $[Q^2_{\min},Q^2_{\max}]$.} 
	\label{fig:Eselection} 
\end{figure} 

Same as the unpolarized scattering process $e^-p\to e^-\Lambda_c$, kinematics of the polarized 
ones is bounded by~\cite{Lai:2021sww} 
\begin{align}
\frac{2E(m_{\Lambda_c}^2-m_{p}^2-2m_{p}E)}{m_{p}+2E}\leq q^2 \leq 0\,. \label{eq:LFC_Q2_range}
\end{align}
This condition indicates that the electron beam energy $E$ determines the maximal $Q^2$ ($Q^2=-q^2$), 
which, in turn, implies that constraints on $Q^2_{\max}$ restrict the $E$ selection. 
An explicit example is that a minimal requirement for $E$ of the scattering process 
can be obtained by using the condition $Q^2_{\max}=Q^2_{\min}=0$; this can also be visualized in Fig.~\ref{fig:Eselection} 
by noting the intersection point of the $E$ axis and the red line that represents the $E$-$Q^2_{\max}$ relation.
Besides the kinematic constraint on $Q^2_{\max}$, 
we also consider the limit from our theoretical framework. As indicated in the previous subsection, our analysis will be carried out in the framework of $\mathcal{L}_{\text{eff}}$ given by Eq.~\eqref{eq:Lag_LQ} at the scale $\mu=2\,\text{GeV}$; to ensure the validity of our results, we require $Q^2_{\max}$ to not exceed $\mu^2=4\,\text{GeV}^2$. Such a requirement, depicted by the blue line in Fig.~\ref{fig:Eselection}, indicates an upper bound $E\lesssim 4.65\,\text{GeV}$, provided that the observables one is interested in, such as the total cross section, involve $Q^2_{\max}$. Otherwise, $E$ is not bounded as above, since one can always concentrate on the lower $Q^2$ region, even though a high $Q^2_{\max}$ is
available due to a high $E$. 

\subsection{Spin asymmetries}
\label{subsec:spin asymmetry}

In terms of the polarized (differential) cross sections, several spin asymmetries can be defined, some of which, as will be shown later, play an important role in disentangling the NP models.

If only the incoming electron beam is polarized, we can define the single-spin parity-violating (PV) longitudinal asymmetry  
\begin{align}
A_{L}^e=\dfrac{\sigma^-_e-\sigma^+_e}{\sigma^-_e+\sigma^+_e}\,, \label{eq:L_single_PV}
\end{align}
where $\sigma^{-}_e$ and $\sigma^{+}_e$ denote the scattering cross sections with the incoming electron beam being right-handed ($e^+_L$) and left-handed ($e^-_L$) polarized, respectively. Here the cross sections can be the total ($\sigma^{\pm}_e$) or the differential ($d\sigma^{\pm}_e/dQ^2$) ones. Similarly, we can also define the single-spin PV asymmetries   
\begin{align}\label{eq:LT_single_PV}
A_{L}^p=\dfrac{\sigma^-_p-\sigma^+_p}{\sigma^-_p+\sigma^+_p}\,, \qquad
A_{T}^p=\dfrac{\widetilde{\sigma}^-_p-\widetilde{\sigma}^+_p}{\widetilde{\sigma}^-_p+\widetilde{\sigma}^+_p}\,,
\end{align}
when only the proton target is longitudinally ($p_L$) and transversely ($p_T$) polarized, respectively. Here $\sigma_p$ and $\widetilde{\sigma}_p$ are the corresponding scattering cross sections.

Concerning the case when the incoming electron beam and the proton target are both longitudinally polarized, we can construct six double-spin asymmetries~\cite{Virey:1998ny,Taxil:1999pf}. Among them, two PV double-spin asymmetries are given, respectively, by 
\begin{align}
A_{L1}^{ep}=\dfrac{\sigma^{--}-\sigma^{++}}{\sigma^{--}+\sigma^{++}}\,,\qquad A_{L2}^{ep}=\dfrac{\sigma^{-+}-\sigma^{+-}}{\sigma^{-+}+\sigma^{+-}}\,, \label{eq:L_double_PV}
\end{align}
while four parity-conserving (PC) ones are defined, respectively, as 
\begin{align}
A_{L3}^{ep}=\dfrac{\sigma^{--}-\sigma^{-+}}{\sigma^{--}+\sigma^{-+}}\,,\qquad A_{L4}^{ep}=\dfrac{\sigma^{--}-\sigma^{+-}}{\sigma^{--}+\sigma^{+-}}\,,\nonumber\\[0.15cm]
A_{L5}^{ep}=\dfrac{\sigma^{++}-\sigma^{-+}}{\sigma^{++}+\sigma^{-+}}\,,\qquad A_{L6}^{ep}=\dfrac{\sigma^{++}-\sigma^{+-}}{\sigma^{++}+\sigma^{+-}}\,, \label{eq:L_double_PC}
\end{align}
where the first superscript of $\sigma$ indicates the polarization direction of the incoming electron beam, whereas the second one denotes that of the proton target. Concerning the case when the proton target is transversely polarized, on the other hand, we can also build six double-spin asymmetries in the same way~\cite{Virey:1998ny,Taxil:1999pf}, with the PV double-spin asymmetries given, respectively, by 
\begin{align}\label{eq:T_double_PV}
A_{T1}^{ep}=\dfrac{\widetilde{\sigma}^{--}-\widetilde{\sigma}^{++}}{\widetilde{\sigma}^{--}+\widetilde{\sigma}^{++}}\,,\qquad A_{T2}^{ep}=\dfrac{\widetilde{\sigma}^{-+}-\widetilde{\sigma}^{+-}}{\widetilde{\sigma}^{-+}+\widetilde{\sigma}^{+-}}\,,
\end{align}
and the PC double-spin asymmetries by
\begin{align}\label{eq:T_double_PC}
A_{T3}^{ep}=\dfrac{\widetilde{\sigma}^{--}-\widetilde{\sigma}^{-+}}{\widetilde{\sigma}^{--}+\widetilde{\sigma}^{-+}}\,,\qquad A_{T4}^{ep}=\dfrac{\widetilde{\sigma}^{--}-\widetilde{\sigma}^{+-}}{\widetilde{\sigma}^{--}+\widetilde{\sigma}^{+-}}\,,\nonumber\\[0.15cm]
A_{T5}^{ep}=\dfrac{\widetilde{\sigma}^{++}-\widetilde{\sigma}^{-+}}{\widetilde{\sigma}^{++}+\widetilde{\sigma}^{-+}}\,,\qquad A_{T6}^{ep}=\dfrac{\widetilde{\sigma}^{++}-\widetilde{\sigma}^{+-}}{\widetilde{\sigma}^{++}+\widetilde{\sigma}^{+-}}\,, 
\end{align}
where $\widetilde{\sigma}$ represents the scattering cross section with the proton target being transversely polarized.

\section{\boldmath Longitudinally polarized scattering processes $e^-p\to e^- \Lambda_c$  }
\label{sec:LPscatter}

\subsection{Observable analyses}
\label{subsec:obs}

The first observable associated with the longitudinally polarized scattering processes $e^-p\to e^- \Lambda_c$ is the single-spin PV asymmetry $A^e_L$, which is defined in terms of the polarized cross sections $\sigma^{\pm}_e$ (cf. Eq.~\eqref{eq:L_single_PV}). Since multiple 
operators in Eq.~\eqref{eq:Lag_LQ} can contribute to the polarized scattering processes, we can generally write $\sigma^{\pm}_e$ as 
\begin{align}\label{eq:dsigma}
d\sigma^{\pm}_e=\sum d\sigma^{\pm}_{\alpha-\beta}\propto\sum g_{\alpha}g^*_{\beta}~|\mathcal{M}|^2_{\alpha-\beta}\,,
\end{align}   
where $g_{\alpha}$ and $g^*_{\beta}$ go through all the WCs in Eq.~\eqref{eq:Lag_LQ}, and $|\mathcal{M}|^2_{\alpha-\beta}$ with a subscript, e.g., $V_{RL}\!-\!V_{RR}$, represents the reduced amplitude squared that is induced by the interference between the operators $j_V^RJ^L_V$ and $j_V^RJ^R_V$. 

\begin{table*}[t]%\small
	\renewcommand*{\arraystretch}{1.8}
	\tabcolsep=0.28cm
	\centering
	\begin{tabular}{ccccccc}
		\hhline{===~===}
		& $\left(g_{V}^{LL}\right)^*$ &  $\left(g_{V}^{LR}\right)^*$ & & & $\left(g_{V}^{RR}\right)^*$ & $\left(g_{V}^{RL}\right)^*$ \\
		\cline{1-3}\cline{5-7}
		$g_{V}^{LL}$ & 2$\overline{|\mathcal{M}|}^2_{V_{LL}-V_{LL}}$ & 2$\overline{|\mathcal{M}|}^2_{V_{LL}-V_{LR}}$ &&$g_{V}^{RR}$  &  2$\overline{|\mathcal{M}|}^2_{V_{LL}-V_{LL}}$  &  2$\overline{|\mathcal{M}|}^2_{V_{LL}-V_{LR}}$\\
		%\cline{1-3}\cline{5-7}
		$g_{V}^{LR}$ &2$\overline{|\mathcal{M}|}^2_{V_{LL}-V_{LR}}$ & 2$\overline{|\mathcal{M}|}^2_{V_{LR}-V_{LR}}$ &&$g_{V}^{RL}$ &2$\overline{|\mathcal{M}|}^2_{V_{LL}-V_{LR}}$  &  2$\overline{|\mathcal{M}|}^2_{V_{LR}-V_{LR}}$\\
		\hhline{===~===} %\cline{1-3}\cline{5-7}		
		& $\left(g_{S}^{L}\right)^*$   &  $\left(g_{T}^{L}\right)^*$  & &	&   $\left(g_{S}^{R}\right)^*$   &  $\left(g_{T}^{R}\right)^*$  \\
		\cline{1-3}\cline{5-7}
		$g_{S}^{L}$ & 2$\overline{|\mathcal{M}|}^2_{S_{L}-S_{L}}$  & 2$\overline{|\mathcal{M}|}^2_{S_{L}-T_{L}}$& &	 $g^{R}_{S}$ &  2$\overline{|\mathcal{M}|}^2_{S_{L}-S_{L}}$   & 2$\overline{|\mathcal{M}|}^2_{S_{L}-T_{L}}$   \\
		%\cline{1-3}\cline{5-7}
		$g_{T}^{L}$ &  2$\overline{|\mathcal{M}|}^2_{S_{L}-T_{L}}$   & 2$\overline{|\mathcal{M}|}^2_{T_{L}-T_{L}}$ &   &  $g^{R}_{T}$ &  2$\overline{|\mathcal{M}|}^2_{S_{L}-T_{L}}$   & 2$\overline{|\mathcal{M}|}^2_{S_{L}-S_{L}}$\\
		\hhline{===~===}
	\end{tabular}
	\caption{Non-zero reduced amplitudes squared of the scattering process with only the electron beam left-handed (left tables) or right-handed (right tables) polarized, where $\overline{|\mathcal{M}|}^2_{\alpha-\beta}$ has been averaged over the spins of the initial electron and proton.}
	\label{tab:crosse}
\end{table*}

Now a few comments about the reduced amplitude squared $|\mathcal{M}|^2_{\alpha-\beta}$ in Eq.~\eqref{eq:dsigma} are in order. First, as shown in Table~\ref{tab:crosse}, the polarization direction of the electron beam selects the proper ($\alpha,\,\beta$) combinations. This can be verified by noting that, for the scattering processes with $e^-_L$, only the operators with left-handed lepton current $j^{L}$ are at work, since the projection operator $(1+\gamma_5\cancel{\xi}_e)/2$ becomes $P_{L}$ in the relativistic limit $m_e/E\to 0$. The same conclusion also holds for the operators with $j^{R}$ when the electron beam is right-handed polarized. Second, the amplitude squared $\overline{|\mathcal{M}|}^2_{\alpha-\beta}$ is obtained by averaging over the initial electron spins, while $|\mathcal{M}|^2_{\alpha-\beta}$ is not---hence differing from the former by a factor of 2. Third, with our convention in Eq.~\eqref{eq:dsigma}, the reduced amplitudes squared are all real, and thus $\overline{|\mathcal{M}|}^2_{\alpha-\beta}$ is identical to $\overline{|\mathcal{M}|}^2_{\beta-\alpha}$. Finally, due to the chiral structures of the lepton and quark currents involved, certain reduced amplitudes squared with different subscripts are identical to each other, e.g., $\overline{|\mathcal{M}|}^2_{V_{LL}-V_{LR}} =\overline{|\mathcal{M}|}^2_{V_{RR}-V_{RL}}$. In this case, only one of them is preserved.    
It is now clear from Table~\ref{tab:crosse} that, if only one operator (or more in certain cases) contributes to the scattering processes, as commonly happens in several LQ models in Table~\ref{tab:LQ}, its associated $A_{L}^e$ is always equal to either $+1$ or $-1$. Such a feature, as will be shown later, can help to disentangle the different LQ models. 

\begin{table*}[t]%\small
	\centering
	\renewcommand*{\arraystretch}{1.8}
	\tabcolsep=0.35cm
	\begin{tabular}{ccc}
		\hline\hline
		& $\left(g^{LL}_{V}\right)^*$ &  $\left(g^{LR}_{V}\right)^*$  \\
		\hline
		$g^{LL}_{V}$   &   2$\overline{|\mathcal{M}|}^2_{V_{LL}-V_{LL}}+|\mathcal{M}^\prime|^{2}_{V_{LL}-V_{LL}}$  & 2$\overline{|\mathcal{M}|}^2_{V_{LL}-V_{LR}}+|\mathcal{M}^\prime|^{2}_{V_{LL}-V_{LR}}$   \\
		%\hline
		$g^{LR}_{V}$ &  2$\overline{|\mathcal{M}|}^2_{V_{LL}-V_{LR}}+|\mathcal{M}^\prime|^{2}_{V_{LL}-V_{LR}}$   &   2$\overline{|\mathcal{M}|}^2_{V_{LR}-V_{LR}}+|\mathcal{M}^\prime|^{2}_{V_{LR}-V_{LR}}$     \\
		\hline\hline 
		&  $\left(g^{L}_{S}\right)^*$ &  $\left(g^{L}_{T}\right)^*$       \\
		\hline
		$g^{L}_{S}$ &  2$\overline{|\mathcal{M}|}^2_{S_{L}-S_{L}}+|\mathcal{M}^\prime|^{2}_{S_{L}-S_{L}}$   & 2$\overline{|\mathcal{M}|}^2_{S_{L}-T_{L}}+|\mathcal{M}^\prime|^{2}_{S_{L}-T_{L}}$   \\
		%\hline
		$g^{L}_{T}$ & 2$\overline{|\mathcal{M}|}^2_{S_{L}-T_{L}}+|\mathcal{M}^\prime|^{2}_{S_{L}-T_{L}}$   & 2$\overline{|\mathcal{M}|}^2_{T_{L}-T_{L}}+|\mathcal{M}^\prime|^{2}_{T_{L}-T_{L}}$ \\
		\hline\hline \\[-0.1cm]
		\hline\hline
		& $\left(g^{RR}_{V}\right)^*$ &  $\left(g^{RL}_{V}\right)^*$ \\
		\hline 
		$g^{RR}_{V}$   &   2$\overline{|\mathcal{M}|}^2_{V_{LL}-V_{LL}}-|\mathcal{M}^\prime|^{2}_{V_{LL}-V_{LL}}$  & 2$\overline{|\mathcal{M}|}^2_{V_{LL}-V_{LR}}-|\mathcal{M}^\prime|^{2}_{V_{LL}-V_{LR}}$   \\
		%\hline 
		$g^{RL}_{V}$ & 2$\overline{|\mathcal{M}|}^2_{V_{LL}-V_{LR}}-|\mathcal{M}^\prime|^{2}_{V_{LL}-V_{LR}}$   &   2$\overline{|\mathcal{M}|}^2_{V_{LR}-V_{LR}}-|\mathcal{M}^\prime|^{2}_{V_{LR}-V_{LR}}$     \\
		\hline\hline
		&  $\left(g^{R}_{S}\right)^*$ &  $\left(g^{R}_{T}\right)^*$       \\
		\hline
		$g^{R}_{S}$ & 2$\overline{|\mathcal{M}|}^2_{S_{L}-S_{L}}-|\mathcal{M}^\prime|^{2}_{S_{L}-S_{L}}$   & 2$\overline{|\mathcal{M}|}^2_{S_{L}-T_{L}}-|\mathcal{M}^\prime|^{2}_{S_{L}-T_{L}}$   \\
		%\hline
		$g^{R}_{T}$ & 2$\overline{|\mathcal{M}|}^2_{S_{L}-T_{L}}-|\mathcal{M}^\prime|^{2}_{S_{L}-T_{L}}$   &2$\overline{|\mathcal{M}|}^2_{T_{L}-T_{L}}-|\mathcal{M}^\prime|^{2}_{T_{L}-T_{L}}$ \\
		\hline\hline		
	\end{tabular}
	\caption{Non-zero reduced amplitudes squared of the polarized scattering processes with $e^{-}_L p^-_L$ (the upper two tables) and $e^{+}_L p^-_L$ (the lower two tables). For the processes with $e^{-}_L p^+_L$ and $e^{+}_L p^+_L$, one simply flips the sign of the $\xi_p$-dependent $|\mathcal{M}^\prime|^2_{\alpha-\beta}$.}
	\label{tab:crossp}
\end{table*}

Features of other spin asymmetries, like $A^p_L$ in Eq.~\eqref{eq:LT_single_PV} 
and $A^{ep}_{L1-L6}$ in Eqs.~\eqref{eq:L_double_PV} and \eqref{eq:L_double_PC}, are even richer. Given that they are all defined in terms of the double-spin polarized cross sections 
(note that $\sigma^{\pm}_p=(\sigma^{+\pm}+\sigma^{-\pm})/2$), it is more convenient to focus on these cross sections. In the framework of the general effective Lagrangian $\mathcal{L}_{\text{eff}}$, we can schematically write the double-spin polarized cross section $\sigma^{ab}$, with $a, b=\pm$, as 
\begin{equation}\label{eq:dsimgaL}
d\sigma^{ab}\!=\!\sum d\sigma^{ab}_{\alpha-\beta}\!\propto\!\sum g_{\alpha}g^*_{\beta}
\left(2\overline{|\mathcal{M}|}^2_{\alpha-\beta}\!+\!c|\mathcal{M}^\prime|^2_{\alpha-\beta}\right)\,,
\end{equation}
where the coefficient $c=\pm$ depends on the subscripts $\alpha$ and $\beta$, as well as the superscripts $a$ and $b$. Here $\overline{|\mathcal{M}|}^2_{\alpha-\beta}$ and $|\mathcal{M}^\prime|^2_{\alpha-\beta}$ represent, respectively, the $\xi_p$-independent and $\xi_p$-dependent reduced amplitudes squared---we refer the readers to Appendix~\ref{app:LpAmplitude} for further details.  
Note that $\overline{|\mathcal{M}|}^2_{\alpha-\beta}$ is identical to 
that in $\sigma^{\pm}_e$ due to the fact that $\sigma^{\pm}_e=(\sigma^{\pm+}+\sigma^{\pm-})/2$.  
The possible ($\alpha,\,\beta$) combinations, together with the non-zero 
reduced amplitudes squared associated with $\sigma^{\pm-}$, are shown in Table~\ref{tab:crossp}. Clearly, more ($\alpha,\,\beta$) combinations are now available for $\sigma^{\pm}_p$ in comparison with $\sigma^{\pm}_e$,  
but $\sigma^{\pm +}$ and $\sigma^{\pm -}$ still share the same set of ($\alpha,\,\beta$) combinations as for $\sigma^{\pm}_e$. 
It is also interesting to observe that, if certain $\xi_p$-independent reduced amplitudes squared with different subscripts are identical to each other, 
so are the $\xi_p$-dependent ones accordingly---as an example, we have  
$\overline{|\mathcal{M}|}^2_{V_{LL}-V_{LR}}\!=\!\overline{|\mathcal{M}|}^2_{V_{RR}-V_{RL}}$ 
and $|\mathcal{M}^\prime|^2_{V_{LL}-V_{LR}}\!=\!|\mathcal{M}^\prime|^2_{V_{RR}-V_{RL}}$. 
This leads to another interesting observation that the overall reduced amplitude squared associated with $\sigma^{\mp +}_{\alpha-\beta}$ is identical to that with $\sigma^{\pm -}_{\alpha^\prime-\beta^\prime}$, 
where the indices $\alpha^\prime$~($\beta^\prime$) and $\alpha$~($\beta$) must be matched so that their associated operators have opposite chiral structures in both the lepton and quark currents. Let us take $V_{LL}-V_{LR}$ again as a demonstration.   
As can be seen from Table~\ref{tab:crossp}, the overall reduced amplitude squared for $\sigma^{-+}_{V_{LL}-V_{LR}}$ is $2\overline{|\mathcal{M}|}^2_{V_{LL}\!-\!V_{LR}}\!-\!|\mathcal{M}^\prime|^2_{V_{LL}\!-\!V_{LR}}$, 
which is found to be identical to that for $\sigma^{+-}_{V_{RR}-V_{RL}}$.   

The presence of $\xi_p$-dependent $|\mathcal{M}^\prime|^2_{\alpha-\beta}$ indicates that, contrary to $A_{L}^e$, $A_{L}^p$ will not take a trivial form, even if only one operator contributes to the scattering processes. It is, however, by no means useless in distinguishing the LQ models. Note that the observations above also lead to some interesting relations 
between $A_{L3}^{ep}$ and $A_{L6}^{ep}$ for certain operators.  
For instance, $A_{L3}^{ep}$ for the operator $j_V^LJ^L_V$ is identical to
$A_{L6}^{ep}$ for the operator $j_V^RJ^R_V$. These relations 
will be very handy in disentangling the LQ models later. 
Finally, it is easy to check that if only one operator is at work, 
$A_{L1}^{ep} $, $A_{L2}^{ep} $, $A_{L4}^{ep} $, and $A_{L5}^{ep}$ 
can only be $\pm$1, just like $A_{L}^e$, making them less interesting to this work.  

\subsection{NP model identification}

In this subsection, we will use the four spin asymmetries, $A_{L}^e$, $A_{L}^p$, $A_{L3}^{ep}$, and $A_{L6}^{ep}$, to identify the various NP models. 
To make our following analyses as general as possible, we will consider all the four vector operators $j_V^LJ_V^L$, $j_V^LJ_V^R$, $j_V^RJ_V^L$, and $j_V^RJ_V^R$, as well as their possible combinations: 
\begin{enumerate}[(1)]
	\item cases with one vector operator: (I) $j^{L}_{V}J^{L}_{V}$, (II) $j^{L}_{V}J^{R}_{V}$, (III) $j^{R}_{V}J^{L}_{V}$, and (IV) $j^{R}_{V}J^{R}_{V}$;
	
	\item cases with two vector operators: (V) $j^{L}_{V}J^{L}_{V}$ and $j^{L}_{V}J^{R}_{V}$, (VI) $j^{L}_{V}J^{L}_{V}$ and $j^{R}_{V}J^{L}_{V}$, (VII) $j^{L}_{V}J^{L}_{V}$ and $j^{R}_{V}J^{R}_{V}$, (VIII) $j^{L}_{V}J^{R}_{V}$ and $j^{R}_{V}J^{L}_{V}$, (IX) $j^{L}_{V}J^{R}_{V}$ and $j^{R}_{V}J^{R}_{V}$, and (X) $j^{R}_{V}J^{L}_{V}$ and $j^{R}_{V}J^{R}_{V}$;
	
	\item cases with three vector operators: (XI) $j^{L}_{V}J^{L}_{V}$, $j^{L}_{V}J^{R}_{V}$ and $j^{R}_{V}J^{L}_{V}$, (XII) $j^{L}_{V}J^{L}_{V}$, $j^{L}_{V}J^{R}_{V}$ and $j^{R}_{V}J^{R}_{V}$, (XIII) $j^{L}_{V}J^{L}_{V}$, $j^{R}_{V}J^{L}_{V}$ and $j^{R}_{V}J^{R}_{V}$, and (XIV) $j^{L}_{V}J^{R}_{V}$, $j^{R}_{V}J^{L}_{V}$ and $j^{R}_{V}J^{R}_{V}$;
	
	\item cases with four vector operators: (XV) $j^{L}_{V}J^{L}_{V}$, $j^{L}_{V}J^{R}_{V}$, $j^{R}_{V}J^{L}_{V}$ and $j^{R}_{V}J^{R}_{V}$. 
\end{enumerate}
In addition, we will neglect contributions from the scalar and tensor operators 
in these NP models, due to the stringent constraints on them from the leptonic $D$-meson decays~\cite{Petric:2010yt,Lai:2021sww}. 
Note that we here have made 
an implicit assumption that the WCs of the scalar and tensor operators are connected in the NP models through, say, Eq.~\eqref{eq:g_denfi_1} as in the LQ models $S_1$ and $R_2$. Otherwise, only contributions from the scalar operators can be neglected, 
and our analyses must be modified accordingly.  

For a simple demonstration, let us consider as a benchmark the beam energy $E=3$~GeV (see the green line in Fig.~\ref{fig:Eselection} for the corresponding $Q^2_{\max}$) and compute $A_{L}^e$, $A_{L}^p$, $A_{L3}^{ep}$, and $A_{L6}^{ep}$ in these different cases.
Clearly, $A_{L}^e$ and $A_{L}^p$ in the cases I--IV are independent of the WCs, since only one operator is involved, whereas those in other cases involving multiple vector operators do depend on their associated WCs. We present in Table~\ref{tab:result1} the resulting $A_{L}^e$ and $A_{L}^p$ for the cases I--IV. By contrast, $A_{L}^e$ and $A_{L}^p$ in other cases can take any number within the range $[-1,1]$. Here we expect that they should not be close to $\pm 1$ in general---an exception can happen if an extreme hierarchy exists among the WCs. Thus, one can separate the four cases I--IV from others by measuring $A_{L}^e$ and $A_{L}^p$. 
Meanwhile, based on the combined results of $A_{L}^e$ and $A_{L}^p$, these four cases are already disentangled from each other as well. 

\begin{table}[t]
	\centering
	\renewcommand\arraystretch{1.6}
	\setlength\tabcolsep{6pt}
	\begin{tabular}{ccccc} \hline\hline
		&I &II &III&IV\\  
		\hline
		$A_{L}^e$& $1$ & $1$ & $-1$ & $-1$ 
		\\ %\hline
		$A_{L}^p$& $-0.998$ & $0.975$ & $-0.975$ & $0.998$ 
		\\ \hline\hline
	\end{tabular}
	\caption{Spin asymmetries $A_{L}^e$ and $A_{L}^p$ in the cases I--IV.}
	\label{tab:result1}
\end{table}

\begin{table*}[t] 
	\centering
	\renewcommand\arraystretch{1.6}
	\setlength\tabcolsep{6pt}
	\begin{tabular}{cccccccccccccccc} \hline\hline
		&V&VI&VII&VIII&IX&X&XI&XII&XIII&XIV&XV\\  
		\hline $A_{L3}^{ep} $& $c_1$ & $-0.998$ & $-0.998$ & $0.975$  & $0.975$& $0$ & $c_1$ & $c_1$  &$-0.998$& $0.975$ & $c_1$  
		\\ %\hline 
		$A_{L6}^{ep} $& $0$ & $0.975$ & $-0.998$ & $0.975$ & $-0.998$ & $c_1$ & $0.975$ & $-0.998$  &$c_1$& $c_1$ & $c_1$  
		\\ \hline\hline
	\end{tabular}
	\caption{Spin asymmetries $A_{L3}^{ep}$ and $A_{L6}^{ep}$ in the cases V--XV, where the constant $c_1$ takes any value within the range $[-1,1]$, but is not expected to be close to $\pm 1$ or 0 in general; see text for details.}
	\label{tab:resultV}
\end{table*}

To further disentangle the remaining 11 cases, one can exploit $A_{L3}^{ep}$ and $A_{L6}^{ep}$. The complete results for the cases V--XV are presented in Table~\ref{tab:resultV}, where the constant $c_1$, depending on $|g_V^{LL}|^2$, $|g_V^{LR}|^2$, and $\text{Re}[g_V^{LL}g_V^{LR*}]$, is within $[-1,1]$, but not expected to be close to $\pm 1$ or 0 in general. It is now clear from Table~\ref{tab:resultV} that measurements of $A_{L3}^{ep}=1, 0, -1, c_1$ can divide the remaining 11 cases into four subgroups, i.e., (VIII, IX, XIV), (X), (VI, VII, XIII), and (V, XI, XII, XV), accordingly. Except the already distinguished case X with $A_{L3}^{ep}=0$, all the other subgroups consist of at least three cases, which can be further disentangled by measurements of $A_{L6}^{ep}=1, 0, -1, c_1$. Thus, all the 15 scenarios are fully identified by measuring the four spin asymmetries, $A_{L}^e$, $A_{L}^p$, $A_{L3}^{ep}$, and $A_{L6}^{ep}$.  
It is also interesting to note that certain repeated entries appear in both Tables~\ref{tab:resultV} and \ref{tab:result1}. 
This is because the electron polarization in $A_{L3}^{ep}$ ($A_{L6}^{ep}$) remains intact, 
so that the operators it automatically selects in the cases V--XV can be identical to the ones in the cases I--IV. Let us illustrate this point with the case VI. Although the two operators $j^L_VJ^L_V$ and $j^R_VJ^L_V$ emerge in this case, only the former contributes to $A_{L3}^{ep}$, rendering $A_{L3}^{ep}$ to be equivalent to $A^p_L$ in the case I.  

We are now ready to apply the model-independent results to the seven concrete LQ models listed in Table~\ref{tab:LQ}, 
$S_1$, $R_2$, $S_3$, $U_3$, $\widetilde V_2$, $\widetilde U_1$, and $V_2$, which correspond to the cases VII, VIII, I, I, II, IV, and III, respectively. 
Obviously, all the LQ models can be fully identified by the four spin asymmetries, except $S_3$ and $U_3$ since they generate the same effective operator (see Table~\ref{tab:coeff}). Thus, other distinct features between them are necessary to distinguish $S_3$ from $U_3$. To this end, an interesting feature is that $S_3$ can mediate proton decays at tree level while $U_3$ does not~\cite{Arnold:2013cva,Gardner:2018azu,Assad:2017iib}. 
Given the strong constraints from proton stability (or generally the $|\Delta B|=1$ processes)~\cite{Zyla:2020zbs}, the 
LQ model $S_3$ could be, therefore, severely constrained, if no additional symmetry is invoked. 

Thus far, our analyses have been carried out with the polarized total cross sections. 
It may be, however, more convenient to work with the polarized differential cross sections. 
As can be inferred from the discussions above, 
$A_{L}^{e}$ in the cases I--IV is always equal to $\pm 1$, irrespective of whether it is formulated in terms of the total or the differential cross sections.
Besides, the repeated entries in Tables~\ref{tab:result1} and \ref{tab:resultV} can be actually understood from $A_{L}^{p}$, which in turn indicates that only $A_{L}^{p}$ in the cases I--IV needs a special attention---certainly, $c_1$ changes as well, but is expected to be still not too close to $0$ or $\pm 1$. In what follows, we will thus explore how $A_{L}^{p}$ and its associated polarized differential cross sections vary within the available kinematic region in the cases I--IV.     

\begin{figure}[t]
	\centering
	\subfigure{\includegraphics[width=0.48\textwidth]{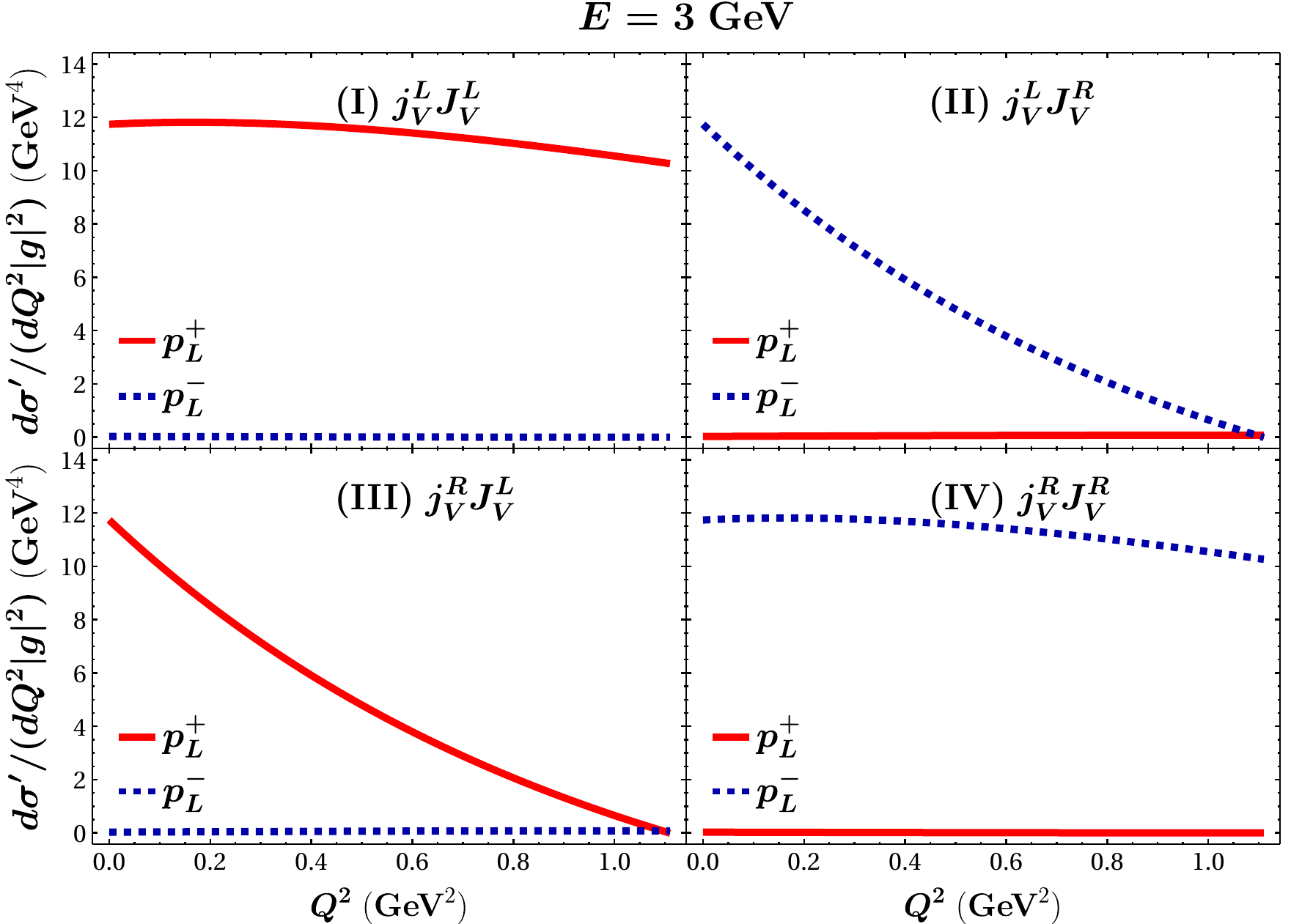}}	
	\caption{Longitudinally polarized differential cross sections $d\sigma^\prime/(dQ^2|g|^2)$ in the cases I--IV, with $d\sigma^\prime=(256\pi m^2_p E^2)d\sigma^{\pm}_p$. Red (solid) and blue (dashed) curves are obtained with a right- and a left-handed polarized proton target, respectively.} 
	\label{fig:dsigmap} 
\end{figure} 

As shown in the left two panels of Fig.~\ref{fig:dsigmap}, when the proton target is left-handed polarized (blue (dashed) curves), the longitudinally polarized differential cross section $d\sigma^\prime/(dQ^2|g|^2)$ is severely suppressed in the cases I and III. When the proton target is right-handed polarized (red (solid) curves), on the other hand, $d\sigma^\prime/(dQ^2|g|^2)$ is roughly a constant in the case I, but decreases rapidly in the case III as $Q^2$ approaches $Q^2_{\text{max}}$. The same observations also hold for the cases II and IV but with the polarization direction of the proton target flipped, as shown in the right two panels of Fig.~\ref{fig:dsigmap}. Based on these observations, two conclusions can be drawn immediately. First, $A_{L}^{p}$ in the case I (IV) remains $-1$ ($+1$) within the whole available kinematic region 
$[Q^2_{\text{min}},\,Q^2_{\text{max}}]$. Second, $A_{L}^{p}$ in the case II (III) remains $+1$ ($-1$) within all the available kinematic regions except that close to $Q^2_{\text{max}}$. 
Thus, to avoid any misidentification, measuring $A_{L}^{p}$ at the low-$Q^2$ region is more favored; besides, more events are expected in the very same region.  

\begin{figure}[t]
	\centering
	\includegraphics[width=0.45\textwidth]{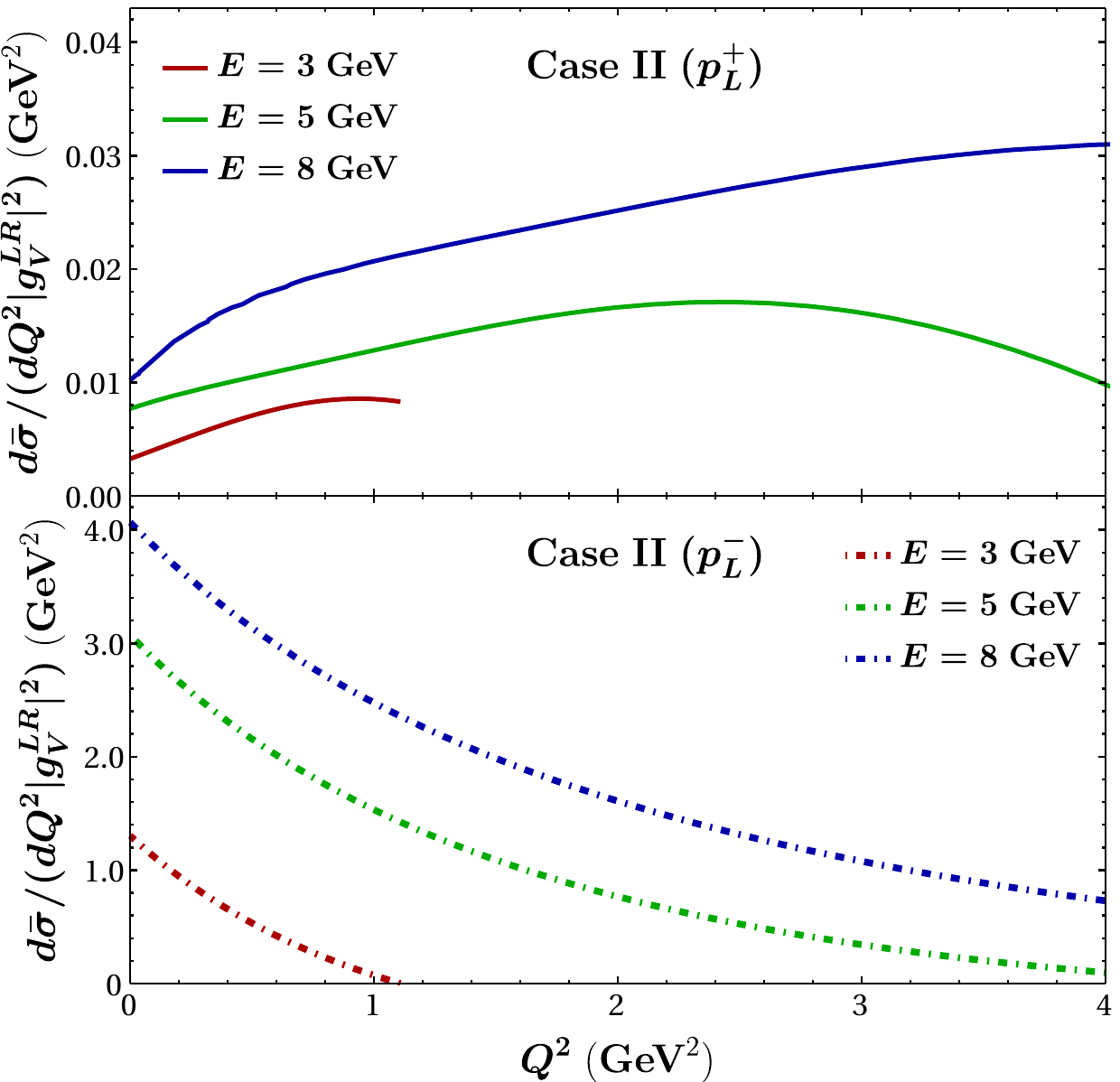}
	\caption{Longitudinally polarized differential cross sections $d\bar{\sigma}/(dQ^2|g^{LL}_V|^2)$ in the case II, with three different electron beam energies, where $d\bar{\sigma} =(256\pi m^2_p)d\sigma^{\pm}_p=d\sigma^\prime/E^2$. }
	\label{fig:Evary} 
\end{figure}

Finally, we explore the dependence of $A_L^p$ on the electron beam energy $E$. To this end, let us take here the case II as an example, due to the distinct feature of its longitudinally polarized differential cross sections (cf. the upper-right plot of Fig.~\ref{fig:dsigmap}). As shown in the upper plot of Fig.~\ref{fig:Evary}, the differential cross section $d\bar{\sigma}/(dQ^2|g^{LR}_V|^2)$, though being enhanced by a small amount with the increase of $E$, still remains severely suppressed, when the proton target is right-handed polarized ($p_L^+$). For the scattering process with a left-handed polarized proton target ($p_L^-$), on the other hand, its longitudinally polarized differential cross section is enhanced as well along with the increase of $E$, but its characteristically decreasing trend with the increase of $Q^2$ persists (see the lower plot of Fig.~\ref{fig:Evary}). As a consequence, the conclusions drawn above about the behaviors of the observable $A_{L}^{p}$ in the case II, as well as in the cases I, III, and IV, 
always hold.  

\subsection{Identification of the scalar and vector contributions in the LQ models}

In the analyses above, we have neglected contributions to the polarized scattering processes 
from the scalar and tensor operators, because they are severely constrained by the leptonic $D$-meson decays~\cite{Petric:2010yt,Lai:2021sww}. Nevertheless, it is still interesting to ask if the constraints on both the vector and scalar contributions reach a similar level in the future---so that both matter---is it possible to fully distinguish them through the low-energy polarized scattering processes $e^-p\to e^-\Lambda_c$ in the LQ models? In what follows, we will demonstrate explicitly that the answer is yes for certain LQ models.

\begin{figure}[t]
	\centering
	\includegraphics[width=0.48\textwidth]{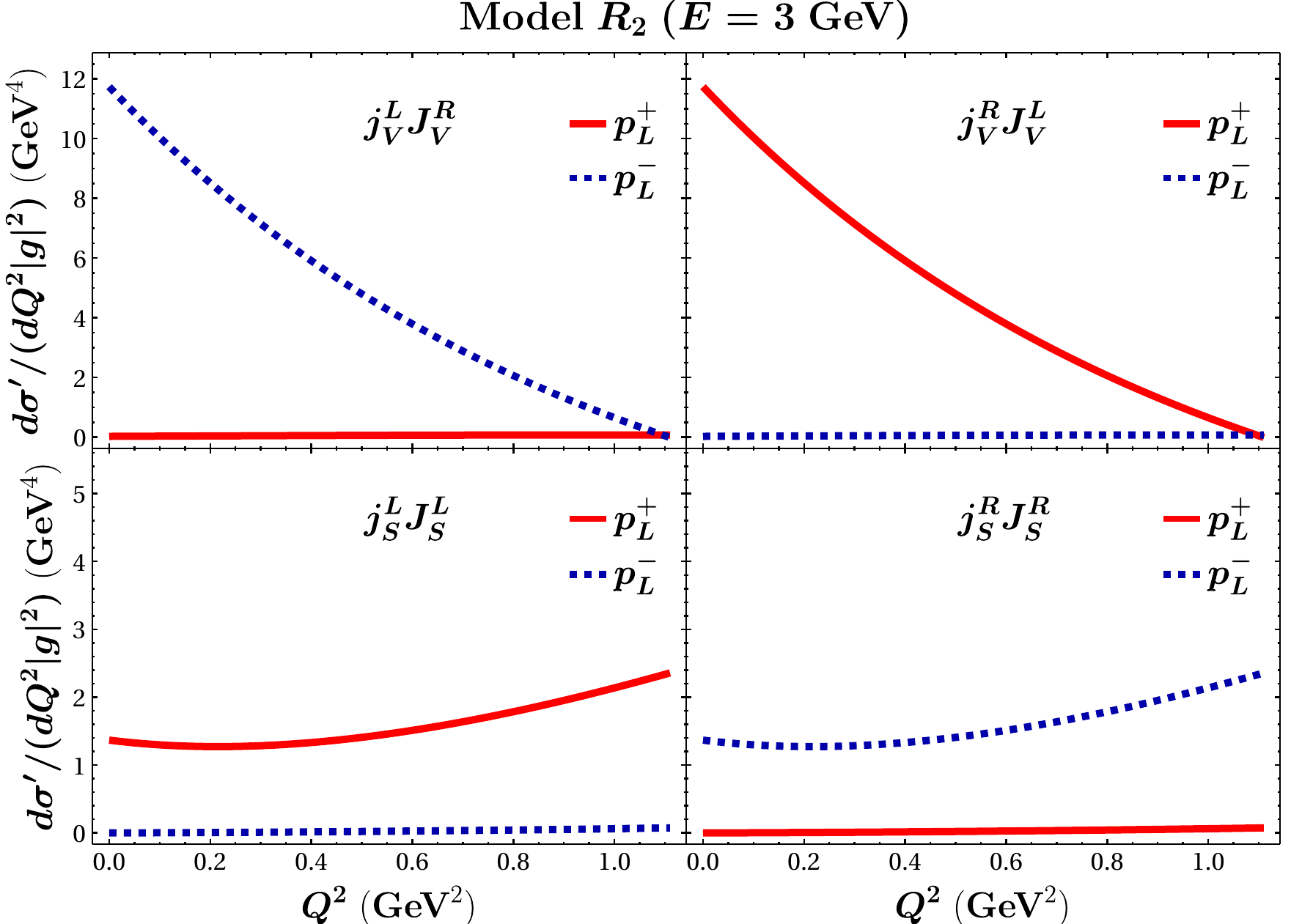}
	\caption{Longitudinally polarized differential cross sections $d\sigma^\prime/(dQ^2|g|^2)$ mediated by the four effective operators in the LQ model $R_2$, with $d\sigma^\prime =(256\pi m^2_p E^2)d\sigma^{\pm}_p$. } 
	\label{fig:s1r2}
\end{figure}

As shown in Table~\ref{tab:coeff}, there are only two LQ models, $S_1$ and $R_2$, that can generate both scalar and vector operators---since contributions from the tensor and scalar operators are connected through Eq.~\eqref{eq:g_denfi_1} at high energy 
or through Eq.~\eqref{eq:RG_R2} at low energy, we will present them universally in terms of the scalar ones. Let us start with the model $R_2$. The resulting differential cross sections $d\sigma^\prime/(dQ^2|g|^2)$ of the longitudinally polarized scattering processes mediated 
by the operators $j^L_VJ^R_V$, $j^R_VJ^L_V$, $j^L_SJ^L_S$ ($j^L_TJ^L_T$), and $j^R_SJ^R_S$ ($j^R_TJ^R_T$) are shown in Fig.~\ref{fig:s1r2}. It can be seen that if the electron beam is left-handed polarized, only the operators $j^L_VJ^R_V$ and $j^L_SJ^L_S$ ($j^L_TJ^L_T$) contribute to the polarized scattering processes, with the corresponding differential cross sections $d\sigma^\prime/(dQ^2|g|^2)$ depicted by the two plots on the left; if the electron beam is right-handed polarized, on the other hand, only the operators $j^R_VJ^L_V$ and $j^R_SJ^R_S$ ($j^R_TJ^R_T$) are at work, whose $d\sigma^\prime/(dQ^2|g|^2)$ are presented by the two plots on the right.

From the two plots on the left of Fig.~\ref{fig:s1r2}, one can see that if the proton target is left-handed polarized (blue (dashed) curves), contribution from the operator $j^L_SJ^L_S$ ($j^L_TJ^L_T$) can be neglected, and thus one actually probes the contribution from the operator $j^L_VJ^R_V$. Then, flipping the polarization direction of the proton target but keeping the electron beam intact, one probes the contribution from the operator $j^L_SJ^L_S$ ($j^L_TJ^L_T$), since the contribution from the operator $j^L_VJ^R_V$ in this case becomes trivial. In this way, the contributions from the two operators $j^L_SJ^L_S$ ($j^L_TJ^L_T$) and $j^L_VJ^R_V$ are distinguished from each other. Following the same procedure but with the polarization direction of the electron beam flipped, one can also identify the contributions from the other two operators $j^R_SJ^R_S$ ($j^R_TJ^R_T$) and $j^R_VJ^L_V$ efficiently. 

Such a scheme, however, cannot be applied to the LQ model $S_1$. This is due to the observation that both the operators $j^L_VJ^L_V$ and $j^L_SJ^L_S$ ($j^L_TJ^L_T$) contribute significantly to the longitudinally polarized scattering process with $e^-_L p^+_L$ (red (solid) curves), as can be inferred from Figs.~\ref{fig:dsigmap} and \ref{fig:s1r2}. On the other hand, both of their contributions become trivial, when the electron beam and proton target are both left-handed polarized (blue (dashed) curves). 

Clearly, this scheme also fails if an extreme hierarchy arises between the scalar (tensor) and vector contributions. Unfortunately, the current experimental constraints on them seem to imply this~\cite{Lai:2021sww}. Thus, to ensure this procedure to be at work, it would be better that the constraints on both the vector and scalar contributions can reach a similar level in the future.

\section{\boldmath Transversely polarized scattering processes $e^-p\to e^- \Lambda_c$}
\label{sec:TPscatter}

We now turn to discussing the low-energy scattering processes $e^-p\to e^- \Lambda_c$ 
with the proton target transversely polarized---the electron beam will always be assumed to be longitudinally polarized throughout this work. As explained already in Sec.~\ref{sec:models}, the polarization vector $\xi_p^{\mu}$ of the transversely polarized proton depends on $\beta$, the azimuthal angle between the proton spin directon and the x axis. In what follows, we will, for simplicity, set that the proton target is polarized along the x axis, so that $\beta=0, \pi$, and $\xi_p^{\mu}=\pm(0,1,0,0)$ accordingly.

\subsection{Observable analyses}

Similar to the longitudinally polarized case presented in the previous section, 
it is also more convenient to discuss the double-spin polarized cross sections $\widetilde{\sigma}^{ab}$ with $a, b=\pm$, since $A^p_T$ in Eq.~\eqref{eq:LT_single_PV} and $A^{ep}_{T1-T6}$ in Eqs.~\eqref{eq:T_double_PV} and \eqref{eq:T_double_PC}  
are all built in terms of them. In the framework of the general low-energy effective Lagrangian $\mathcal{L}_{\text{eff}}$, $\widetilde{\sigma}^{ab}$ can be conveniently defined in the same way as for $\sigma^{ab}$ in Eq.~\eqref{eq:dsimgaL}, except that  $|\mathcal{M}^\prime|^2$ should be now replaced by $|\widetilde{\mathcal{M}}|^2$, 
where the latter represents the $\xi_p$-dependent reduced amplitude squared.  
Thus, we will dive into the reduced amplitudes squared immediately.

\begin{table*}[t]
	\centering
	\renewcommand*{\arraystretch}{1.8}
	\tabcolsep=0.35cm
	\begin{tabular}{ccc}
		\hline\hline
		& $\left(g^{LL}_{V}\right)^*$ &  $\left(g^{LR}_{V}\right)^*$  \\
		\hline
		$g^{LL}_{V}$   &   2$\overline{|\mathcal{M}|}^2_{V_{LL}-V_{LL}}+|\mathcal{\widetilde{M}}|^{2}_{V_{LL}-V_{LL}}$  & 2$\overline{|\mathcal{M}|}^2_{V_{LL}-V_{LR}}+|\mathcal{\widetilde{M}}|^{2}_{V_{LL}-V_{LR}}$   \\
		%\hline
		$g^{LR}_{V}$ &  2$\overline{|\mathcal{M}|}^2_{V_{LL}-V_{LR}}+|\mathcal{\widetilde{M}}|^{2*}_{V_{LL}-V_{LR}}$   &   2$\overline{|\mathcal{M}|}^2_{V_{LR}-V_{LR}}+|\mathcal{\widetilde{M}}|^{2}_{V_{LR}-V_{LR}}$     \\
		\hline\hline
		&  $\left(g^{L}_{S}\right)^*$ &  $\left(g^{L}_{T}\right)^*$       \\
		\hline
		$g^{L}_{S}$ &  2$\overline{|\mathcal{M}|}^2_{S_{L}-S_{L}}+|\mathcal{\widetilde{M}}|^{2}_{S_{L}-S_{L}}$   & 2$\overline{|\mathcal{M}|}^2_{S_{L}-T_{L}}+|\mathcal{\widetilde{M}}|^{2}_{S_{L}-T_{L}}$   \\
		%\hline
		$g^{L}_{T}$ &2$\overline{|\mathcal{M}|}^2_{S_{L}-T_{L}}+|\mathcal{\widetilde{M}}|^{2*}_{S_{L}-T_{L}}$   & 2$\overline{|\mathcal{M}|}^2_{T_{L}-T_{L}}+|\mathcal{\widetilde{M}}|^{2}_{T_{L}-T_{L}}$ \\
		\hline\hline  \\[-0.1cm]
		\hline\hline
		& $\left(g^{RR}_{V}\right)^*$ &  $\left(g^{RL}_{V}\right)^*$ \\
		\hline 
		$g^{RR}_{V}$   &   2$\overline{|\mathcal{M}|}^2_{V_{LL}-V_{LL}}-|\mathcal{\widetilde{M}}|^{2}_{V_{LL}-V_{LL}}$  &  2$\overline{|\mathcal{M}|}^2_{V_{LL}-V_{LR}}-|\mathcal{\widetilde{M}}|^{2*}_{V_{LL}-V_{LR}}$   \\
		%\hline 
		$g^{RL}_{V}$ &  2$\overline{|\mathcal{M}|}^2_{V_{LL}-V_{LR}}-|\mathcal{\widetilde{M}}|^{2}_{V_{LL}-V_{LR}}$   &   2$\overline{|\mathcal{M}|}^2_{V_{LR}-V_{LR}}-|\mathcal{\widetilde{M}}|^{2}_{V_{LR}-V_{LR}}$     \\
		\hline\hline		
		&  $\left(g^{R}_{S}\right)^*$ &  $\left(g^{R}_{T}\right)^*$       \\
		\hline
		$g^{R}_{S}$ & 2$\overline{|\mathcal{M}|}^2_{S_{L}-S_{L}}-|\mathcal{\widetilde{M}}|^{2}_{S_{L}-S_{L}}$   & 2$\overline{|\mathcal{M}|}^2_{S_{L}-T_{L}}-|\mathcal{\widetilde{M}}|^{2*}_{S_{L}-T_{L}}$   \\
		\hline
		$g^{R}_{T}$ &  2$\overline{|\mathcal{M}|}^2_{S_{L}-T_{L}}-|\mathcal{\widetilde{M}}|^{2}_{S_{L}-T_{L}}$   & 2$\overline{|\mathcal{M}|}^2_{T_{L}-T_{L}}-|\mathcal{\widetilde{M}}|^{2}_{T_{L}-T_{L}}$ \\
		\hline\hline		
	\end{tabular}
	\caption{Non-zero reduced amplitudes squared of the polarized scattering processes with $e^{-}_L p^-_T$ (the upper two tables) and $e^{+}_Lp^-_T$ (the lower two tables). Flipping the sign of the $\xi_p$-dependent $|\mathcal{\widetilde{M}}|^2_{\alpha-\beta}$ yields the corresponding amplitudes squared for the processes with $e^{-}_Lp^+_T$ and $e^{+}_Lp^+_T$. }
	\label{tab:crossT}
\end{table*}

Compared with the reduced amplitudes squared in Table~\ref{tab:crossp}, these in Table~\ref{tab:crossT} exhibit certain similarities. For instance, the $\xi_p$-independent parts of the reduced amplitudes squared induced by the operators $j^{L}_{V}J^{L}_{V}$ and $j^{R}_{V}J^{R}_{V}$ are the same, whereas the $\xi_p$-dependent parts are opposite in sign, 
rendering that the observable $A^p_T$ induced by the operator $j^{L}_{V}J^{L}_{V}$ is opposite in sign to that by the operator $j^{R}_{V}J^{R}_{V}$. The same conclusion also applies to other pairs, like ($j^{L}_{V}J^{R}_{V}$, $j^{R}_{V}J^{L}_{V}$), 
($j^{L}_{S}J^{L}_{S}$, $j^{R}_{S}J^{R}_{S}$), and ($j^{L}_{T}J^{L}_{T}$, $j^{R}_{T}J^{R}_{T}$).
Besides, the $\xi_p$-independent parts induced by the same operators in both cases 
are identical to each other, due to $\sigma^{a}_e=(\sigma^{a+}+\sigma^{a-})/2
=(\widetilde{\sigma}^{a+}+\widetilde{\sigma}^{a-})/2$, with $a=\pm$. 

Of course, there exist some differences between the two cases. In contrast to the longitudinally polarized case, the $\xi_p$-dependent reduced amplitudes squared induced by the interference between two different operators, e.g., ($j^{L}_{V}J^{L}_{V}$, $j^{L}_{V}J^{R}_{V}$), ($j^{R}_{V}J^{R}_{V}$,  $j^{R}_{V}J^{L}_{V}$), ($j^{L}_{S}J^{L}_{S}$, $j^{L}_{T}J^{L}_{T}$), 
and ($j^{R}_{S}J^{R}_{S}$, $j^{R}_{T}J^{R}_{T}$), are complex---though the $\xi_p$-independent ones are still real. Such a distinct feature leads to an interesting application, as will be shown later. Another difference is that $|\widetilde{\mathcal{M}}|^2$ depends explicitly on the trigonometric functions of $\phi$, the azimuthal angle between the direction of the scattered electrons and the x axis in the lab frame, whereas $|\mathcal{M}^\prime|^2$ 
does not. This means that $A^p_T$ and the other six spin asymmetries, as well as their associated 
transversely polarized differential cross sections, all depend on $\phi$ as well---here we have implicitly concentrated on the differential cross sections, since integrating over $\phi$ will obliterate the polarization effects.

\begin{figure}[t]
	\centering
    \subfigure{\begin{minipage}{0.238\textwidth}
    \centering
	\includegraphics[width=1.68in]{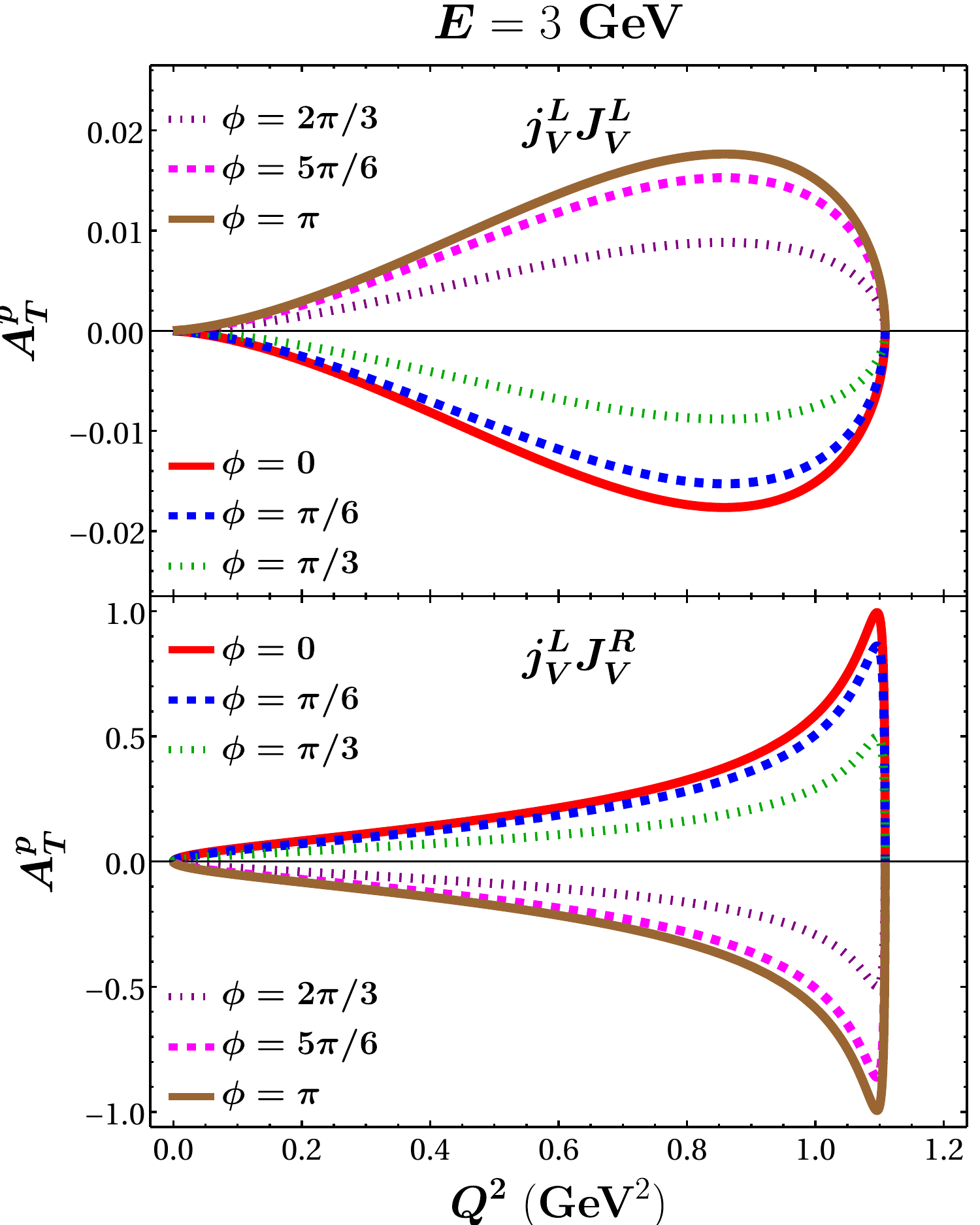}
    \end{minipage}}\hspace{0.018cm}
    \subfigure{\begin{minipage}{0.238\textwidth}
    \centering
	\includegraphics[width=1.60in]{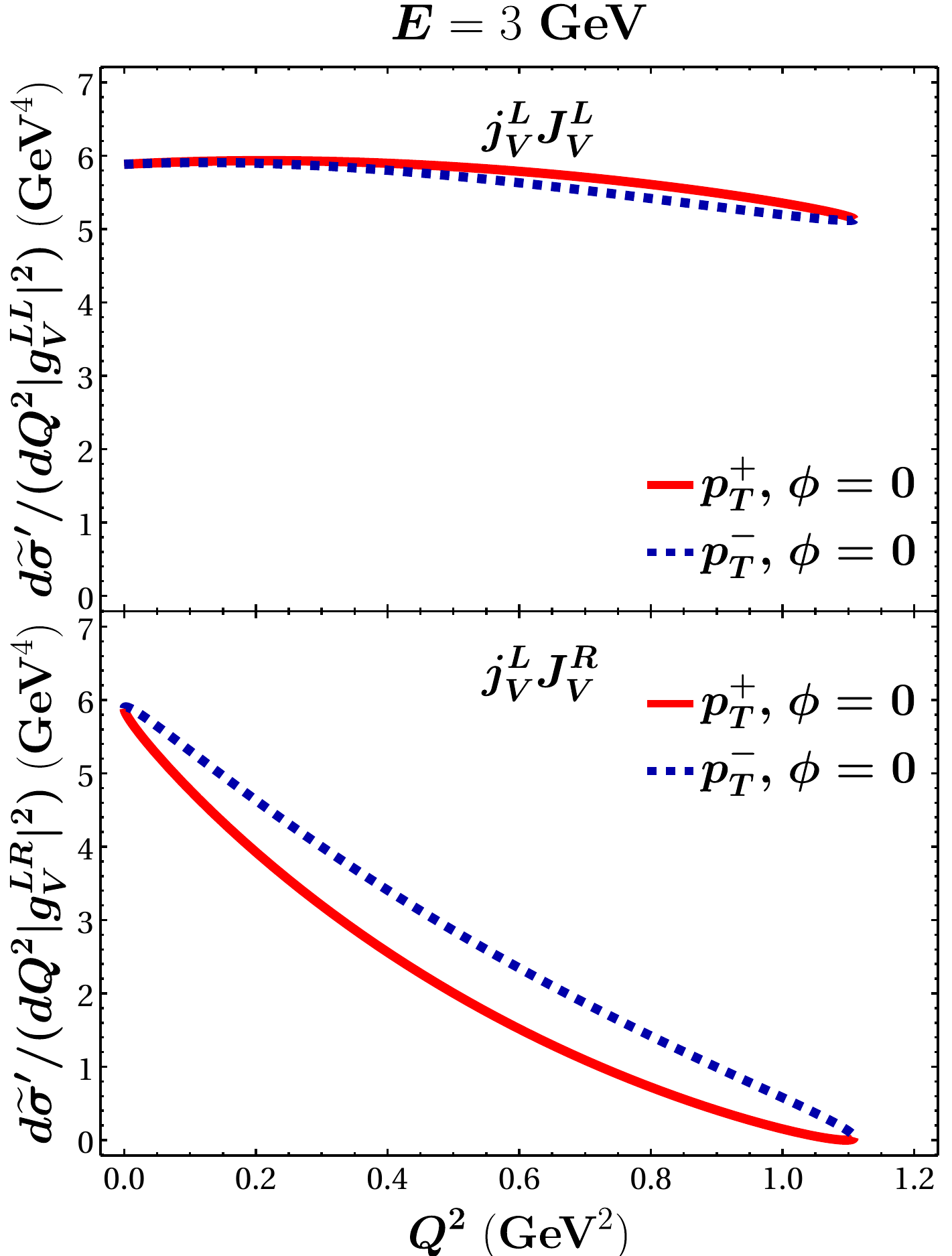}
    \end{minipage}}
    \caption{Transversely polarized spin asymmetry $A^p_T$ (the left panel) and differential cross sections $d\widetilde{\sigma}/(dQ^2|g|^2)$ (the right panel) for the operators $j^L_VJ^L_V$ (the upper plots) and $j^L_VJ^R_V$ (the lower plots), with $d\widetilde{\sigma} =(256\pi m^2_p E^2)d\widetilde{\sigma}^{\pm}_p$.}
    \label{fig:ATp}
\end{figure}

With the reduced amplitudes squared (or the differential cross sections) at hand, we can 
now explore the prospect for identifying NP models by using the low-energy transversely polarized $ep$ scattering processes. Following the same procedure as in the longitudinally polarized case, we first compute $A^p_T$ induced by the vector operators $j^L_VJ^L_V$, $j^L_VJ^R_V$, $j^R_VJ^L_V$, and $j^R_VJ^R_V$ (i.e., the cases I--IV) with the benchmark $E=3$ GeV. Due to the interesting relations among $A^p_T$ induced by the four vector operators, only two spin asymmetries are independent. In what follows, we will thus take the operators $j^L_VJ^L_V$ (case I) and $j^L_VJ^R_V$ (case II) for an illustration. It should be noted that 
contributions from the scalar and tensor operators will not be considered 
due to the same reason as explained in the longitudinally polarized case.

The variations of $A^p_T$ in the cases I and II with respect to $Q^2$ are shown, respectively, by the upper and the lower plot in the left panel of Fig.~\ref{fig:ATp}. From these two figures, one can immediately make two interesting observations. 
First, $A^p_T$ in both cases exhibits a periodic pattern with respect to $\phi$.   
This is because the $\xi_p$-dependent $|\widetilde{\mathcal{M}}|^2$
is proportional to $\cos\phi$, as shown in Appendix~\ref{app:TpAmplitude}. 
Such a periodic characteristic indicates that one cannot explore $A^p_T$ by exploiting the  polarized total cross sections, as already pointed out above.  
Second, $A^p_T$ in both cases are generally suppressed,  
particularly in the former. It can be seen that $|A^p_T|$ is always trivial, practically irrespective of $\phi$ and $Q^2$ in the case I, while is below $0.5$ in most of the available kinematic region in the case II, even with $\phi=0, \pi$ where the polarization effects are maximal. 

\begin{figure}[t]
	\centering
	\includegraphics[width=0.45\textwidth]{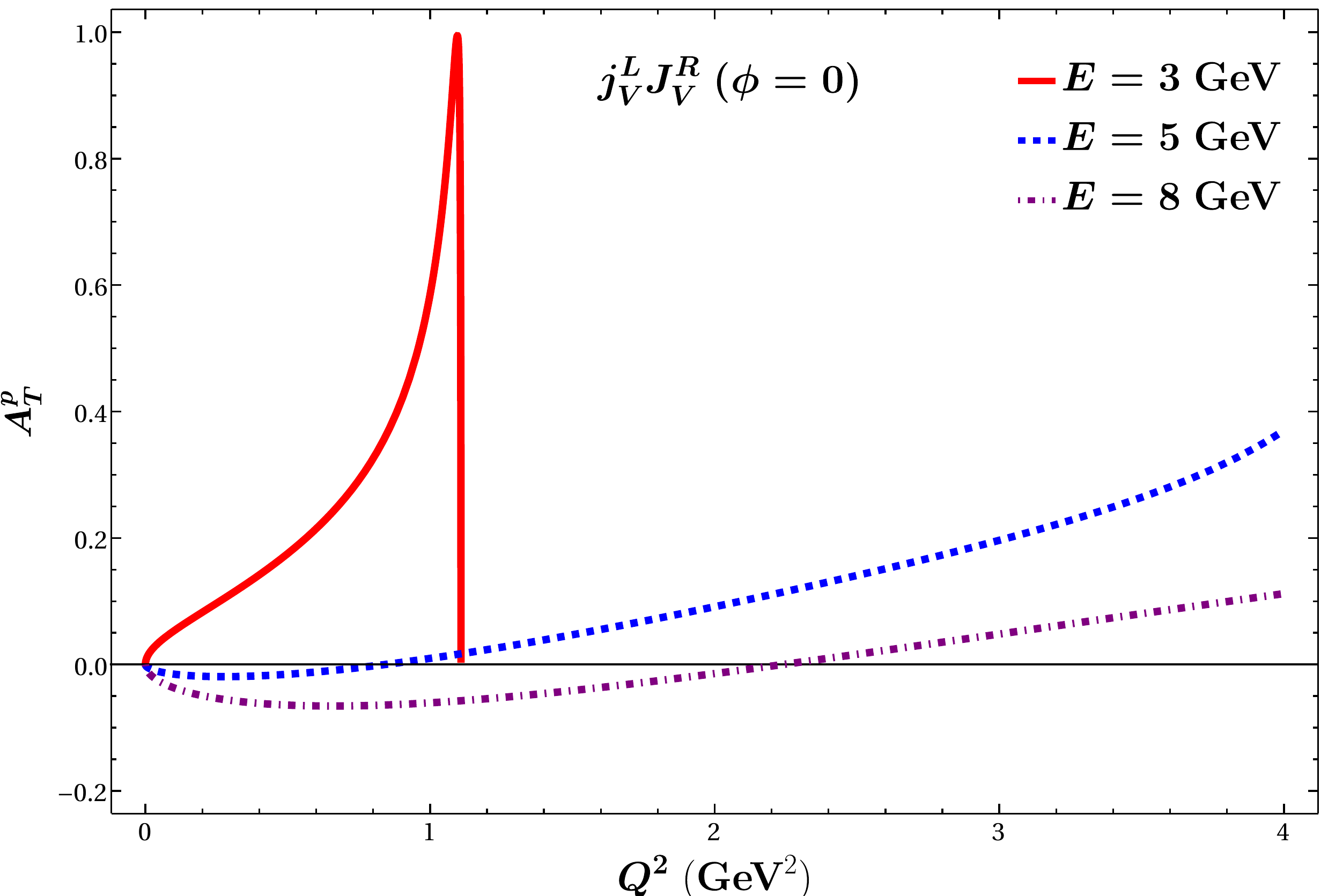}
	\caption{Transversely polarized spin asymmetry $A^p_T$ in the case II with different electron beam energies.}
	\label{fig:caseIIEvary} 
\end{figure} 

Given that the polarization effects are far less prominent in the transversely polarized $ep$ scattering processes in the cases I--IV, the mechanism previously used to distinguish the NP models (the cases I--XV) becomes less applicable now. One may argue about this by pointing out that $|A^p_T|$ in the case II can reach a relatively high value in the high-$Q^2$ region. 
However, such a region could easily exceed the upper limit of $Q^2$, where the theoretical framework fails---particularly for the high electron beam energy $E$~\cite{Lai:2021sww}. 
Indeed, as shown in Fig.~\ref{fig:caseIIEvary}, high $|A^p_T|$ can no longer be reached in the 
available $Q^2$ region as $E$ increases. Furthermore, the transversely polarized differential cross sections $d\widetilde{\sigma}^\prime$ 
decrease rapidly as $Q^2$ increases, as explicitly demonstrated (especially the lower plot) in the right panel of Fig.~\ref{fig:ATp}.    
Thus, measuring a non-zero $A^p_T$ becomes very demanding, which, nevertheless, makes the 
mechanism less appealing to this work.

\subsection{Direct probe of the imaginary part of the WCs}

Although the transversely polarized $ep$ scattering processes may be less convenient to distinguish the NP models, they can directly probe into the imaginary part of the WCs, 
which is currently not possible (or at least not optimal) for the rare FCNC decays of charmed hadrons, the resonant searches in the high-$p_T$ dilepton invariant mass tails, as well as the low-energy unpolarized (and longitudinally polarized) $ep$ scattering processes, since their associated reduced amplitudes squared are all real.  

As mentioned before and also explicitly shown in Appendix~\ref{app:TpAmplitude}, 
the $\xi_p$-dependent reduced amplitudes squared in Table~\ref{tab:crossT}
induced by the interference between a pair of different operators, 
($j^{L}_{V}J^{L}_{V}$, $j^{L}_{V}J^{R}_{V}$), 
($j^{R}_{V}J^{R}_{V}$, $j^{R}_{V}J^{L}_{V}$), 
($j^{L}_{S}J^{L}_{S}$, $j^{L}_{T}J^{L}_{T}$), 
and ($j^{R}_{S}J^{R}_{S}$, $j^{R}_{T}J^{R}_{T}$), 
consist of both the real and imaginary parts, with the former proportional to $\cos\phi$, while the latter to $\sin\phi$. To be as general as possible, we will consider the transversely polarized $ep$ scattering processes in 
the framework of the general low-energy effective Lagrangian $\mathcal{L}_{\text{eff}}$, 
but with contributions from the scalar operators neglected due to the stringent constraints 
from the leptonic $D$-meson decays~\cite{Petric:2010yt,Lai:2021sww}.

We will first consider the polarized $ep$ scattering process with $e^-_L p^-_T$. Its amplitude squared $|\mathcal{M}|^2$ can be written as
\begin{align}
|\mathcal{M}|^2=&|\mathcal{M}_{V_{LL}}|^2+|\mathcal{M}_{V_{LR}}|^2+|\mathcal{M}_{T_{L}}|^2\nonumber\\[0.12cm]
&+\mathcal{M}_{V_{LL}}\mathcal{M}^*_{V_{LR}}+\mathcal{M}^*_{V_{LL}}\mathcal{M}_{V_{LR}}\,, 
\end{align}
where the first three non-interference terms are given, respectively, by 
\begin{align}
|\mathcal{M}_{V_{LL}}|^2&=|g_{V}^{LL}|^2\left(
2\overline{|\mathcal{M}|}^2_{V_{LL}-V_{LL}}+|\mathcal{\widetilde{M}}|^{2}_{V_{LL}-V_{LL}}\right)\,,\nonumber\\[0.12cm]
|\mathcal{M}_{V_{LR}}|^2&=|g_{V}^{LR}|^2\left( 2\overline{|\mathcal{M}|}^2_{V_{LR}-V_{LR}}+|\mathcal{\widetilde{M}}|^{2}_{V_{LR}-V_{LR}}\right)\,,\nonumber\\[0.12cm]
|\mathcal{M}_{T_{L}}|^2&=|g_{T}^{L}|^2\left(2
\overline{|\mathcal{M}|}^2_{T_{L}-T_{L}}+|\mathcal{\widetilde{M}}|^{2}_{T_{L}-T_{L}}\right)\,,
\end{align}
which depend on $q^2$ and $\cos\phi$. The last two interference terms, on the other hand, can be written as 
\begin{align}
&\mathcal{M}_{V_{LL}}\mathcal{M}^*_{V_{LR}}+\mathcal{M}^*_{V_{LL}}\mathcal{M}_{V_{LR}}\nonumber \\[0.12cm]
&=2\,\text{Re}[g_V^{LL}g_V^{LR*}]\mathcal{A}-2\,\text{Im}[g_V^{LL}g_V^{LR*}]\mathcal{A}^\prime \sin\phi\,,
\end{align}
where $\mathcal{A}$, which consists of $2\overline{|\mathcal{M}|}^2_{V_{LL}-V_{LR}}$ and 
the real part of $|\mathcal{\widetilde{M}}|^{2}_{V_{LL}-V_{LR}}$, is also dependent on $q^2$ and $\cos\phi$, whereas $\mathcal{A}^\prime$, the imaginary part of  $|\mathcal{\widetilde{M}}|^{2}_{V_{LL}-V_{LR}}$ with $\sin\phi$ being factored out, depends exclusively on $q^2$. Focusing on the small $\phi$ region, we can then write the amplitude squared $|\mathcal{M}|^2$ approximately as
\begin{align}
|\mathcal{M}|^2\approx \mathcal{A}_1(q^2)-\mathcal{A}_2(q^2)\,\phi\,, \label{eq:Mline}
\end{align}
with 
\begin{align}
\mathcal{A}_1&\!=\!|\mathcal{M}_{V_{LL}}|^2\!+\!|\mathcal{M}_{V_{LR}}|^2\!+\!|\mathcal{M}_{T_{L}}|^2\!
+\!2\text{Re}[g_V^{LL}g_V^{LR*}]\mathcal{A}\,,\nonumber\\[0.12cm]
\mathcal{A}_2&\!=\!2\text{Im}[g_V^{LL}g_V^{LR*}]\mathcal{A}^\prime\,,
\end{align}
where $\mathcal{A}_{1,2}$ depend only on $q^2$ due to $\cos\phi\approx 1$.  
Now it can be seen that measuring $\mathcal{A}_2$, the slope of $|\mathcal{M}|^2$ in Eq.~\eqref{eq:Mline}, directly probes into the imaginary part of $g_V^{LL}g_V^{LR*}$. 
Similarly, the imaginary part of $g_V^{RR}g_V^{RL*}$ can be measured by flipping the polarization direction of the electron beam. Certainly, the same procedure can be again applied to $g_V^{LL}g_V^{LR*}$ and $g_V^{RR}g_V^{RL*}$ after flipping the polarization direction of the proton target. This is, however, unnecessary, since neither $\mathcal{A}_{1}$ nor $\mathcal{A}_{2}$ will change too much, as indicated by Fig.~\ref{fig:ATp}, 
and only the sign before $\mathcal{A}_{2}$ in Eq.~\eqref{eq:Mline} will be flipped. 

\begin{figure}[t]
	\centering
	\includegraphics[width=0.47\textwidth]{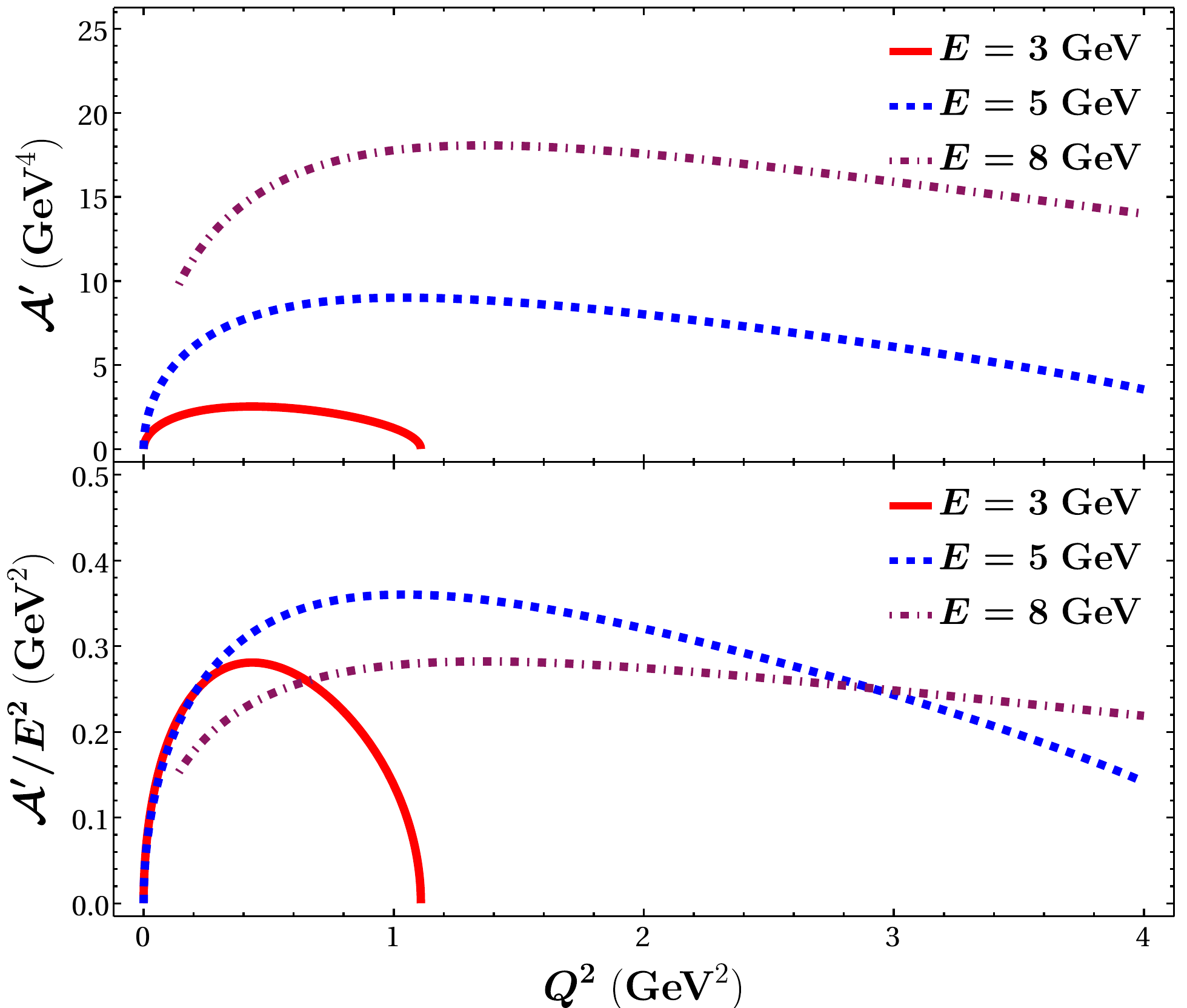}
	\caption{Slopes $\mathcal{A}^\prime$ and $\mathcal{A}^\prime/E^2$ with respect to the kinematics $Q^2$ and beam energy $E$, where both the electron beam and proton target are left-handed polarized.}
	\label{fig:slope} 
\end{figure} 

Given that the differential cross section $d^2\tilde{\sigma}/dQ^2d\phi$
can be conveniently connected to $|\mathcal{M}|^2$ through $d^2\tilde{\sigma}/dQ^2d\phi\!=\!|\mathcal{M}|^2/(128\pi^2 m^2_pE^2)$ (see Appendix~\ref{app:expquantity}), we thus explore in the following how the slopes $\mathcal{A}^\prime$ and $\mathcal{A}^\prime/E^2$ vary with respect to the kinematics $Q^2$ and the beam energy $E$. It can be seen from Fig.~\ref{fig:slope} that 
relatively low $Q^2$ is certainly favored. In particular, $Q^2\approx 0.4$~GeV$^2$ and $Q^2\approx 1$~GeV$^2$ are the two best kinematic regimes to measure $\mathcal{A}^\prime/E^2$ with $E=3$~GeV and $E=5,\,8$~GeV, respectively. It is also interesting to note that increasing $E$ will not be of as much benefit to $\mathcal{A}^\prime/E^2$ as to $\mathcal{A}^\prime$.   

Let us finally conclude this section by making the following comment. Since none of the LQ models, as shown in Table~\ref{tab:coeff}, can produce the interference between two different vector operators, a clear non-zero measurement of $\mathcal{A}^\prime/E^2$ will certainly exclude all the LQ models, unless at least two of them arise simultaneously in a given model.   

\section{Prospect and constraints}
\label{sec:Prospect}

All the mechanisms proposed in the previous two sections will be less appealing, if the polarized scattering processes $e^-p\to e^-\Lambda_c$ cannot be observed at all. In this section, we will show that, with our properly designed experimental setups, very promising event rates associated with the polarized scattering processes can be expected in the LQ models. Even if, unfortunately, no event were observed with the designed experimental setups in the future, the polarized $ep$ scattering processes would still yield competitive constraints, compared with the charmed-hadron weak decays and high-$p_T$ dilepton invariant mass tails.

\subsection{Experimental setups}

We will consider the fixed-target polarized scattering experiments. The event rate of the double-spin polarized scattering process $\vec{e}^{\,-} \vec{p} \to e^- \Lambda_c$ is given by~\cite{Ohlsen:1972zz,Donnelly:1985ry}
\begin{equation}\label{eq:double-spin-event}
\frac{dN}{dt}\!=\!\mathcal{L}a(\sigma_0\!+\! h_e P_e\sigma_e\!+\! fh_p P_p\sigma_p\!+\! fh_e h_p P_p P_e\sigma_{ep})\,,
\end{equation}
where $\sigma_0$ and $\sigma_{e,p,ep}$ denote the spin-independent and the spin-dependent cross sections of the process, respectively. The luminosity $\mathcal{L}$ is given by $\mathcal{L}=IL\kappa \rho_T$, where $I$ is the beam intensity, and $\kappa$, $L$, and $\rho_T$ are the packing factor, the length, and the number density of the proton target, respectively.~\footnote{Note that $\rho_T=\rho/m_p$, where $\rho$ is the density of the proton target.} For simplicity, the apparatus acceptance $a$ will be assumed to be $1$ in our later analyses. Note that $h_{e(p)}=\pm 1$ represent the right- and left-handed polarized electron beam (proton target), with $P_{e(p)}$ denoting the corresponding degree of polarization and $f$ the dilution factor of the target~\cite{SpinMuonSMC:1997mkb,Barone:2003fy}. Based on Eq.~\eqref{eq:double-spin-event}, the event rates of the single-spin polarized scattering processes, $\vec{e}^{\,-} p \to e^- \Lambda_c$ and $e^{\,-} \vec{p} \to e^- \Lambda_c$, can be obtained straightforwardly; we refer the readers to Appendix~\ref{app:expquantity} for further details. 

We now introduce the experimental setups designed for the polarized $ep$ scattering processes. Our desired longitudinally polarized electron beam is required to have a beam energy $E=3.48$~GeV, an intensity up to 100~$\mu$A, and a polarization of $89.4\%$ approximately; such a beam has been used by the HAPPEX collaboration at Jefferson Laboratory (JLab)~\cite{HAPPEX:2011xlw}. For the unpolarized proton target, we choose a liquid hydrogen (LH) one with a density $\rho=71.3\times 10^{-3}\text{g}/\text{cm}^3$. Such a target has been utilized in the Qweak experiment at JLab
with a $3$-$\text{kW}$ cooling power applied to break the target length limit induced by the heating problem~\cite{Allison:2014tpu}. Applying the same cooling system to our case, i.e., exposing the LH target to our desired electron beam, we find that the maximal LH target length is about $70$~cm. For the polarized proton target, on the other hand, we favor the UVa/JLab polarized solid ammonia (SNH$_3$) target, which has been extensively used in many experiments at JLab (see, e.g., Refs.~\cite{J.Arrington,A.Camsonne}). The target can be polarized both longitudinally and transversely with its polarization up to $80\%$~\cite{Allada2011}. Although it can only tolerate a polarized beam of $10$--$100$~nA at the moment, potential upgrades of the Continuous Electron Beam Accelerator Facility at JLab in the upcoming era indicate that a SNH$_3$ target that can tolerate a much intense beam ($I>100\,\mu$A) will be available~\cite{Arrington:2021alx}. For a simple event-rate estimation of the polarized $ep$ scattering processes, we here assume that the same target parameters, i.e., its polarization and packing factor, can be achieved in the future. We summarize in Table~\ref{table:target_parameters} our preferred experimental parameters of the proton targets for the low-energy polarized $ep$ scattering processes.
 
\begin{table}[t]
	\renewcommand*\arraystretch{1.6}
	\setlength\tabcolsep{2.5pt}
	\centering
	\begin{tabular}{cccccc} 
		\hline\hline
		& Packing & Length & Density & Dilution &Polarization \\[-0.1cm]
		&  factor $\kappa$ & $L\,(\text{cm})$ &$\rho\,(g/\text{cm}^3)$  & factor $f$ &$P_p$ \\
		\hline
		LH & 1 & 70  & $71.3\times 10^{-3} $ &1  & 0\\
		%\hline
		SNH$_3$ & 0.5  & 6  & 0.917  & 0.15&80\% \\
		\hline\hline
	\end{tabular}
	\caption{Experimental parameters of the proton targets for the low-energy polarized $ep$ scattering processes.} 
	\label{table:target_parameters}
\end{table}

To maximize the chances of observing the polarized $ep$ scattering processes, we have been thus far collecting proper electron beams and proton targets that have been or will be utilized in different experiments. However, it is interesting to note that certain polarized scattering processes can already be conducted with the same beam and target as used in some previous experiments. For instance, the polarized electron beam and unpolarized LH target utilized by the HAPPEX collaboration~\cite{HAPPEX:2011xlw} can also be used to detect the longitudinally polarized $\vec{e}^{\,-} p \to e^- \Lambda_c$ process. In addition, if the expected higher luminosity at JLab can be reached in the upcoming era~\cite{Arrington:2021alx}, it may be possible to explore other polarized scattering processes at the same facility. There is nonetheless a caveat here: besides the proper beams and targets, a sophisticated detecting system for the produced particles is also necessary. Especially for the $\Lambda_c$ baryon, since it is not easy to keep track of all its decay products, one may focus only on one of its decay channels, such as $\Lambda_c\to p K^-\pi^+$ with its branching fraction of about $6.28\%$~\cite{Zyla:2020zbs}.  For the corresponding detecting system, one may draw inspiration from the recent measurements of the branching fraction of $\Lambda_c\to p K^-\pi^+$ through $e^+e^-$ collisions (see, e.g., Refs.~\cite{Belle:2013jfq,BESIII:2015bjk}).

\subsection{\boldmath Polarized $e^-p\to e^-\Lambda_c$ in the LQ models}

With the aforementioned experimental setups, let us first
evaluate the modified total cross sections with the degrees of beam and target polarizations, $P_{e,p}$, taken into account, in the framework of the general low-energy effective Lagrangian $\mathcal{L}_{\text{eff}}$.  

We will start with the longitudinally polarized scattering process $\vec{e}^{\,-} p \to e^- \Lambda_c$. In the $e^+_L$ case, the modified total cross section reads 
\begin{align}\label{eq:sigmaELP}
\sigma_{e}^{+}=&\Big\{82.95\,|g^{RR}_{V}|^2+36.40\,|g^{RL}_V|^2+4.68\,\text{Re}\left[g^{RR}_{V}g^{RL*}_{V}\right]\nonumber \\[0.12cm]
&+4.64\,|g^{LL}_{V}|^2+2.04\,|g^{LR}_V|^2+0.26\,\text{Re}\left[g^{LL}_{V}g^{LR*}_{V}\right]\nonumber \\[0.12cm] &+16.80\,|g^{R}_S|^2+0.94\,|g^{L}_S|^2+533.39\,|g^{R}_{T}|^2\nonumber \\[0.12cm]
&+29.85\,|g^{L}_{T}|^2-109.42\,\text{Re}\left[g^{R}_Sg^{R*}_T\right]\nonumber \\[0.12cm]
&-6.12\,\text{Re}\left[g^{L}_Sg^{L*}_T\right]
\Big\}\times 10^{-4}\,\text{GeV}^2\,,
\end{align}
while in the $e^-_L$ case,  
\begin{align} \label{eq:sigmaELM}
\sigma_{e}^{-}=&\Big\{82.95\,|g^{LL}_{V}|^2+36.40\,|g^{LR}_V|^2+4.68\,\text{Re}\left[g^{LL}_{V}g^{LR*}_{V}\right]\nonumber \\[0.12cm]
&+4.64\,|g^{RR}_{V}|^2+2.04\,|g^{RL}_V|^2+0.26\,\text{Re}\left[g^{RR}_{V}g^{RL*}_{V}\right]\nonumber \\[0.12cm] &+16.80\,|g^{L}_S|^2+0.94\,|g^{R}_S|^2+533.39\,|g^{L}_{T}|^2\nonumber \\[0.12cm]
&+29.85\,|g^{R}_{T}|^2-109.42\,\text{Re}\left[g^{L}_Sg^{L*}_T\right]\nonumber \\[0.12cm]
&-6.12\,\text{Re}\left[g^{R}_Sg^{R*}_T\right]
\Big\}\times 10^{-4}\,\text{GeV}^2\,. 
\end{align}
Note that the WCs associated with the operators involving the lepton current $j^L$ ($j^R$) 
appear in Eq.~\eqref{eq:sigmaELP} (Eq.~\eqref{eq:sigmaELM}) due to the imperfect $P_e$, i.e., $P_e\neq 1$. 

For the longitudinally polarized scattering process $e^- \vec{p} \to e^- \Lambda_c$, we have the modified total cross sections 
\begin{align} \label{eq:sigmaPL}
\sigma_{p}^{+}=&\Big\{49.04\,|g^{LL}_{V}|^2+16.96\,|g^{LR}_V|^2+2.35\,\text{Re}\left[g^{LL}_{V}g^{LR*}_{V}\right]\nonumber \\[0.12cm]
&+38.56\,|g^{RR}_{V}|^2+21.48\,|g^{RL}_V|^2+2.60\,\text{Re}\left[g^{RR}_{V}g^{RL*}_{V}\right]\nonumber \\[0.12cm] &+9.91\,|g^{L}_S|^2+7.83\,|g^{R}_S|^2+315.07\,|g^{L}_{T}|^2\nonumber \\[0.12cm]
&+248.17\,|g^{R}_{T}|^2-64.81\,\text{Re}\left[g^{L}_Sg^{L*}_T\right]\nonumber \\[0.12cm]
&-50.73\,\text{Re}\left[g^{R}_Sg^{R*}_T\right]\Big\}\times 10^{-4}\,\text{GeV}^2\,,\nonumber \\[0.12cm]
\sigma_{p}^{-}=&\Big\{49.04\,|g^{RR}_{V}|^2+16.96\,|g^{RL}_V|^2+2.35\,\text{Re}\left[g^{RR}_{V}g^{RL*}_{V}\right]\nonumber \\[0.12cm]
&+38.56\,|g^{LL}_{V}|^2+21.48\,|g^{LR}_V|^2+2.60\,\text{Re}\left[g^{LL}_{V}g^{LR*}_{V}\right]\nonumber \\[0.12cm] &+9.91\,|g^{R}_S|^2+7.83\,|g^{L}_S|^2+315.07\,|g^{R}_{T}|^2\nonumber \\[0.12cm]
&+248.17\,|g^{L}_{T}|^2-64.81\,\text{Re}\left[g^{R}_Sg^{R*}_T\right]\nonumber \\[0.12cm]
&-50.73\,\text{Re}\left[g^{L}_Sg^{L*}_T\right]\Big\}\times 10^{-4}\,\text{GeV}^2\,,
\end{align}
in the cases of $p^+_L$ and $p^-_L$, respectively. It should be noted that we have, for simplicity, used the experimental parameters ($E$ and $I$) of the polarized electron beam 
to give a simple estimation of this process, although an unpolarized electron beam with similar $E$ and $I$ is already available, and has been utilized in the APEX experiment at JLab 
for hunting sub-GeV dark vector bosons~\cite{Abrahamyan:2011gv,Essig:2010xa}. 

We also compute the modified total cross section of the double-spin polarized scattering 
process $\vec{e}^{\,-} \vec{p} \to e^- \Lambda_c$. As an illustration, let us consider the two sets of modified cross sections, ($\sigma^{-+}$, $\sigma^{--}$) and  ($d\widetilde{\sigma}^{-+}/d\phi$, $d\widetilde{\sigma}^{--}/d\phi$), which are related to the measured double-spin asymmetries $(A^{ep}_{L3})_{\text{exp}}$ and $(A^{ep}_{T3})_{\text{exp}}$, respectively. Numerically, the former two are given by
\begin{align} \label{eq:sigmaEPL}
\sigma^{-+}=&\Big\{92.88\,|g^{LL}_{V}|^2+32.13\,|g^{LR}_V|^2+4.44\,\text{Re}\left[g^{LL}_{V}g^{LR*}_{V}\right]\nonumber \\[0.12cm]
&+4.09\,|g^{RR}_{V}|^2+2.28\,|g^{RL}_V|^2+0.27\,\text{Re}\left[g^{RR}_{V}g^{RL*}_{V}\right]\nonumber \\[0.12cm] &+18.77\,|g^{L}_S|^2+0.83\,|g^{R}_S|^2+596.74\,|g^{L}_{T}|^2\nonumber \\[0.12cm]
&+26.31\,|g^{R}_{T}|^2-122.76\,\text{Re}\left[g^{L}_Sg^{L*}_T\right]\nonumber \\[0.12cm]
&-5.38\,\text{Re}\left[g^{R}_Sg^{R*}_T\right]\Big\}\times 10^{-4}\,\text{GeV}^2\,,\nonumber \\[0.2cm]
\sigma^{--}=&\Big\{73.03\,|g^{LL}_{V}|^2+40.68\,|g^{LR}_V|^2+4.91\,\text{Re}\left[g^{LL}_{V}g^{LR*}_{V}\right]\nonumber \\[0.12cm]
&+5.20\,|g^{RR}_{V}|^2+1.80\,|g^{RL}_V|^2+0.25\,\text{Re}\left[g^{RR}_{V}g^{RL*}_{V}\right]\nonumber \\[0.12cm] &+14.83\,|g^{L}_S|^2+1.05\,|g^{R}_S|^2+470.04\,|g^{L}_{T}|^2\nonumber \\[0.12cm]
&+33.40\,|g^{R}_{T}|^2-96.09\,\text{Re}\left[g^{L}_Sg^{L*}_T\right]\nonumber \\[0.12cm]
&-6.87\,\text{Re}\left[g^{R}_Sg^{R*}_T\right]\Big\}\times 10^{-4}\,\text{GeV}^2\,,
\end{align} 
while the latter two read
\begin{align} \label{eq:sigmaEPT}
\frac{d\widetilde{\sigma}^{-+}}{d\phi}\bigg|_{\phi=0}=&\Big\{13.20\,|g^{LL}_{V}|^2+5.73\,|g^{LR}_V|^2+0.74\,|g^{RR}_{V}|^2\nonumber \\[0.12cm]
&\hspace{-1.9cm}+0.33\,|g^{RL}_V|^2+1.76\,\text{Re}\left[g^{LL}_{V}g^{LR*}_{V}\right]-0.02\,\text{Re}\left[g^{RR}_{V}g^{RL*}_{V}\right]\nonumber \\[0.12cm] &+2.64\,|g^{L}_S|^2+0.15\,|g^{R}_S|^2+85.06\,|g^{L}_{T}|^2\nonumber \\[0.12cm]
&+4.74\,|g^{R}_{T}|^2-17.04\,\text{Re}\left[g^{L}_Sg^{L*}_T\right]\nonumber \\[0.12cm]
&-1.00\,\text{Re}\left[g^{R}_Sg^{R*}_T\right]\Big\}\times 10^{-4}\,\text{GeV}^2\,,\nonumber \\[0.2cm]
\frac{d\widetilde{\sigma}^{--}}{d\phi}\bigg|_{\phi=0}=&\Big\{13.20\,|g^{LL}_{V}|^2+5.86\,|g^{LR}_V|^2+0.74\,|g^{RR}_{V}|^2\nonumber \\[0.12cm]
&\hspace{-1.9cm}+0.32\,|g^{RL}_V|^2-0.27\,\text{Re}\left[g^{LL}_{V}g^{LR*}_{V}\right]+0.10\,\text{Re}\left[g^{RR}_{V}g^{RL*}_{V}\right]\nonumber \\[0.12cm] &+2.71\,|g^{L}_S|^2+0.15\,|g^{R}_S|^2+84.72\,|g^{L}_{T}|^2\nonumber \\[0.12cm]
&+4.76\,|g^{R}_{T}|^2-17.79\,\text{Re}\left[g^{L}_Sg^{L*}_T\right]\nonumber \\[0.12cm]
&-0.95\,\text{Re}\left[g^{R}_Sg^{R*}_T\right]\Big\}\times 10^{-4}\,\text{GeV}^2\,,
\end{align}  
where the angle $\phi$ has been set to 0 for simplicity. 
 
Before moving on to the event-rate estimation in different LQ models, as we did in Ref.~\cite{Lai:2021sww}, let us briefly summarize in Table~\ref{table:constraints_LFC} the most relevant and stringent constraints on the effective coefficients $[g]^{ee,cu}$ from the
charmed-hadron weak decays and the high-$p_T$ dilepton invariant mass tails. It can be seen that contributions from the scalar operators are stringently constrained by the measurement of the branching ratio $\mathcal{B}(D^0\to e^+ e^-)$, which, through the low-energy WC relation in Eq.~\eqref{eq:RG_R2}, in turn leads to severe constraints on the tensor contributions in the LQ models $S_1$ and $R_2$, the only two LQ models in Table~\ref{tab:LQ} that can generate 
non-vanishing scalar and tensor effective operators. We thus neglect these contributions and focus on the vector ones, whose most severe upper bounds clearly come from the high-$p_T$ dilepton invariant mass tails, as shown in Table~\ref{table:constraints_LFC}. 

\begin{table}[t]
	\renewcommand\arraystretch{1.6} 
	\tabcolsep=0.08cm
	\centering
	\begin{tabular}{ccccc}
		\hline\hline
		Processes &$\big|g^{LL,RR}_V\big|^2$ &$\big|g^{LR,RL}_V\big|^2$ & $\big|g^{L,R}_S\big|^2$& $\big|g^{L,R}_T\big|^2$  \\ \hline
		$D^0\to e^+e^-$~\cite{Petric:2010yt} &$\backslash$  &$\backslash$  & 0.062& $\backslash$ \\ 
		$D^+\to \pi^+e^+e^-$~\cite{Lees:2011hb} & 14& 14& 6.3 & 13 \\ 
		$pp(q\bar{q})\to e^+e^-$~\cite{CMS:2021ctt} & 3.6& 3.6 & 22&0.57 \\
		\hline \hline
	\end{tabular}
	\caption{Summary of the upper bounds on the WCs $[g]^{ee,cu}$ at $90\%$ confidence level from the (semi)leptonic $D$-meson decays and the high-$p_T$ dilepton invariant mass tails in the framework of the general low-energy effective Lagrangian (for more details, see Ref.~\cite{Lai:2021sww} and references therein). Note that we have factored out the common factor $G^2_F\alpha^2_e/\pi^2$. The entries with ``$\backslash$'' mean that the processes in the first column put no constraints on the corresponding WCs.}
	\label{table:constraints_LFC} 
\end{table} 

Taking the upper limits on $g_V$ from Table~\ref{table:constraints_LFC}, focusing on the decay channel $\Lambda_c\to p K^-\pi^+$, and supposing a run time of one year, we now evaluate the expected event rates of the various polarized $ep$ scattering processes in units of number per year ($N$/yr) in the LQ models. The final results are presented in Table~\ref{tab:Event}, from which the following interesting observations can be made. First, the expected $(A^e_L)_{\text{exp}}$ and $(A^p_L)_{\text{exp}}$ in the models $S_1$ and $R_2$ will be equal to 0, because the same upper limits on $g^{LL}_V$ and $g^{RR}_V$ have been taken. Due to the same reason, the expected event rate of the single-spin polarized scattering process with $e^+_L$ ($p^+_L$) in the model $V_2$ is identical to that with $e^-_L$ ($p^-_L$) in the model $\widetilde{V}_2$ (see Eqs.~\eqref{eq:sigmaELP}, \eqref{eq:sigmaELM}, and \eqref{eq:sigmaPL}). 
Second, identical event rates will always be expected in the LQ models $S_3$ and $U_3$ 
for any polarized scattering processes listed in Table~\ref{tab:Event}, since these two models are indistinguishable from each other in this work. Third, processes with $e_L^-p_T^+$ and $e_L^-p_T^- $ will have the same (or very close) event rates, mainly because
the $\xi_p$-dependent $|\widetilde{\mathcal{M}}|^2$ is much smaller than 
the $\xi_p$-independent $\overline{|\mathcal{M}|}^2$, as has been explicitly demonstrated 
in Fig.~\ref{fig:ATp}. Such an observation once again demonstrates that it becomes less applicable to use the transversely polarized $ep$ scattering processes to distinguish the NP models.

\begin{table}[t]
	\centering
	\renewcommand\arraystretch{1.6}
	\setlength\tabcolsep{4.8pt}
	\begin{tabular}{lccccccc} \hline\hline
		&$S_1$&$S_3$&$R_2$&$U_3$&$\widetilde{U}_1$&$V_2$&$\widetilde{V}_2$\\  \hline 
		$e_L^+ $& 33.41 & 1.76 & 14.70 &1.76  &31.65& 13.88 & 0.75 \\ 
		$e_L^- $& 33.41 & 31.65 & 14.70 &31.65  &1.76& 0.75 & 13.88 \\  
		$p_L^+ $& 15.00 & 10.30 & 8.10 &10.30  &8.10& 4.52 & 3.58 \\  
		$p_L^- $& 15.00 & 8.10 & 8.10 &8.10  &10.30& 3.58 & 4.52 \\  
		$e_L^-p_L^+ $& 20.41 &19.53 & 7.22 & 19.53 & 0.88 & 0.50 & 6.78 	\\ 
		$e_L^-p_L^- $& 16.45 & 15.39 & 8.92 & 15.39 & 1.07& 0.38 & 8.54 \\ 
		$e_L^-p_T^+ $& 2.95 &2.76 & 1.26 & 2.76 & 0.13 & 0.06 & 1.19 	\\  
		$e_L^-p_T^- $& 2.95 & 2.76 & 1.32 & 2.76 & 0.13 & 0.06 & 1.26 \\ \hline\hline
	\end{tabular}
	\caption{Summary of event-rate estimations for the polarized $e^-p\to e^-\Lambda_c\,(\to p K^-\pi^+)$ scattering processes in different LQ models, where the event rate is given in units of $N$/yr, and $\Delta\phi=1$ has been taken for the cases $e_L^-p_T^+$ and $e_L^-p_T^-$.}
	\label{tab:Event}
\end{table}

\subsection{\boldmath Competitive constraints from the polarized scattering process $\vec{e}^{\,-} p \to e^- \Lambda_c$}

The various event rates in Table~\ref{tab:Event} are what one can expect in the best scenario, 
because they are just carried out with the currently available upper limits of the WCs. Thus, it is still possible that no event of the polarized $ep$ scattering processes is observed
with our properly designed experimental setups in the future. In what follows, we will show that, even in such a worst-case scenario, investigating experimentally the various polarized $ep$ scattering processes in the future will not be in vain.

From the numerical coefficients of the various WCs $g$ in Eqs.~\eqref{eq:sigmaELP}--\eqref{eq:sigmaEPT}, one can expect that the polarized scattering process $\vec{e}^{\,-} p \to e^- \Lambda_c$ with the designed experimental setups will set the most severe constraints on them. Thus, let us concentrate on this process. Assuming that only one WC contributes to the scattering process at a time and the produced $\Lambda_c$ is solely detected through the decay channel $\Lambda_c\to p K^-\pi^+$, we obtain the resulting constraints on the WCs $g$ in Table~\ref{tab:constraint} with the $100\%$ detecting efficiency of the final particles and the run time of one year. Note that the constraints on $|g^{RR}_V|$, $|g^{RL}_V|$, $|g^{R}_S|$, and $|g^{R}_T|$ in the $e^-_L$ case have not been given, mainly because their contributions are much smaller in comparison with those from $|g^{LL}_V|$, $|g^{LR}_V|$, $|g^{L}_S|$, and $|g^{L}_T|$. Besides, the presence of $|g^{RR}_V|$, $|g^{RL}_V|$, $|g^{R}_S|$, and $|g^{R}_T|$ in the $e^-_L$ case, as has been pointed out before, results from $P_e\neq 1$, and is expected to disappear as $P_e\to 1$. The same arguments have been applied to the $e^+_L$ case as well.

\begin{table}[t] 
	\centering
	\renewcommand\arraystretch{1.6}
	\setlength\tabcolsep{2.5pt}
	\begin{tabular}{ccccccccc} \hline \hline
		&$|g_V^{LL}|^2$&$|g_V^{RR}|^2$&$|g_V^{LR}|^2$ &$|g_V^{RL}|^2$&$|g_S^{L}|^2$&$|g_S^{R}|^2$&$|g_T^{L}|^2$&$|g_T^{R}|^2$\\  
		\hline
		$e_L^-$& $0.111$ & $\backslash$ & $0.255$ & $\backslash$ & $0.557$ & $\backslash$ & $0.016$ & $\backslash$
		\\
		$e_L^+$& $\backslash$ & $0.111$ & $\backslash$ & $0.255$ & $\backslash$ & $0.557$ & $\backslash$ &$0.016$
		\\ \hline \hline
	\end{tabular}
	\caption{Constraints on the WCs $[g]^{ee,cu}$ from the polarized scattering processes $\vec{e}^{\,-}p\to e^-\Lambda_c\,(\to p K^-\pi^+)$ in the framework of the general low-energy effective Lagrangian. The entries with ``$\backslash$'' mean that the processes with the electron beam polarization specified by the first column set less interesting constraints on the corresponding WCs than from those with an opposite electron beam polarization. Same as in Table~\ref{table:constraints_LFC}, the common factor $G^2_F\alpha^2_e/\pi^2$ has been factored out as well. }
	\label{tab:constraint}
\end{table}

From the numerical results presented in Tables~\ref{tab:constraint} and \ref{table:constraints_LFC}, it can be seen that, compared with other processes except the leptonic $D$-meson decay, significant improvements in constraining the effective WCs can be made through the low-energy polarized scattering process $\vec{e}^{\,-} p \to e^- \Lambda_c$, even by assuming one specific decay channel $\Lambda_c\to p K^-\pi^+$. This also indicates that the polarized scattering processes can provide a further complementarity to the charmed-hadron weak decays and the high-$p_T$ dilepton invariant mass tails. It should be, however, mentioned that our results can be strengthened, if more decay channels of the $\Lambda_c$ baryon are considered. On the other hand, these observations would be weakened by the non-$100\%$ detecting efficiency of the final particles and by the various uncertainties, such as the theoretical (both statistical and systematic) ones of the $\Lambda_c\to p$ form factors estimated by the lattice QCD calculations~\cite{Meinel:2017ggx}.

Let us conclude this section by pointing out another merit of the low-energy polarized $ep$ scattering processes. As shown in Table~\ref{tab:coeff}, two pairs of effective vector operators, ($j^L_VJ^L_V$, $j^R_VJ^R_V$) and ($j^L_VJ^R_V$, $j^R_VJ^L_V$), are generated in the LQ modes $S_1$ and $R_2$, respectively. When setting constraints on the corresponding effective WCs $g$ from the charmed-hadron weak decays, the high-$p_T$ dilepton invariant mass tails, as well as the unpolarized $ep$ scattering processes~\cite{Lai:2021sww}, one meets the following constraining formulas: 
\begin{equation}\label{eq:constraining_formula}
	a|g^{LL}_V|^2+b|g^{RR}_V|^2\leq c\,, \quad a^\prime|g^{LR}_V|^2+b^\prime|g^{RL}_V|^2\leq c^\prime\,,
\end{equation}
for the LQ models $S_1$ and $R_2$, respectively. Here contributions from the scalar and tensor operators have been neglected, and the non-zero constants $a$, $b$, $c$, $a^\prime$, $b^\prime$, and $c^\prime$ are in general distinct for different processes. It is interesting to note that $a=b$ and $a^\prime=b^\prime$ always hold for the charmed-hadron weak decays and the high-$p_T$ dilepton invariant mass tails, while $a\neq b$ and $a^\prime\neq b^\prime$ for the unpolarized $ep$ scattering processes~\cite{Lai:2021sww}. Now it is clear from Eq.~\eqref{eq:constraining_formula} that setting constraints on the individual WCs $g_V$ from  these processes becomes very nontrivial without adopting any special treatments---the most common one is probably by saturating the process with one non-vanishing WC at a time. For the process $\vec{e}^{\,-} p \to e^- \Lambda_c$, on the other hand, such a treatment becomes not necessary, since only one $g_V$ appears in each constraining formula in Eq.~\eqref{eq:constraining_formula}. 
Certainly, small $b$ and $b^\prime$ do appear in practice due to $P_e\neq 1$, but can be reasonably neglected, after taking into account the currently available high degree of the electron beam polarization~\cite{HAPPEX:2011xlw}. 

\section{Conclusion}
\label{sec:con}

In this paper, we have investigated the potential for searching and identifying 
the LQ effects in the charm sector through the low-energy polarized scattering processes $e^-p\to e^-\Lambda_c$. Specifically, we have considered the single-spin polarized scattering processes, $\vec{e}^{\,-}p\to e^-\Lambda_c$ and $e^-\vec{p}\to e^-\Lambda_c$, as well as the double-spin polarized one $\vec{e}^{\,-}\vec{p}\to e^-\Lambda_c$. Based on these polarized processes, together with their associated polarized cross sections, we have defined several spin asymmetries, which are found to be very efficient in disentangling the different LQ models from each other.

Focusing first on the longitudinal polarized scattering processes, we have shown in an almost model-independent way that the 15 NP scenarios, including the seven concrete LQ models, can be effectively disentangled from each other by measuring the four spin asymmetries, $A_{L}^e$, $A_{L}^p$, $A_{L3}^{ep}$, and $A_{L6}^{ep}$. To be thorough, we have also examined this mechanism by considering the polarized differential cross sections and exploring how these spin asymmetries behave with respect to the electron beam energy $E$ and the kinematics $Q^2$. 
It is found that, to make the procedure most efficient, both the low $Q^2$ and the high $E$ regime are favored. Besides this appealing application, we have discovered that the scalar (tensor) and vector contributions in the LQ model $R_2$ can be distinguished through the longitudinally polarized scattering processes as well.  

We have then investigated the transversely polarized scattering processes in various aspects. 
In contrast to the longitudinal case, it would be very challenging to identify the NP models 
through these transversely polarized scattering processes, mainly because the polarization effects are far less prominent. Far from being in vain, however, measurements of these transversely polarized scattering processes offer us a unique opportunity to probe directly into the imaginary part of the effective WCs. 

Given that all the mechanisms we have proposed are based on the premise 
that the LQ signals can be detected, we have finally performed a simple event-rate 
estimation for all the polarized scattering processes with the properly designed experimental setups, demonstrating that promising even rates can be expected for these processes. 
On the other hand, even in the worst-case scenario---no LQ signals are observed at all, we have shown in a model-independent way that the low-energy polarized scattering processes can provide more competitive constraints, in comparison with the charmed-hadron weak decays and the high-$p_T$ dilepton invariant mass tails. Furthermore, we have pointed out that, by maneuvering the electron beam polarization in the process $\vec{e}^{\,-}p\to e^-\Lambda_c$, one can directly set constraints on the WCs $g_V$ in the LQ models $S_1$ and $R_2$, which, by contrast, will be tricky for other processes without taking any special treatments.

\section*{Acknowledgments}
This work is supported by the National Natural Science Foundation of China under Grant 
Nos.~12135006, 12075097, 12047527, 11675061 and 11775092.

\appendix

\section{\boldmath Definitions and parametrization of the $\Lambda_c\to p$ form factors}
\label{appendix:form factor}

The form factors for $\Lambda_c\to p$ transition can be conveniently parametrized in 
the helicity basis~\cite{Feldmann:2011xf,Meinel:2017ggx,Das:2018sms}. For the vector and axial-vector currents, their hadronic matrix elements are defined, respectively, by 
\begin{align}
&\langle N^{+}(p,s)|\bar{u}\gamma^{\mu}c|\Lambda_c(p',s')\rangle \nonumber \\[0.15cm]
&\ \ \!=\!\bar{u}_N(p,s)\bigg[f_0(q^2)(m_{\Lambda_c}\!-\!m_N)\frac{q^{\mu}}{q^2}\nonumber \\[0.15cm]
&\ \  \ \ \ \!+\!f_+(q^2)\frac{m_{\Lambda_c}\!+\!m_N}{s_+}
\Big(p'^{\mu}\!+\!p^{\mu}\!-\!(m^2_{\Lambda_c}\!-\!m^2_N)\frac{q^{\mu}}{q^2}\Big) \nonumber \\[0.15cm]
&\ \ \ \ \  \!+\! f_{\perp}(q^2)\Big(\gamma^{\mu}\!-\!\frac{2m_N}{s_+}p'^{\mu}\!-\!\frac{2m_{\Lambda_c}}{s_+}p^{\mu}\Big)\bigg]u_{\Lambda_c}(p',s'), \label{eq:vect}
\end{align}
and 
\begin{align}
&\langle N^{+}(p,s)|\bar{u}\gamma^{\mu}\gamma^5c|\Lambda_c(p',s')\rangle \nonumber \\[0.15cm]
&\ \ \!=\!-\bar{u}_N(p,s)\gamma^5\bigg[g_0(q^2)(m_{\Lambda_c}\!+\! m_N)\frac{q^{\mu}}{q^2}\nonumber \\[0.15cm]
&\ \ \ \ \ \!+\! g_+(q^2)\frac{m_{\Lambda_c}\!-\!m_N}{s_-}
\Big(p'^{\mu}\!+\!p^{\mu}\!-\!(m^2_{\Lambda_c}\!-\! m^2_N)\frac{q^{\mu}}{q^2}\Big) \nonumber \\[0.15cm]
&\ \ \ \ \ \!+\!g_{\perp}(q^2)\Big(\gamma^{\mu}\!+\!\frac{2m_N}{s_-}p'^{\mu}\!-\!\frac{2m_{\Lambda_c}}{s_-}p^{\mu}\Big)\bigg]u_{\Lambda_c}(p',s'), \label{eq:psvect}
\end{align}
where $q=p^\prime-p$ and $s_{\pm}=(m_{\Lambda_c}\pm m_N)^2-q^2$. Note that we have denoted the proton by $N^+$ instead of $p$ to avoid possible confusion with the proton's momentum. 
From Eqs.~\eqref{eq:vect} and \eqref{eq:psvect}, we can obtain the hadronic matrix elements of the scalar and pseudo-scalar currents through the equation of motion,
\begin{align}
&\langle N^{+}(p,s)|\bar{u}c|\Lambda_c(p',s')\rangle \nonumber \\[0.15cm]
&\ \ =\frac{(m_{\Lambda_c}- m_N)}{m_{c}- m_{u}}f_0(q^2)\bar{u}_N(p,s)u_{\Lambda_c}(p',s') , \\[0.2cm]
&\langle N^{+}(p,s)|\bar{u}\gamma^5	c|\Lambda_c(p',s')\rangle \nonumber \\[0.15cm]
&\ \  =\frac{(m_{\Lambda_c}+m_N)}{m_{c}+m_{u}}g_0(q^2)\bar{u}_N(p,s)\gamma^5 u_{\Lambda_c}(p',s'),
\end{align}
where $m_{u(c)}$ denotes the $u(c)$-quark running mass. Finally, the hadronic matrix element for the tensor current is given by 
\begin{align}
&\langle N^{+}(p,s)|\bar{u}i\sigma_{\mu\nu} c|\Lambda_c(p',s')\rangle \nonumber \\[0.15cm]
&\ \ \!=\!\bar{u}_N(p,s) \bigg[2h_+ \frac{p'_{\mu}p_{\nu}\! -\! p'_{\nu}p_{\mu}}{s_+}\!
+\! h_{\perp}\Big(\frac{m_{\Lambda_c}\! +\! m_N}{q^2}\nonumber \\[0.15cm]
&\ \ \ \ \ \! \times\! (q_{\mu}\gamma_{\nu}\!-\! q_{\nu}\gamma_{\mu})\!-\! 2\left(\frac{1}{q^2}\!
+\!\frac{1}{s_+}\right)(p'_{\mu}p_{\nu}\!-\! p'_{\nu}p_{\mu})\Big)\nonumber \\[0.15cm]
&\ \ \ \ \ \!+\! \tilde{h}_+\Big(i\sigma_{\mu\nu}\!-\!\frac{2}{s_-}[m_{\Lambda_c}(p_{\mu}\gamma_{\nu}\!-\!p_{\nu}\gamma_{\mu})\nonumber \\[0.15cm]
&\ \ \ \ \ \!-\!m_N(p'_{\mu}\gamma_{\nu}\!-\!p'_{\nu}\gamma_{\mu})+\!p'_{\mu}p_{\nu}\!-\!p'_{\nu}p_{\mu}]\Big)\nonumber \\[0.15cm]
&\ \ \ \ \ \!+\tilde{h}_{\perp}\frac{m_{\Lambda_c}-m_N}{q^2s_-}\Big((m_{\Lambda_c}^2-m_N^2-q^2)(\gamma_{\mu}p'_{\nu}-\gamma_{\nu}p'_{\mu})\nonumber \\[0.15cm]
&\ \ \ \ \ \!-(m_{\Lambda_c}^2-m_N^2+q^2)(\gamma_{\mu}p_{\nu}-\gamma_{\nu}p_{\mu})\nonumber \\[0.15cm]
&\ \ \ \ \ \!+\!2(m_{\Lambda_c}-m_N)(p'_{\mu}p_{\nu}-p'_{\nu}p_{\mu})\Big)\bigg]u_{\Lambda_c}(p',s'), \label{eq:tensor}
\end{align}
where $\sigma_{\mu\nu}=i[\gamma_{\mu},\gamma_{\nu}]/2$. These form factors satisfy the endpoint relations
\begin{align}
& f_+(0)=f_0(0), \quad && g_+(0)=g_0(0), \nonumber \\[0.2cm]
& g_+(q_{\max}^2)=g_\perp(q_{\max}^2), \quad && \tilde{h}_{+}(q_{\max}^2)=\tilde{h}_{\perp}(q_{\max}^2),
\end{align}
with $q_{\max}^2=(m_{\Lambda_c} - m_N)^2$. 

The parametrization of the $\Lambda_c\to p$ form factors takes the form~\cite{Bourrely:2008za,Meinel:2017ggx}
\begin{align}\label{eq:formfactorpara}
f(q^{2})=\frac{1}{1-q^{2} /(m_{\text{pole}}^{f})^{2}} \sum_{n=0}^{n_{\max}} a_{n}^{f}\left[z(q^{2})\right]^{n},
\end{align}
with the expansion variable defined by
\begin{align}
z(q^{2})=\frac{\sqrt{t_{+}-q^{2}}-\sqrt{t_{+}-t_{0}}} {\sqrt{t_{+}-q^{2}}+\sqrt{t_{+}-t_{0}}},
\end{align}
where $t_{+}=(m_{D}+m_{\pi})^{2}$ is set equal to the threshold of $D\pi$ two-particle states, and $t_{0}=q_{\max}^{2}$ determines which value of $q^{2}$ gets mapped to $z=0$. In this way, one maps the complex $q^{2}$ plane, cut along the real axis for $q^{2} \geq t_{+}$, onto the disk $|z|<1$. The central values and statistical uncertainties of $a^f_{0,1,2}$ in Eq.~\eqref{eq:formfactorpara} for different form factors $f(q^2)$ 
have been evaluated in Ref.~\cite{Meinel:2017ggx} by the nominal fit ($n_{\max}=2$),
while their systematic uncertainties can be obtained by a combined analysis of both the nominal and higher-order ($n_{\max}=3$) fits. We refer the readers to Ref.~\cite{Meinel:2017ggx} for further details.

\section{Polarized cross sections and experimental quantities}
\label{app:expquantity}

In this appendix, we clarify the relations among the polarized cross sections, the spin asymmetries, and the experimentally measurable quantities relevant to this work.   

Starting with the double-spin longitudinally polarized scattering process $\vec{e}^{\,-}(k)+\vec{p}(P)\to e^-(k^\prime)+\Lambda_c(P^\prime)$ mediated by the general low-energy effective Lagrangian $\mathcal{L}_{\text{eff}}$, we can write its cross section in the lab frame as 
\begin{align}\label{eq:double-spin-Lcs}
\frac{d\sigma^{h_eh_p}}{dq^2}\!&=\!\frac{1}{64\pi m^2_p E^2}\big(\mathcal{A}_0\!+\!h_e\mathcal{A}_e\!+\!h_p\mathcal{A}_p\!+\!h_eh_p\mathcal{A}_{ep}\big) \nonumber \\
&=\!\frac{d\sigma_0+h_ed\sigma_e+h_pd\sigma_p+h_eh_pd\sigma_{ep}}{dq^2}\,,
\end{align} 
where $h_{e,p}=\pm 1$ originate from the polarization four-vectors $\xi^{\mu}_{e,p}$, such that the latter now take the forms
\begin{align}\label{eq:polarizationvectors}
	\xi^{\mu}_e=h_e\Big(\dfrac{|\pmb{k}|}{m_e},\dfrac{k^0\pmb{k}}{m_e |\pmb{k}|}\Big)\,, \qquad \xi_p^{\mu}=h_p(0,0,0,1)\,.
\end{align}
The whole amplitude squared $|\mathcal{M}|^2$ in Eq.~\eqref{eq:double-spin-Lcs}
is divided into four pieces, among which $\mathcal{A}_0$ denotes the $\xi^{\mu}_{e,p}$-independent one, $h_e\mathcal{A}_e$ ($h_p\mathcal{A}_p$) the $\xi^{\mu}_{e}$ ($\xi^{\mu}_{p}$)-dependent one, while $h_eh_p\mathcal{A}_{ep}$ the $\xi^{\mu}_{e}\xi^{\mu}_{e}$-dependent one. It is important to remind that all these amplitudes squared are not averaged over the spins of initial electron and proton.

Based on the double-spin cross section in Eq.~\eqref{eq:double-spin-Lcs}, the cross sections of other longitudinally polarized scattering processes can be obtained straightforwardly. For the single-spin polarized scattering process $\vec{e}^{\,-}p\to e^-\Lambda_c$, its cross section is given by 
\begin{align}\label{eq:single-spin-Lecs}
	\frac{d\sigma^{h_e}_e}{dq^2}=\frac{1}{2}\sum_{h_p} \frac{d\sigma^{h_eh_p}}{dq^2}
	=\frac{d\sigma_0+h_ed\sigma_e}{dq^2}\,,
\end{align} 
where the factor $1/2$ arises from the spin average of the initial proton---since it is unpolarized. Similarly, one can get the cross section of $e^{\,-}\vec{p}\to e^-\Lambda_c$,
\begin{align}\label{eq:single-spin-Lpcs}
\frac{d\sigma^{h_p}_p}{dq^2}=\frac{1}{2}\sum_{h_e} \frac{d\sigma^{h_eh_p}}{dq^2}
=\frac{d\sigma_0+h_pd\sigma_p}{dq^2}\,.
\end{align} 
It is also easy to verify that the cross section of the unpolarized
scattering process $e^{\,-}p\to e^-\Lambda_c$ is given by 
\begin{align}
\frac{d\sigma}{dq^2}=\frac{1}{4}\sum_{h_eh_p} \frac{d\sigma^{h_eh_p}}{dq^2}
=\frac{d\sigma_0}{dq^2}\,.
\end{align} 

The cross section of the double-spin transversely polarized process can be written in 
a similar way as 
\begin{align}\label{eq:double-spin-Tcs}
\frac{d^2\tilde{\sigma}^{h_eh_p}}{dq^2d\phi}\!&=\!\frac{1}{128\pi^2 m^2_p E^2}\big(\mathcal{A}_0\!+\!h_e\mathcal{A}^\prime_e\!+\!h_p\mathcal{A}^\prime_p\!+\!h_eh_p\mathcal{A}^\prime_{ep}\big) \nonumber \\
&=\!\frac{d\tilde{\sigma}_0+h_ed\tilde{\sigma}_e+h_pd\tilde{\sigma}_p+h_eh_pd\tilde{\sigma}_{ep}}{dq^2}\,.
\end{align} 
Following the same procedure as above, one can easily obtain the cross sections of other transversely polarized processes. From Eqs.~\eqref{eq:single-spin-Lecs} and \eqref{eq:single-spin-Lpcs}, one can see that 
\begin{equation}
	A^e_{L}=-\frac{d\sigma_{e}}{d\sigma_{0}}\,, \qquad A^p_{L}=-\frac{d\sigma_{p}}{d\sigma_{0}}\,,
\end{equation}
which in turn lead to 
\begin{align}
	\sigma^{h_eh_p}&=\sigma_0+h_e \sigma_{e}+h_p\sigma_p+h_eh_p\sigma_{ep} \nonumber \\[0.12cm]
	&=\sigma_0\left(1-h_eA^e_{L}-h_pA^p_{L}+h_eh_p\mathcal{C}_{ep}\right)\,,
\end{align}
where $\mathcal{C}_{ep}=\sigma_{ep}/\sigma_{0}$ represents the analyzing power of the reaction~\cite{Ohlsen:1972zz}. With the replacements $L\to T$ and $\sigma \to \tilde{\sigma}$, all the formulas above hold for the transversely polarized processes as well.

Thus far, we have been focusing on the theoretical analyses. In reality, given that neither the degree of the electron beam polarization ($P_e$) nor the degree of the proton target polarization ($P_p$) can reach $100\%$, one must take such a deficiency into account during the event-rate estimations or the data analyses in experiment. To this end, we can conveniently replace the quantities $h_{e,p}$ by $P_{e,p}h_{e,p}$ in the polarized cross sections accordingly (see, e.g., Refs.~\cite{SpinMuonSMC:1997mkb,Barone:2003fy} for explicit examples). 

With the modified cross sections at hand, it can be clearly seen that the measured single-spin asymmetries $A_{\text{exp}}$ are related to $A^e_L$ and $A^p_{L,T}$ through $A_{\text{exp}}=P A$. For the measured double-spin asymmetries, on the other hand, 
no simple relations are available in general. For example, the measured $(A^{ep}_{L3})_{\text{exp}}$ is given by
\begin{align}\label{eq:AL3}
  	\left(A^{ep}_{L3}\right)_{\text{exp}}=P_p\frac{P_e\sigma_{ep}-\sigma_{p}}{\sigma_{0}-P_e\sigma_{e}}\,,
\end{align}
which is non-trivially connected to $A^{ep}_{L3}$, 
\begin{align}
A^{ep}_{L3}=\frac{\sigma_{ep}-\sigma_{p}}{\sigma_{0}-\sigma_{e}}\,.
\end{align}
It can be seen that $A^{ep}_{\text{exp}}=P_pP_eA^{ep}$, if both $\sigma_{e}$ and $\sigma_{p}$ 
vanish, as happens in the polarized deep-inelastic lepton-nucleon inclusive scattering process in the one-photon-exchange approximation (see, e.g., Ref.~\cite{Anselmino:1994gn} and references therein). It is also interesting to note that $A^{ep}_{\text{exp}}=P_pA^{ep}$ in the limit of $P_e=1$, irrespective of whether $\sigma_{e}$ and $\sigma_{p}$ vanish or not. %Since a high degree of the electron beam polarization with $P_e\simeq 0.9$ has been achieved~\cite{HAPPEX:2011xlw}, we expect that $A^{ep}$ can be extracted from $A^{ep}_{\text{exp}}$ in this way within a tolerable accuracy.

Finally, we make a comment about the proton target polarization. Usually, the proton polarization is realized through the polarization of nucleus. However, since only a fraction of the target nucleons can be polarized, a dilution factor $f$ is often introduced to take account of this fact. Thus, $P_p$ in various differential cross sections is often accompanied by the factor $f$~\cite{SpinMuonSMC:1997mkb,Barone:2003fy}.

\section{\boldmath Amplitude squared of longitudinally polarized scattering process 
	$e^-\vec{p}\to e^-\Lambda_c$}
\label{app:LpAmplitude}

For convenience of future discussions, we provide here the explicit expression of the
amplitude squared $|\mathcal{M}|^2 $ of the longitudinally polarized scattering process $e^-(k)+\vec{p}(P)\to e^-(k^\prime)+\Lambda_c(P^\prime)$ mediated by the general low-energy effective Lagrangian $\mathcal{L}_{\text{eff}}$; note that the explicit expression of the corresponding cross sections can be obtained from Eq.~\eqref{eq:single-spin-Lpcs}. With all the operators in Eq.~\eqref{eq:Lag_LQ} taken into account, the amplitude squared $|\mathcal{M}|^2 $
with a left-handed polarized proton target ($p^-_L$) is given by 
\begin{widetext}
	\begin{align} \label{eq:Lpamplitude}
	   |\mathcal{M}|^2&=(|g_{V}^{LL}|^2+|g_{V}^{RR}|^2)\overline{|\mathcal{M}|}^2_{V_{LL}-V_{LL}}+\frac{1}{2}(|g_{V}^{LL}|^2-|g_{V}^{RR}|^2)|\mathcal{M}^\prime|^2_{V_{LL}-V_{LL}}\nonumber\\[0.12cm]
	&+(|g_{V}^{LR}|^2+|g_{V}^{RL}|^2)\overline{|\mathcal{M}|}^2_{V_{LR}-V_{LR}}+\frac{1}{2}(|g_{V}^{LR}|^2-|g_{V}^{RL}|^2)|\mathcal{M}^\prime|^2_{V_{LR}-V_{LR}}\nonumber\\[0.12cm]
	&+(|g_{S}^{L}|^2+|g_{S}^{R}|^2)\overline{|\mathcal{M}|}^2_{S_{L}-S_{L}}+\frac{1}{2}(|g_{S}^{L}|^2-|g_{S}^{R}|^2)|\mathcal{M}^\prime|^2_{S_{L}-S_{L}}\nonumber\\[0.12cm]
	&+(|g_{T}^{L}|^2+|g_{T}^{R}|^2)\overline{|\mathcal{M}|}^2_{T_{L}-T_{L}}+\frac{1}{2}(|g_{T}^{L}|^2-|g_{T}^{R}|^2)|\mathcal{M}^\prime|^2_{T_{L}-T_{L}}\nonumber\\[0.12cm]
	&+2\text{Re}[g_{V}^{LR}g_{V}^{LL*}\!+g_{V}^{RL}g_{V}^{RR*}]\overline{|\mathcal{M}|}^2_{V_{LR}-V_{LL}}\!+\text{Re}[g_{V}^{LR}g_{V}^{LL*}\!-g_{V}^{RL}g_{V}^{RR*}]|\mathcal{M}^\prime|^2_{V_{LR}-V_{LL}}\nonumber\\[0.12cm]
	&+2\text{Re}[g_{S}^{L}g_{T}^{L*}+g_{S}^{R}g_{T}^{R*}]\overline{|\mathcal{M}|}^2_{S_{L}-T_{L}}+\text{Re}[g_{S}^{L}g_{T}^{L*}-g_{S}^{R}g_{T}^{R*}]|\mathcal{M}^\prime|^2_{S_{L}-T_{L}}\,,
	\end{align}
\end{widetext}	
where a factor $2$ accounting for the average over the initial electron spins
has been taken care of, and $\overline{|\mathcal{M}|}_{\alpha-\beta}^2 $ and $|\mathcal{M}^\prime|_{\alpha-\beta}^2 $ on the right-hand side 
represent the reduced $\xi_p$-independent and $\xi_p$-dependent amplitudes squared, respectively. One can easily obtain from Eq.~\eqref{eq:Lpamplitude} the amplitude squared $|\mathcal{M}|^2$ with a right-handed polarized proton target ($p^+_L$)
by flipping the sign of $|\mathcal{M}^\prime|_{\alpha-\beta}^2 $ while keeping $\overline{|\mathcal{M}|}_{\alpha-\beta}^2$ intact. 
Since the explicit expressions of the $\xi_p$-independent $\overline{|\mathcal{M}|}_{\alpha-\beta}^2$ have already been given in Ref.~\cite{Lai:2021sww}, here we only present the $\xi_p$-dependent $|\mathcal{M}^\prime|_{\alpha-\beta}^2$, which read, respectively, as  
\begin{widetext}
	\begin{align}
	|\mathcal{M}^\prime|^2_{V_{LL}-V_{LL}} =&-\left\{\frac{q^2 f_\perp^2}{2E \big[(m_{\Lambda_c} + m_p)^2 - q^2\big]}+\frac{q^2 g_\perp^2}{2E \big[(m_{\Lambda_c} - m_p)^2 - q^2\big]}\right\}\nonumber\\[0.12cm]
	& \times(m_p^2-m_{\Lambda_c}^2 + 4 E m_p +  q^2) \big[m_p q^2 + E ( m_p^2-m_{\Lambda_c}^2+ q^2)\big]\nonumber\\[0.12cm]
	& -\frac{m_p q^2 (m_p^2\!-m_{\Lambda_c}^2 + 4 E m_p+ q^2) \big[4 E^2 m_p + m_p q^2 + 
		2 E (m_p^2\!-m_{\Lambda_c}^2 + q^2)\big]}{E \big[m_{\Lambda_c}^4 + (m_p^2 - q^2)^2 - 2 m_{\Lambda_c}^2 (m_p^2 + q^2)\big]}\nonumber\\[0.12cm]
	&\times \big[(m_{\Lambda_c} - m_p)f_\perp g_+-(m_{\Lambda_c}+ m_p)f_+ g_\perp\big]	\nonumber\\[0.12cm]
	&+\left\{\frac{q^2m_p(m_{\Lambda_c}+ m_p) f_+f_\perp}{E \big[(m_{\Lambda_c} + m_p)^2 - q^2\big]}-\frac{q^2m_p(m_{\Lambda_c} - m_p) g_+g_\perp}{E \big[(m_{\Lambda_c} - m_p)^2 - q^2\big]}\right\}\nonumber\\[0.12cm]
	&\times\big[4 E^2 m_p + m_p q^2 + 2 E (m_p^2\!-m_{\Lambda_c}^2 + q^2)\big]\nonumber\\[0.12cm]
	&-\frac{E (m_{\Lambda_c}^2 - m_p^2) - (E + m_p) q^2}{E\big[m_{\Lambda_c}^4 + (m_p^2 - q^2)^2 - 2 m_{\Lambda_c}^2 (m_p^2 + q^2)\big]}\Big\{2m_p(m_{\Lambda_c}^2 - m_p^2) \nonumber\\[0.12cm]
	&\times\big[4 E^2 m_p + m_p q^2	+ 2 E (m_p^2\!-m_{\Lambda_c}^2 + q^2)\big]f_+ g_+ -q^2\big[m_{\Lambda_c}^4 + 8 E^2 m_p^2 \nonumber\\[0.12cm]
	&+ m_p^4 + q^4 + 4 E m_p (m_p^2 + q^2) - 
	2 m_{\Lambda_c}^2 (2 E m_p + m_p^2 + q^2)\big]f_\perp g_\perp\Big\}
	\,, \\[0.2cm]
	%-------------------------------------------------------------------
	|\mathcal{M}^\prime|^2_{V_{LR}-V_{LR}}=&-\left\{\frac{q^2 f_\perp^2}{2E \big[(m_{\Lambda_c} + m_p)^2 - q^2\big]}+\frac{q^2 g_\perp^2}{2E \big[(m_{\Lambda_c} - m_p)^2 - q^2\big]}\right\}\nonumber\\[0.12cm]
	& \times(m_p^2-m_{\Lambda_c}^2 + 4 E m_p +  q^2) \big[m_p q^2 + E ( m_p^2-m_{\Lambda_c}^2+ q^2)\big]\nonumber\\[0.12cm]
	& +\frac{m_p q^2 (m_p^2\!-m_{\Lambda_c}^2 + 4 E m_p+ q^2) \big[4 E^2 m_p + m_p q^2 + 
		2 E (m_p^2\!-m_{\Lambda_c}^2 + q^2)\big]}{E \big[m_{\Lambda_c}^4 + (m_p^2 - q^2)^2 - 2 m_{\Lambda_c}^2 (m_p^2 + q^2)\big]}\nonumber\\[0.12cm]
	&\times \big[(m_{\Lambda_c} - m_p)f_\perp g_+-(m_{\Lambda_c}+ m_p)f_+ g_\perp\big]	\nonumber\\[0.12cm]
	&+\left\{\frac{q^2m_p(m_{\Lambda_c} + m_p) f_+f_\perp}{E \big[(m_{\Lambda_c} + m_p)^2 - q^2\big]}-\frac{q^2m_p(m_{\Lambda_c} - m_p) g_+g_\perp}{E \big[(m_{\Lambda_c} - m_p)^2 - q^2\big]}\right\}\nonumber\\[0.12cm]
	&\times\big[4 E^2 m_p + m_p q^2 + 2 E (m_p^2\!-m_{\Lambda_c}^2 + q^2)\big]\nonumber\\[0.12cm]
	&+\frac{E (m_{\Lambda_c}^2 - m_p^2) - (E + m_p) q^2}{E \big[m_{\Lambda_c}^4 + (m_p^2 - q^2)^2 - 2 m_{\Lambda_c}^2 (m_p^2 + q^2)\big]}\Big\{2m_p(m_{\Lambda_c}^2 - m_p^2) \nonumber\\[0.12cm]
	&\times\big[4 E^2 m_p + m_p q^2	+ 2 E (m_p^2\!-m_{\Lambda_c}^2 + q^2)\big]f_+ g_+ -q^2\big[m_{\Lambda_c}^4 + 8 E^2 m_p^2 \nonumber\\[0.12cm]
	&+ m_p^4 + q^4 + 4 E m_p (m_p^2 + q^2) - 
	2 m_{\Lambda_c}^2 (2 E m_p + m_p^2 + q^2)\big]f_\perp g_\perp\Big\}
	\,, \\[0.2cm]
	%--------------------------------------------------------------------
	|\mathcal{M}^\prime|^2_{S_{L}-S_{L}}=&\frac{q^2 f_0 \,g_0(m_{\Lambda_c}^2 - m_p^2)\big[E (m_{\Lambda_c}^2 - m_p^2 - q^2) - m_p q^2\big]}{2E m_c^2}\,,\\[0.2cm]
	%--------------------------------------------
	|\mathcal{M}^\prime|^2_{T_{L}-T_{L}}=&-\frac{16m_pq^2\big[(m_{\Lambda_c}+m_p)h_\perp \tilde{h}_+\!-\!(m_{\Lambda_c}\!-\!m_p)h_+\tilde{h}_\perp\big]}{E\big[(m_{\Lambda_c}\!-\!m_p)^2-q^2\big]\big[(m_{\Lambda_c}+m_p)^2\!-\!q^2\big]}(m_{\Lambda_c}^2\!-\!4E m_p\!-\!m_p^2\!-\!q^2)\nonumber\\[0.12cm]
	& \times\! \big[4 E^2 m_p\!+\!m_p q^2\!\!-\!2 E (m_{\Lambda_c}^2\!\!-\!m_p^2\!\!-\!q^2)\big]-\Big\{8 h_\perp \tilde{h}_\perp (m_{\Lambda_c}^2-m_p^2) m_p\big[4 E^2 m_p \nonumber\\[0.12cm]
	& + m_p q^2 + 2 E (m_p^2-m_{\Lambda_c}^2 + q^2)\big]-2 h_+ \tilde{h}_+q^2 (m_p^2-m_{\Lambda_c}^2 + 4 E m_p + q^2)^2\Big\}\nonumber\\[0.12cm]
	& \times\frac{4\big[E (m_{\Lambda_c}^2 - m_p^2 - q^2) - m_p q^2\big]}{E\big[(m_{\Lambda_c}\!\!-\!m_p)^2\!\!-\!q^2\big]\big[(m_{\Lambda_c}\!+\!m_p)^2\!\!-\!q^2\big]}\,, \\[0.2cm]
	%--------------------------------------------
	|\mathcal{M}^\prime|^2_{V_{LR}-V_{LL}}=&-\left\{\frac{q^2 f_\perp^2}{2 E \big[(m_{\Lambda_c} + m_p)^2 - q^2\big]}-\frac{q^2 g_\perp^2}{2E \big[(m_{\Lambda_c} - m_p)^2 - q^2\big]}\right\}\nonumber\\[0.12cm]
	& \times(m_p^2-m_{\Lambda_c}^2 + 4 E m_p +  q^2) \big[m_p q^2 + E ( m_p^2-m_{\Lambda_c}^2+ q^2)\big]\nonumber\\[0.12cm]
	&+\left\{\frac{q^2m_p(m_{\Lambda_c} + m_p) f_+f_\perp}{E \big[(m_{\Lambda_c} + m_p)^2 - q^2\big]}+\frac{q^2m_p(m_{\Lambda_c} - m_p) g_+g_\perp}{E \big[(m_{\Lambda_c} - m_p)^2 - q^2\big]}\right\}\nonumber\\[0.12cm]
	&\times\big[4 E^2 m_p + m_p q^2 + 2 E (m_p^2\!-m_{\Lambda_c}^2 + q^2)\big]
	\,,\\[0.2cm]
	%--------------------------------------------
	|\mathcal{M}^\prime|^2_{S_{L}-T_{L}}=&\left\{\frac{q^2(m_{\Lambda_c} + m_p)g_0\, h_+}{E m_c\big[(m_{\Lambda_c}+m_p)^2-q^2\big]}+\frac{q^2(m_{\Lambda_c} - m_p)f_0 \,\tilde{h}_+}{E m_c\big[(m_{\Lambda_c}-m_p)^2-q^2\big]}\right\}\nonumber\\[0.12cm]
	&\times( m_p^2-m_{\Lambda_c}^2 + 4 E m_p +q^2) \big[m_p q^2 + E ( m_p^2-m_{\Lambda_c}^2 + q^2)\big]\nonumber\\[0.12cm]
	&-2\left\{\frac{q^2m_p(m_{\Lambda_c} + m_p)^2g_0\, h_\perp}{E m_c\big[(m_{\Lambda_c}+m_p)^2-q^2\big]}-\frac{q^2m_p(m_{\Lambda_c} - m_p)^2f_0 \,\tilde{h}_\perp}{E m_c\big[(m_{\Lambda_c}-m_p)^2-q^2\big]}\right\}\nonumber\\[0.12cm]
	&\times	\big[4 E^2 m_p + m_p q^2 + 2 E (m_p^2-m_{\Lambda_c}^2+ q^2)\big]\,.
	\end{align}
\end{widetext}

\section{\boldmath Amplitude squared of transversely polarized scattering process 
	$e^-\vec{p}\to e^-\Lambda_c$}
\label{app:TpAmplitude}

Let us now present the explicit amplitude squared $|\mathcal{M}|^2 $ 
of the transversely polarized scattering process $e^-(k)+\vec{p}(P)\to e^-(k^\prime)+\Lambda_c(P^\prime)$ 
in the framework of the general low-energy effective Lagrangian $\mathcal{L}_{\text{eff}}$. Taking into account of all the operators in Eq.~\eqref{eq:Lag_LQ}, we write the amplitude squared $|\mathcal{M}|^2$ with a left-handed polarized proton target ($p^-_T$) as 
\begin{widetext}
	\begin{align} \label{eq:Tpamplitude}
	\overline{|\mathcal{M}|}^2&=(|g_{V}^{LL}|^2+|g_{V}^{RR}|^2)\overline{|\mathcal{M}|}^2_{V_{LL}-V_{LL}}+\frac{1}{2}(|g_{V}^{LL}|^2-|g_{V}^{RR}|^2)|\widetilde{\mathcal{M}}|^2_{V_{LL}-V_{LL}}\nonumber\\[0.12cm]
	&+(|g_{V}^{LR}|^2+|g_{V}^{RL}|^2)\overline{|\mathcal{M}|}^2_{V_{LR}-V_{LR}}+\frac{1}{2}(|g_{V}^{LR}|^2-|g_{V}^{RL}|^2)|\widetilde{\mathcal{M}}|^2_{V_{LR}-V_{LR}}\nonumber\\[0.12cm]
	&+(|g_{S}^{L}|^2+|g_{S}^{R}|^2)\overline{|\mathcal{M}|}^2_{S_{L}-S_{L}}+\frac{1}{2}(|g_{S}^{L}|^2-|g_{S}^{R}|^2)|\widetilde{\mathcal{M}}|^2_{S_{L}-S_{L}}\nonumber\\[0.12cm]
	&+(|g_{T}^{L}|^2+|g_{T}^{R}|^2)\overline{|\mathcal{M}|}^2_{T_{L}-T_{L}}+\frac{1}{2}(|g_{T}^{L}|^2-|g_{T}^{R}|^2)|\widetilde{\mathcal{M}}|^2_{T_{L}-T_{L}}\nonumber\\[0.12cm]
	&+2\text{Re}[g_{V}^{LL}g_{V}^{LR*}+g_{V}^{RL}g_{V}^{RR*}]\overline{|\mathcal{M}|}^2_{V_{LL}-V_{LR}}+\text{Re}\Big[\left(g_{V}^{LL}g_{V}^{LR*}\!-g_{V}^{RL}g_{V}^{RR*}\right)|\widetilde{\mathcal{M}}|^2_{V_{LL}-V_{LR}}\Big]\nonumber\\[0.12cm]
	&+2\text{Re}[g_{S}^{L}g_{T}^{L*}+g_{T}^{R}g_{S}^{R*}]\overline{|\mathcal{M}|}^2_{S_{L}-T_{L}}
	+\text{Re}\Big[\left(g_{S}^{L}g_{T}^{L*}-g_{T}^{R}g_{S}^{R*}\right)|\widetilde{\mathcal{M}}|^2_{S_{L}-T_{L}}\Big]\,,
	\end{align}
\end{widetext}	
where the average over the initial electron spins has been taken into account, and $|\widetilde{\mathcal{M}}|_{\alpha-\beta}^2$ on the right-hand side represents the reduced $\xi_p$-dependent amplitude squared. Same as for the previous case discussed in Appendix~\ref{app:LpAmplitude}, flipping the sign of $|\widetilde{\mathcal{M}}|_{\alpha-\beta}^2$ 
while keeping $\overline{|\mathcal{M}|}_{\alpha-\beta}^2 $ intact in Eq.~\eqref{eq:Tpamplitude}
yields the amplitude squared $|\mathcal{M}|^2$ of the scattering process with a right-handed polarized proton target ($p^+_T$). For convenience, we give the $\xi_p$-dependent $|\widetilde{\mathcal{M}}|_{\alpha-\beta}^2$, respectively, as
\begin{widetext}
	\begin{align}
	|\widetilde{\mathcal{M}}|^2_{V_{LL}-V_{LL}} =&\left\{\frac{ 2m_p \left(m_{\Lambda_c}^2-m_p^2\right) \left[4 E^2 m_p+2 E \left(m_p^2-m_{\Lambda_c}^2+q^2\right)+m_p q^2\right]f_+\, g_+}{E \big[(m_{\Lambda_c} - m_p)^2 - q^2\big] \big[(m_{\Lambda_c} + m_p)^2 - q^2\big]}\right.\nonumber\\[0.12cm]
	&-\frac{f_\perp^2 \left(4 E m_p+m_p^2-m_{\Lambda_c}^2+q^2\right)}{2 E \big[(m_{\Lambda_c} + m_p)^2 - q^2\big]}+\big[f_+\, g_\perp (m_p+m_{\Lambda_c})+f_\perp \,g_+ (m_p-m_{\Lambda_c})\big]\nonumber\\[0.12cm]
	&\times\frac{\left(4 E m_p+m_p^2-m_{\Lambda_c}^2+q^2\right) \left[E \left(m_p^2-m_{\Lambda_c}^2+q^2\right)+m_p q^2\right]}{E \big[(m_{\Lambda_c}- m_p)^2 - q^2\big] \big[(m_{\Lambda_c} + m_p)^2 - q^2\big]}\nonumber\\[0.12cm]
	&+\left\{\frac{f_+\, f_\perp (m_p+m_{\Lambda_c})}{E \big[(m_{\Lambda_c} + m_p)^2 - q^2\big]}+\frac{g_+\,g_\perp (m_p-m_{\Lambda_c})}{E\big[(m_{\Lambda_c} - m_p)^2 - q^2\big]}\right\}\left[E \left(m_p^2-m_{\Lambda_c}^2+q^2\right)+m_p q^2\right]\nonumber\\[0.12cm]
	&-\frac{q^2 \left[8 E^2 m_p^2+4 E m_p \left(m_p^2-m_{\Lambda_c}^2+q^2\right)+m_p^4-2 m_p^2 m_{\Lambda_c}^2+\left(m_{\Lambda_c}^2-q^2\right)^2\right]f_\perp\,g_\perp}{E \big[(m_{\Lambda_c}-m_p)^2 - q^2\big] \big[(m_{\Lambda_c} + m_p)^2 - q^2\big]}\nonumber\\[0.12cm]
	&\left.-\frac{g_\perp^2 \left(4 E m_p+m_p^2-m_{\Lambda_c}^2+q^2\right)}{2E \big[(m_{\Lambda_c}-m_p)^2 - q^2\big]}\right\}\sqrt{-m_p q^2 \left(2 E \left(2 E m_p+m_p^2-m_{\Lambda_c}^2+q^2\right)+m_p q^2\right)} \cos (\phi )
	\,, \\[0.2cm]
   %-------------------------------------------------------------------
   |\widetilde{\mathcal{M}}|^2_{V_{LR}-V_{LR}}=&\left\{-\frac{2m_p \left(m_{\Lambda_c}^2-m_p^2\right) \left[4 E^2 m_p+2 E \left(m_p^2-m_{\Lambda_c}^2+q^2\right)+m_p q^2\right]f_+\, g_+}{E \big[(m_{\Lambda_c} - m_p)^2 - q^2\big] \big[(m_{\Lambda_c} + m_p)^2 - q^2\big]}\right.\nonumber\\[0.12cm]
   &-\frac{f_\perp^2 \left(4 E m_p+m_p^2-m_{\Lambda_c}^2+q^2\right)}{2E \big[(m_{\Lambda_c} + m_p)^2 - q^2\big]}-\big[f_+\, g_\perp (m_p+m_{\Lambda_c})+f_\perp \,g_+ (m_p-m_{\Lambda_c})\big]\nonumber\\[0.12cm]
   &\times\frac{\left(4 E m_p+m_p^2-m_{\Lambda_c}^2+q^2\right) \left[E \left(m_p^2-m_{\Lambda_c}^2+q^2\right)+m_p q^2\right]}{E \big[(m_{\Lambda_c}- m_p)^2 - q^2\big] \big[(m_{\Lambda_c} + m_p)^2 - q^2\big]}\nonumber\\[0.12cm]
   &+\left\{\frac{f_+\, f_\perp (m_p+m_{\Lambda_c})}{E \big[(m_{\Lambda_c} + m_p)^2 - q^2\big]}+\frac{g_+\,g_\perp (m_p-m_{\Lambda_c})}{E\big[(m_{\Lambda_c} - m_p)^2 - q^2\big]}\right\}\left[E \left(m_p^2-m_{\Lambda_c}^2+q^2\right)+m_p q^2\right]\nonumber\\[0.12cm]
   &+\frac{q^2 \left[8 E^2 m_p^2+4 E m_p \left(m_p^2-m_{\Lambda_c}^2+q^2\right)+m_p^4-2 m_p^2  m_{\Lambda_c}^2+\left(m_{\Lambda_c}^2-q^2\right)^2\right]f_\perp\,g_\perp}{E \big[(m_{\Lambda_c}-m_p)^2 - q^2\big] \big[(m_{\Lambda_c} + m_p)^2 - q^2\big]}\nonumber\\[0.12cm]
   &\left.-\frac{g_\perp^2 \left(4 E m_p+m_p^2-m_{\Lambda_c}^2+q^2\right)}{2E \big[(m_{\Lambda_c}-m_p)^2 - q^2\big]}\right\}\sqrt{-m_p q^2 \left(2 E \left(2 E m_p+m_p^2-m_{\Lambda_c}^2+q^2\right)+m_p q^2\right)} \cos (\phi )
   \,, \\[0.2cm]
   %--------------------------------------------------------------------
   |\widetilde{\mathcal{M}}|^2_{S_{L}-S_{L}}=&-\frac{q^2 f_0 \,g_0(m_{\Lambda_c}^2 - m_p^2)  \sqrt{-m_p q^2 \left[2 E \left(2 E m_p+m_p^2-m_{\Lambda_c}^2+q^2\right)+m_p q^2\right]} \cos (\phi )}{2E m_c^2}\,,\\[0.2cm]
   %--------------------------------------------	
   |\widetilde{\mathcal{M}}|^2_{T_{L}-T_{L}}=&4\Big\{8\left(m_{\Lambda_c}^2-m_p^2\right) m_p  \left[(4 E^2 m_p+2 E \left(m_p^2-m_{\Lambda_c}^2+q^2\right)+m_p q^2\right]h_\perp\, \tilde{h}_\perp \nonumber\\[0.12cm]
   &-2q^2\left(4Em_p+m_p^2-m_{\Lambda_c}^2+q^2\right)^2\tilde{h}_+\, h_+ +4\left[h_+\,\tilde{h}_\perp (m_p-m_{\Lambda_c})+h_\perp \tilde{h}_+ (m_p+m_{\Lambda_c})\right]\nonumber\\[0.12cm]	
   &\times \left(4 E m_p+m_p^2-m_{\Lambda_c}^2+q^2\right) \left[E  \left(m_p^2-m_{\Lambda_c}^2+q^2\right)+m_p q^2\right]\Big\}\nonumber\\[0.12cm]
   &\times\frac{\sqrt{-m_p q^2 \left(2 E \left(2 E m_p+m_p^2-m_{\Lambda_c}^2+q^2\right)+m_p  q^2\right)}\cos (\phi )}{E\left[(m_{\Lambda_c} - m_p)^2-q^2\right] \left[(m_{\Lambda_c} + m_p)^2-q^2\right]}\,,\\[0.2cm]
   %--------------------------------------------
   |\widetilde{\mathcal{M}}|^2_{V_{LL}-V_{LR}}=&\left\{\left\{\frac{q^2 g_\perp^2}{2 E  \big[(m_{\Lambda_c} - m_p)^2 - q^2\big]}-\frac{q^2 f_\perp^2}{2 E \big[(m_{\Lambda_c} + m_p)^2 - q^2\big]}\right\}(4 E m_p + m_p^2 - m_{\Lambda_c}^2 + q^2)\cos(\phi)\right.\nonumber\\[0.12cm]
   &\hspace{-0.2cm} -\left\{\frac{(m_p -m_{\Lambda_c}) g_+\, g_\perp}{E \big[(m_{\Lambda_c}- m_p)^2 - q^2\big]}-\frac{(m_p + m_{\Lambda_c}) f_p\, f_\perp}{E \big[(m_{\Lambda_c}+ m_p)^2 - q^2\big]}\right\}\left[m_p q^2 + E (m_p^2 - m_{\Lambda_c}^2 + q^2)\right]\cos(\phi)\nonumber\\[0.12cm]
   &\hspace{-0.2cm} -\!i\left[f_\perp\, g_+ (m_p \!-\! m_{\Lambda_c}) \!-\! f_+\, g_\perp (m_p  \!+\!m_{\Lambda_c})\right] \sin (\phi ) \Bigg\}\sqrt{-m_p q^2 \left(2 E \left(2 E m_p\!+\!m_p^2\!-\!m_{\Lambda_c}^2\!+\!q^2\right)\!+\!m_p q^2\right)}
   \,,\\[0.2cm]
   %--------------------------------------------
   |\widetilde{\mathcal{M}}|^2_{S_{L}-T_{L}}&=\left\{\left\{\frac{(m_{\Lambda_c}-m_p)f_0\,\tilde{h}_+}{E m_c \big[(m_{\Lambda_c} - m_p)^2 - q^2\big]}+\frac{(m_p+m_{\Lambda_c})g_0\, h_+}{E m_c \big[(m_{\Lambda_c} + m_p)^2 - q^2\big]}\right\} q^2(4 E m_p + m_p^2 - m_{\Lambda_c}^2 + q^2)\cos (\phi )\right.\nonumber\\[0.12cm]
   &+2\left\{\frac{(m_p-m_{\Lambda_c})^2f_0\,\tilde{h}_\perp}{E m_c \big[(m_{\Lambda_c} -  m_p)^2 - q^2\big]}-\frac{(m_p+m_{\Lambda_c})^2g_0\, h_\perp}{E m_c \big[(m_{\Lambda_c} + m_p)^2 - q^2\big]}\right\}\left[E \left(m_p^2-m_{\Lambda_c}^2+t\right)+m_p q^2\right]\cos (\phi )\nonumber\\[0.12cm]
   &\left.+\frac{2i \left(m_p^2-m_{\Lambda_c}^2\right)(f_0\,h_\perp-g_0\,  \tilde{h}_\perp)}{m_c} \sin (\phi ) \right\}\sqrt{-m_p q^2 \left(2 E \left(2 E m_p+m_p^2-m_{\Lambda_c}^2+q^2\right)+m_p q^2\right)}\,.
   \end{align}
\end{widetext}
It can be seen that $|\widetilde{\mathcal{M}}|_{\alpha-\beta}^2$ induced by the interference between two different operators consists of both the real and imaginary terms. Moreover, 
the real terms are proportional to $\cos(\phi)$, while the imaginary ones to $\sin(\phi)$.  
By contrast, $|\widetilde{\mathcal{M}}|_{\alpha-\alpha}^2$ induced by the same operator 
only has the real terms, which are proportional to $\cos(\phi)$.

\bibliographystyle{apsrev4-1}
\bibliography{reference}

\begin{thebibliography}{123}%
\makeatletter
\providecommand \@ifxundefined [1]{%
 \@ifx{#1\undefined}
}%
\providecommand \@ifnum [1]{%
 \ifnum #1\expandafter \@firstoftwo
 \else \expandafter \@secondoftwo
 \fi
}%
\providecommand \@ifx [1]{%
 \ifx #1\expandafter \@firstoftwo
 \else \expandafter \@secondoftwo
 \fi
}%
\providecommand \natexlab [1]{#1}%
\providecommand \enquote  [1]{``#1''}%
\providecommand \bibnamefont  [1]{#1}%
\providecommand \bibfnamefont [1]{#1}%
\providecommand \citenamefont [1]{#1}%
\providecommand \href@noop [0]{\@secondoftwo}%
\providecommand \href [0]{\begingroup \@sanitize@url \@href}%
\providecommand \@href[1]{\@@startlink{#1}\@@href}%
\providecommand \@@href[1]{\endgroup#1\@@endlink}%
\providecommand \@sanitize@url [0]{\catcode `\\12\catcode `\$12\catcode
  `\&12\catcode `\#12\catcode `\^12\catcode `\_12\catcode `\%12\relax}%
\providecommand \@@startlink[1]{}%
\providecommand \@@endlink[0]{}%
\providecommand \url  [0]{\begingroup\@sanitize@url \@url }%
\providecommand \@url [1]{\endgroup\@href {#1}{\urlprefix }}%
\providecommand \urlprefix  [0]{URL }%
\providecommand \Eprint [0]{\href }%
\providecommand \doibase [0]{http://dx.doi.org/}%
\providecommand \selectlanguage [0]{\@gobble}%
\providecommand \bibinfo  [0]{\@secondoftwo}%
\providecommand \bibfield  [0]{\@secondoftwo}%
\providecommand \translation [1]{[#1]}%
\providecommand \BibitemOpen [0]{}%
\providecommand \bibitemStop [0]{}%
\providecommand \bibitemNoStop [0]{.\EOS\space}%
\providecommand \EOS [0]{\spacefactor3000\relax}%
\providecommand \BibitemShut  [1]{\csname bibitem#1\endcsname}%
\let\auto@bib@innerbib\@empty
%</preamble>
\bibitem [{\citenamefont {Pati}\ and\ \citenamefont
  {Salam}(1974)}]{Pati:1974yy}%
  \BibitemOpen
  \bibfield  {author} {\bibinfo {author} {\bibfnamefont {J.~C.}\ \bibnamefont
  {Pati}}\ and\ \bibinfo {author} {\bibfnamefont {A.}~\bibnamefont {Salam}},\
  }\href {\doibase 10.1103/PhysRevD.10.275} {\bibfield  {journal} {\bibinfo
  {journal} {Phys. Rev. D}\ }\textbf {\bibinfo {volume} {10}},\ \bibinfo
  {pages} {275} (\bibinfo {year} {1974})},\ \bibinfo {note} {[Erratum:
  Phys.Rev.D 11, 703--703 (1975)]}\BibitemShut {NoStop}%
\bibitem [{\citenamefont {Georgi}\ and\ \citenamefont
  {Glashow}(1974)}]{Georgi:1974sy}%
  \BibitemOpen
  \bibfield  {author} {\bibinfo {author} {\bibfnamefont {H.}~\bibnamefont
  {Georgi}}\ and\ \bibinfo {author} {\bibfnamefont {S.~L.}\ \bibnamefont
  {Glashow}},\ }\href {\doibase 10.1103/PhysRevLett.32.438} {\bibfield
  {journal} {\bibinfo  {journal} {Phys. Rev. Lett.}\ }\textbf {\bibinfo
  {volume} {32}},\ \bibinfo {pages} {438} (\bibinfo {year} {1974})}\BibitemShut
  {NoStop}%
\bibitem [{\citenamefont {Georgi}\ \emph {et~al.}(1974)\citenamefont {Georgi},
  \citenamefont {Quinn},\ and\ \citenamefont {Weinberg}}]{Georgi:1974yf}%
  \BibitemOpen
  \bibfield  {author} {\bibinfo {author} {\bibfnamefont {H.}~\bibnamefont
  {Georgi}}, \bibinfo {author} {\bibfnamefont {H.~R.}\ \bibnamefont {Quinn}}, \
  and\ \bibinfo {author} {\bibfnamefont {S.}~\bibnamefont {Weinberg}},\ }\href
  {\doibase 10.1103/PhysRevLett.33.451} {\bibfield  {journal} {\bibinfo
  {journal} {Phys. Rev. Lett.}\ }\textbf {\bibinfo {volume} {33}},\ \bibinfo
  {pages} {451} (\bibinfo {year} {1974})}\BibitemShut {NoStop}%
\bibitem [{\citenamefont {Fritzsch}\ and\ \citenamefont
  {Minkowski}(1975)}]{Fritzsch:1974nn}%
  \BibitemOpen
  \bibfield  {author} {\bibinfo {author} {\bibfnamefont {H.}~\bibnamefont
  {Fritzsch}}\ and\ \bibinfo {author} {\bibfnamefont {P.}~\bibnamefont
  {Minkowski}},\ }\href {\doibase 10.1016/0003-4916(75)90211-0} {\bibfield
  {journal} {\bibinfo  {journal} {Annals Phys.}\ }\textbf {\bibinfo {volume}
  {93}},\ \bibinfo {pages} {193} (\bibinfo {year} {1975})}\BibitemShut
  {NoStop}%
\bibitem [{\citenamefont {Georgi}(1975)}]{Georgi:1974my}%
  \BibitemOpen
  \bibfield  {author} {\bibinfo {author} {\bibfnamefont {H.}~\bibnamefont
  {Georgi}},\ }\href {\doibase 10.1063/1.2947450} {\bibfield  {journal}
  {\bibinfo  {journal} {AIP Conf. Proc.}\ }\textbf {\bibinfo {volume} {23}},\
  \bibinfo {pages} {575} (\bibinfo {year} {1975})}\BibitemShut {NoStop}%
\bibitem [{\citenamefont {Senjanovic}\ and\ \citenamefont
  {Sokorac}(1983)}]{Senjanovic:1982ex}%
  \BibitemOpen
  \bibfield  {author} {\bibinfo {author} {\bibfnamefont {G.}~\bibnamefont
  {Senjanovic}}\ and\ \bibinfo {author} {\bibfnamefont {A.}~\bibnamefont
  {Sokorac}},\ }\href {\doibase 10.1007/BF01574858} {\bibfield  {journal}
  {\bibinfo  {journal} {Z. Phys. C}\ }\textbf {\bibinfo {volume} {20}},\
  \bibinfo {pages} {255} (\bibinfo {year} {1983})}\BibitemShut {NoStop}%
\bibitem [{\citenamefont {Witten}(1985)}]{Witten:1985xc}%
  \BibitemOpen
  \bibfield  {author} {\bibinfo {author} {\bibfnamefont {E.}~\bibnamefont
  {Witten}},\ }\href {\doibase 10.1016/0550-3213(85)90603-0} {\bibfield
  {journal} {\bibinfo  {journal} {Nucl. Phys. B}\ }\textbf {\bibinfo {volume}
  {258}},\ \bibinfo {pages} {75} (\bibinfo {year} {1985})}\BibitemShut
  {NoStop}%
\bibitem [{\citenamefont {Frampton}\ and\ \citenamefont
  {Lee}(1990)}]{Frampton:1989fu}%
  \BibitemOpen
  \bibfield  {author} {\bibinfo {author} {\bibfnamefont {P.~H.}\ \bibnamefont
  {Frampton}}\ and\ \bibinfo {author} {\bibfnamefont {B.-H.}\ \bibnamefont
  {Lee}},\ }\href {\doibase 10.1103/PhysRevLett.64.619} {\bibfield  {journal}
  {\bibinfo  {journal} {Phys. Rev. Lett.}\ }\textbf {\bibinfo {volume} {64}},\
  \bibinfo {pages} {619} (\bibinfo {year} {1990})}\BibitemShut {NoStop}%
\bibitem [{\citenamefont {Murayama}\ and\ \citenamefont
  {Yanagida}(1992)}]{Murayama:1991ah}%
  \BibitemOpen
  \bibfield  {author} {\bibinfo {author} {\bibfnamefont {H.}~\bibnamefont
  {Murayama}}\ and\ \bibinfo {author} {\bibfnamefont {T.}~\bibnamefont
  {Yanagida}},\ }\href {\doibase 10.1142/S0217732392000070} {\bibfield
  {journal} {\bibinfo  {journal} {Mod. Phys. Lett. A}\ }\textbf {\bibinfo
  {volume} {7}},\ \bibinfo {pages} {147} (\bibinfo {year} {1992})}\BibitemShut
  {NoStop}%
\bibitem [{\citenamefont {Dorsner}\ and\ \citenamefont
  {Fileviez~Perez}(2005)}]{Dorsner:2005fq}%
  \BibitemOpen
  \bibfield  {author} {\bibinfo {author} {\bibfnamefont {I.}~\bibnamefont
  {Dorsner}}\ and\ \bibinfo {author} {\bibfnamefont {P.}~\bibnamefont
  {Fileviez~Perez}},\ }\href {\doibase 10.1016/j.nuclphysb.2005.06.016}
  {\bibfield  {journal} {\bibinfo  {journal} {Nucl. Phys. B}\ }\textbf
  {\bibinfo {volume} {723}},\ \bibinfo {pages} {53} (\bibinfo {year} {2005})},\
  \Eprint {http://arxiv.org/abs/hep-ph/0504276} {arXiv:hep-ph/0504276}
  \BibitemShut {NoStop}%
\bibitem [{\citenamefont {Dor\v{s}ner}\ \emph {et~al.}(2016)\citenamefont
  {Dor\v{s}ner}, \citenamefont {Fajfer}, \citenamefont {Greljo}, \citenamefont
  {Kamenik},\ and\ \citenamefont {Ko\v{s}nik}}]{Dorsner:2016wpm}%
  \BibitemOpen
  \bibfield  {author} {\bibinfo {author} {\bibfnamefont {I.}~\bibnamefont
  {Dor\v{s}ner}}, \bibinfo {author} {\bibfnamefont {S.}~\bibnamefont {Fajfer}},
  \bibinfo {author} {\bibfnamefont {A.}~\bibnamefont {Greljo}}, \bibinfo
  {author} {\bibfnamefont {J.~F.}\ \bibnamefont {Kamenik}}, \ and\ \bibinfo
  {author} {\bibfnamefont {N.}~\bibnamefont {Ko\v{s}nik}},\ }\href {\doibase
  10.1016/j.physrep.2016.06.001} {\bibfield  {journal} {\bibinfo  {journal}
  {Phys. Rept.}\ }\textbf {\bibinfo {volume} {641}},\ \bibinfo {pages} {1}
  (\bibinfo {year} {2016})},\ \Eprint {http://arxiv.org/abs/1603.04993}
  {arXiv:1603.04993 [hep-ph]} \BibitemShut {NoStop}%
\bibitem [{\citenamefont {Buchmuller}\ \emph {et~al.}(1987)\citenamefont
  {Buchmuller}, \citenamefont {Ruckl},\ and\ \citenamefont
  {Wyler}}]{Buchmuller:1986zs}%
  \BibitemOpen
  \bibfield  {author} {\bibinfo {author} {\bibfnamefont {W.}~\bibnamefont
  {Buchmuller}}, \bibinfo {author} {\bibfnamefont {R.}~\bibnamefont {Ruckl}}, \
  and\ \bibinfo {author} {\bibfnamefont {D.}~\bibnamefont {Wyler}},\ }\href
  {\doibase 10.1016/0370-2693(87)90637-X} {\bibfield  {journal} {\bibinfo
  {journal} {Phys. Lett. B}\ }\textbf {\bibinfo {volume} {191}},\ \bibinfo
  {pages} {442} (\bibinfo {year} {1987})},\ \bibinfo {note} {[Erratum:
  Phys.Lett.B 448, 320--320 (1999)]}\BibitemShut {NoStop}%
\bibitem [{\citenamefont {Bifani}\ \emph {et~al.}(2019)\citenamefont {Bifani},
  \citenamefont {Descotes-Genon}, \citenamefont {Romero~Vidal},\ and\
  \citenamefont {Schune}}]{Bifani:2018zmi}%
  \BibitemOpen
  \bibfield  {author} {\bibinfo {author} {\bibfnamefont {S.}~\bibnamefont
  {Bifani}}, \bibinfo {author} {\bibfnamefont {S.}~\bibnamefont
  {Descotes-Genon}}, \bibinfo {author} {\bibfnamefont {A.}~\bibnamefont
  {Romero~Vidal}}, \ and\ \bibinfo {author} {\bibfnamefont {M.-H.}\
  \bibnamefont {Schune}},\ }\href {\doibase 10.1088/1361-6471/aaf5de}
  {\bibfield  {journal} {\bibinfo  {journal} {J. Phys. G}\ }\textbf {\bibinfo
  {volume} {46}},\ \bibinfo {pages} {023001} (\bibinfo {year} {2019})},\
  \Eprint {http://arxiv.org/abs/1809.06229} {arXiv:1809.06229 [hep-ex]}
  \BibitemShut {NoStop}%
\bibitem [{\citenamefont {Bernlochner}\ \emph {et~al.}(2022)\citenamefont
  {Bernlochner}, \citenamefont {Sevilla}, \citenamefont {Robinson},\ and\
  \citenamefont {Wormser}}]{Bernlochner:2021vlv}%
  \BibitemOpen
  \bibfield  {author} {\bibinfo {author} {\bibfnamefont {F.~U.}\ \bibnamefont
  {Bernlochner}}, \bibinfo {author} {\bibfnamefont {M.~F.}\ \bibnamefont
  {Sevilla}}, \bibinfo {author} {\bibfnamefont {D.~J.}\ \bibnamefont
  {Robinson}}, \ and\ \bibinfo {author} {\bibfnamefont {G.}~\bibnamefont
  {Wormser}},\ }\href {\doibase 10.1103/RevModPhys.94.015003} {\bibfield
  {journal} {\bibinfo  {journal} {Rev. Mod. Phys.}\ }\textbf {\bibinfo {volume}
  {94}},\ \bibinfo {pages} {015003} (\bibinfo {year} {2022})},\ \Eprint
  {http://arxiv.org/abs/2101.08326} {arXiv:2101.08326 [hep-ex]} \BibitemShut
  {NoStop}%
\bibitem [{\citenamefont {Albrecht}\ \emph {et~al.}(2021)\citenamefont
  {Albrecht}, \citenamefont {van Dyk},\ and\ \citenamefont
  {Langenbruch}}]{Albrecht:2021tul}%
  \BibitemOpen
  \bibfield  {author} {\bibinfo {author} {\bibfnamefont {J.}~\bibnamefont
  {Albrecht}}, \bibinfo {author} {\bibfnamefont {D.}~\bibnamefont {van Dyk}}, \
  and\ \bibinfo {author} {\bibfnamefont {C.}~\bibnamefont {Langenbruch}},\
  }\href {\doibase 10.1016/j.ppnp.2021.103885} {\bibfield  {journal} {\bibinfo
  {journal} {Prog. Part. Nucl. Phys.}\ }\textbf {\bibinfo {volume} {120}},\
  \bibinfo {pages} {103885} (\bibinfo {year} {2021})},\ \Eprint
  {http://arxiv.org/abs/2107.04822} {arXiv:2107.04822 [hep-ex]} \BibitemShut
  {NoStop}%
\bibitem [{\citenamefont {London}\ and\ \citenamefont
  {Matias}(2021)}]{London:2021lfn}%
  \BibitemOpen
  \bibfield  {author} {\bibinfo {author} {\bibfnamefont {D.}~\bibnamefont
  {London}}\ and\ \bibinfo {author} {\bibfnamefont {J.}~\bibnamefont
  {Matias}},\ }\href {\doibase 10.1146/annurev-nucl-102020-090209} {\
  (\bibinfo {year} {2021}),\ 10.1146/annurev-nucl-102020-090209},\ \Eprint
  {http://arxiv.org/abs/2110.13270} {arXiv:2110.13270 [hep-ph]} \BibitemShut
  {NoStop}%
\bibitem [{\citenamefont {Aoyama}\ \emph {et~al.}(2020)\citenamefont {Aoyama}
  \emph {et~al.}}]{Aoyama:2020ynm}%
  \BibitemOpen
  \bibfield  {author} {\bibinfo {author} {\bibfnamefont {T.}~\bibnamefont
  {Aoyama}} \emph {et~al.},\ }\href {\doibase 10.1016/j.physrep.2020.07.006}
  {\bibfield  {journal} {\bibinfo  {journal} {Phys. Rept.}\ }\textbf {\bibinfo
  {volume} {887}},\ \bibinfo {pages} {1} (\bibinfo {year} {2020})},\ \Eprint
  {http://arxiv.org/abs/2006.04822} {arXiv:2006.04822 [hep-ph]} \BibitemShut
  {NoStop}%
\bibitem [{\citenamefont {Alonso}\ \emph {et~al.}(2015)\citenamefont {Alonso},
  \citenamefont {Grinstein},\ and\ \citenamefont
  {Martin~Camalich}}]{Alonso:2015sja}%
  \BibitemOpen
  \bibfield  {author} {\bibinfo {author} {\bibfnamefont {R.}~\bibnamefont
  {Alonso}}, \bibinfo {author} {\bibfnamefont {B.}~\bibnamefont {Grinstein}}, \
  and\ \bibinfo {author} {\bibfnamefont {J.}~\bibnamefont {Martin~Camalich}},\
  }\href {\doibase 10.1007/JHEP10(2015)184} {\bibfield  {journal} {\bibinfo
  {journal} {JHEP}\ }\textbf {\bibinfo {volume} {10}},\ \bibinfo {pages} {184}
  (\bibinfo {year} {2015})},\ \Eprint {http://arxiv.org/abs/1505.05164}
  {arXiv:1505.05164 [hep-ph]} \BibitemShut {NoStop}%
\bibitem [{\citenamefont {Bauer}\ and\ \citenamefont
  {Neubert}(2016)}]{Bauer:2015knc}%
  \BibitemOpen
  \bibfield  {author} {\bibinfo {author} {\bibfnamefont {M.}~\bibnamefont
  {Bauer}}\ and\ \bibinfo {author} {\bibfnamefont {M.}~\bibnamefont
  {Neubert}},\ }\href {\doibase 10.1103/PhysRevLett.116.141802} {\bibfield
  {journal} {\bibinfo  {journal} {Phys. Rev. Lett.}\ }\textbf {\bibinfo
  {volume} {116}},\ \bibinfo {pages} {141802} (\bibinfo {year} {2016})},\
  \Eprint {http://arxiv.org/abs/1511.01900} {arXiv:1511.01900 [hep-ph]}
  \BibitemShut {NoStop}%
\bibitem [{\citenamefont {Barbieri}\ \emph {et~al.}(2016)\citenamefont
  {Barbieri}, \citenamefont {Isidori}, \citenamefont {Pattori},\ and\
  \citenamefont {Senia}}]{Barbieri:2015yvd}%
  \BibitemOpen
  \bibfield  {author} {\bibinfo {author} {\bibfnamefont {R.}~\bibnamefont
  {Barbieri}}, \bibinfo {author} {\bibfnamefont {G.}~\bibnamefont {Isidori}},
  \bibinfo {author} {\bibfnamefont {A.}~\bibnamefont {Pattori}}, \ and\
  \bibinfo {author} {\bibfnamefont {F.}~\bibnamefont {Senia}},\ }\href
  {\doibase 10.1140/epjc/s10052-016-3905-3} {\bibfield  {journal} {\bibinfo
  {journal} {Eur. Phys. J. C}\ }\textbf {\bibinfo {volume} {76}},\ \bibinfo
  {pages} {67} (\bibinfo {year} {2016})},\ \Eprint
  {http://arxiv.org/abs/1512.01560} {arXiv:1512.01560 [hep-ph]} \BibitemShut
  {NoStop}%
\bibitem [{\citenamefont {Das}\ \emph {et~al.}(2016)\citenamefont {Das},
  \citenamefont {Hati}, \citenamefont {Kumar},\ and\ \citenamefont
  {Mahajan}}]{Das:2016vkr}%
  \BibitemOpen
  \bibfield  {author} {\bibinfo {author} {\bibfnamefont {D.}~\bibnamefont
  {Das}}, \bibinfo {author} {\bibfnamefont {C.}~\bibnamefont {Hati}}, \bibinfo
  {author} {\bibfnamefont {G.}~\bibnamefont {Kumar}}, \ and\ \bibinfo {author}
  {\bibfnamefont {N.}~\bibnamefont {Mahajan}},\ }\href {\doibase
  10.1103/PhysRevD.94.055034} {\bibfield  {journal} {\bibinfo  {journal} {Phys.
  Rev. D}\ }\textbf {\bibinfo {volume} {94}},\ \bibinfo {pages} {055034}
  (\bibinfo {year} {2016})},\ \Eprint {http://arxiv.org/abs/1605.06313}
  {arXiv:1605.06313 [hep-ph]} \BibitemShut {NoStop}%
\bibitem [{\citenamefont {Be\v{c}irevi\'c}\ \emph {et~al.}(2016)\citenamefont
  {Be\v{c}irevi\'c}, \citenamefont {Fajfer}, \citenamefont {Ko\v{s}nik},\ and\
  \citenamefont {Sumensari}}]{Becirevic:2016yqi}%
  \BibitemOpen
  \bibfield  {author} {\bibinfo {author} {\bibfnamefont {D.}~\bibnamefont
  {Be\v{c}irevi\'c}}, \bibinfo {author} {\bibfnamefont {S.}~\bibnamefont
  {Fajfer}}, \bibinfo {author} {\bibfnamefont {N.}~\bibnamefont {Ko\v{s}nik}},
  \ and\ \bibinfo {author} {\bibfnamefont {O.}~\bibnamefont {Sumensari}},\
  }\href {\doibase 10.1103/PhysRevD.94.115021} {\bibfield  {journal} {\bibinfo
  {journal} {Phys. Rev. D}\ }\textbf {\bibinfo {volume} {94}},\ \bibinfo
  {pages} {115021} (\bibinfo {year} {2016})},\ \Eprint
  {http://arxiv.org/abs/1608.08501} {arXiv:1608.08501 [hep-ph]} \BibitemShut
  {NoStop}%
\bibitem [{\citenamefont {Sahoo}\ \emph {et~al.}(2017)\citenamefont {Sahoo},
  \citenamefont {Mohanta},\ and\ \citenamefont {Giri}}]{Sahoo:2016pet}%
  \BibitemOpen
  \bibfield  {author} {\bibinfo {author} {\bibfnamefont {S.}~\bibnamefont
  {Sahoo}}, \bibinfo {author} {\bibfnamefont {R.}~\bibnamefont {Mohanta}}, \
  and\ \bibinfo {author} {\bibfnamefont {A.~K.}\ \bibnamefont {Giri}},\ }\href
  {\doibase 10.1103/PhysRevD.95.035027} {\bibfield  {journal} {\bibinfo
  {journal} {Phys. Rev. D}\ }\textbf {\bibinfo {volume} {95}},\ \bibinfo
  {pages} {035027} (\bibinfo {year} {2017})},\ \Eprint
  {http://arxiv.org/abs/1609.04367} {arXiv:1609.04367 [hep-ph]} \BibitemShut
  {NoStop}%
\bibitem [{\citenamefont {Popov}\ and\ \citenamefont
  {White}(2017)}]{Popov:2016fzr}%
  \BibitemOpen
  \bibfield  {author} {\bibinfo {author} {\bibfnamefont {O.}~\bibnamefont
  {Popov}}\ and\ \bibinfo {author} {\bibfnamefont {G.~A.}\ \bibnamefont
  {White}},\ }\href {\doibase 10.1016/j.nuclphysb.2017.08.007} {\bibfield
  {journal} {\bibinfo  {journal} {Nucl. Phys. B}\ }\textbf {\bibinfo {volume}
  {923}},\ \bibinfo {pages} {324} (\bibinfo {year} {2017})},\ \Eprint
  {http://arxiv.org/abs/1611.04566} {arXiv:1611.04566 [hep-ph]} \BibitemShut
  {NoStop}%
\bibitem [{\citenamefont {Chen}\ \emph {et~al.}(2017)\citenamefont {Chen},
  \citenamefont {Nomura},\ and\ \citenamefont {Okada}}]{Chen:2017hir}%
  \BibitemOpen
  \bibfield  {author} {\bibinfo {author} {\bibfnamefont {C.-H.}\ \bibnamefont
  {Chen}}, \bibinfo {author} {\bibfnamefont {T.}~\bibnamefont {Nomura}}, \ and\
  \bibinfo {author} {\bibfnamefont {H.}~\bibnamefont {Okada}},\ }\href
  {\doibase 10.1016/j.physletb.2017.10.005} {\bibfield  {journal} {\bibinfo
  {journal} {Phys. Lett. B}\ }\textbf {\bibinfo {volume} {774}},\ \bibinfo
  {pages} {456} (\bibinfo {year} {2017})},\ \Eprint
  {http://arxiv.org/abs/1703.03251} {arXiv:1703.03251 [hep-ph]} \BibitemShut
  {NoStop}%
\bibitem [{\citenamefont {Crivellin}\ \emph {et~al.}(2017)\citenamefont
  {Crivellin}, \citenamefont {M{\"u}ller},\ and\ \citenamefont
  {Ota}}]{Crivellin:2017zlb}%
  \BibitemOpen
  \bibfield  {author} {\bibinfo {author} {\bibfnamefont {A.}~\bibnamefont
  {Crivellin}}, \bibinfo {author} {\bibfnamefont {D.}~\bibnamefont
  {M{\"u}ller}}, \ and\ \bibinfo {author} {\bibfnamefont {T.}~\bibnamefont
  {Ota}},\ }\href {\doibase 10.1007/JHEP09(2017)040} {\bibfield  {journal}
  {\bibinfo  {journal} {JHEP}\ }\textbf {\bibinfo {volume} {09}},\ \bibinfo
  {pages} {040} (\bibinfo {year} {2017})},\ \Eprint
  {http://arxiv.org/abs/1703.09226} {arXiv:1703.09226 [hep-ph]} \BibitemShut
  {NoStop}%
\bibitem [{\citenamefont {Alok}\ \emph {et~al.}(2019)\citenamefont {Alok},
  \citenamefont {Kumar}, \citenamefont {Kumar},\ and\ \citenamefont
  {Sharma}}]{Alok:2017jaf}%
  \BibitemOpen
  \bibfield  {author} {\bibinfo {author} {\bibfnamefont {A.~K.}\ \bibnamefont
  {Alok}}, \bibinfo {author} {\bibfnamefont {J.}~\bibnamefont {Kumar}},
  \bibinfo {author} {\bibfnamefont {D.}~\bibnamefont {Kumar}}, \ and\ \bibinfo
  {author} {\bibfnamefont {R.}~\bibnamefont {Sharma}},\ }\href {\doibase
  10.1140/epjc/s10052-019-7219-0} {\bibfield  {journal} {\bibinfo  {journal}
  {Eur. Phys. J. C}\ }\textbf {\bibinfo {volume} {79}},\ \bibinfo {pages} {707}
  (\bibinfo {year} {2019})},\ \Eprint {http://arxiv.org/abs/1704.07347}
  {arXiv:1704.07347 [hep-ph]} \BibitemShut {NoStop}%
\bibitem [{\citenamefont {Buttazzo}\ \emph {et~al.}(2017)\citenamefont
  {Buttazzo}, \citenamefont {Greljo}, \citenamefont {Isidori},\ and\
  \citenamefont {Marzocca}}]{Buttazzo:2017ixm}%
  \BibitemOpen
  \bibfield  {author} {\bibinfo {author} {\bibfnamefont {D.}~\bibnamefont
  {Buttazzo}}, \bibinfo {author} {\bibfnamefont {A.}~\bibnamefont {Greljo}},
  \bibinfo {author} {\bibfnamefont {G.}~\bibnamefont {Isidori}}, \ and\
  \bibinfo {author} {\bibfnamefont {D.}~\bibnamefont {Marzocca}},\ }\href
  {\doibase 10.1007/JHEP11(2017)044} {\bibfield  {journal} {\bibinfo  {journal}
  {JHEP}\ }\textbf {\bibinfo {volume} {11}},\ \bibinfo {pages} {044} (\bibinfo
  {year} {2017})},\ \Eprint {http://arxiv.org/abs/1706.07808} {arXiv:1706.07808
  [hep-ph]} \BibitemShut {NoStop}%
\bibitem [{\citenamefont {Bordone}\ \emph {et~al.}(2018)\citenamefont
  {Bordone}, \citenamefont {Cornella}, \citenamefont {Fuentes-Mart{\'\i}n},\
  and\ \citenamefont {Isidori}}]{Bordone:2018nbg}%
  \BibitemOpen
  \bibfield  {author} {\bibinfo {author} {\bibfnamefont {M.}~\bibnamefont
  {Bordone}}, \bibinfo {author} {\bibfnamefont {C.}~\bibnamefont {Cornella}},
  \bibinfo {author} {\bibfnamefont {J.}~\bibnamefont {Fuentes-Mart{\'\i}n}}, \
  and\ \bibinfo {author} {\bibfnamefont {G.}~\bibnamefont {Isidori}},\ }\href
  {\doibase 10.1007/JHEP10(2018)148} {\bibfield  {journal} {\bibinfo  {journal}
  {JHEP}\ }\textbf {\bibinfo {volume} {10}},\ \bibinfo {pages} {148} (\bibinfo
  {year} {2018})},\ \Eprint {http://arxiv.org/abs/1805.09328} {arXiv:1805.09328
  [hep-ph]} \BibitemShut {NoStop}%
\bibitem [{\citenamefont {Be\v{c}irevi\'c}\ \emph {et~al.}(2018)\citenamefont
  {Be\v{c}irevi\'c}, \citenamefont {Dor\v{s}ner}, \citenamefont {Fajfer},
  \citenamefont {Ko\v{s}nik}, \citenamefont {Faroughy},\ and\ \citenamefont
  {Sumensari}}]{Becirevic:2018afm}%
  \BibitemOpen
  \bibfield  {author} {\bibinfo {author} {\bibfnamefont {D.}~\bibnamefont
  {Be\v{c}irevi\'c}}, \bibinfo {author} {\bibfnamefont {I.}~\bibnamefont
  {Dor\v{s}ner}}, \bibinfo {author} {\bibfnamefont {S.}~\bibnamefont {Fajfer}},
  \bibinfo {author} {\bibfnamefont {N.}~\bibnamefont {Ko\v{s}nik}}, \bibinfo
  {author} {\bibfnamefont {D.~A.}\ \bibnamefont {Faroughy}}, \ and\ \bibinfo
  {author} {\bibfnamefont {O.}~\bibnamefont {Sumensari}},\ }\href {\doibase
  10.1103/PhysRevD.98.055003} {\bibfield  {journal} {\bibinfo  {journal} {Phys.
  Rev. D}\ }\textbf {\bibinfo {volume} {98}},\ \bibinfo {pages} {055003}
  (\bibinfo {year} {2018})},\ \Eprint {http://arxiv.org/abs/1806.05689}
  {arXiv:1806.05689 [hep-ph]} \BibitemShut {NoStop}%
\bibitem [{\citenamefont {Kumar}\ \emph {et~al.}(2019)\citenamefont {Kumar},
  \citenamefont {London},\ and\ \citenamefont {Watanabe}}]{Kumar:2018kmr}%
  \BibitemOpen
  \bibfield  {author} {\bibinfo {author} {\bibfnamefont {J.}~\bibnamefont
  {Kumar}}, \bibinfo {author} {\bibfnamefont {D.}~\bibnamefont {London}}, \
  and\ \bibinfo {author} {\bibfnamefont {R.}~\bibnamefont {Watanabe}},\ }\href
  {\doibase 10.1103/PhysRevD.99.015007} {\bibfield  {journal} {\bibinfo
  {journal} {Phys. Rev. D}\ }\textbf {\bibinfo {volume} {99}},\ \bibinfo
  {pages} {015007} (\bibinfo {year} {2019})},\ \Eprint
  {http://arxiv.org/abs/1806.07403} {arXiv:1806.07403 [hep-ph]} \BibitemShut
  {NoStop}%
\bibitem [{\citenamefont {Angelescu}\ \emph {et~al.}(2018)\citenamefont
  {Angelescu}, \citenamefont {Be\v{c}irevi\'c}, \citenamefont {Faroughy},\ and\
  \citenamefont {Sumensari}}]{Angelescu:2018tyl}%
  \BibitemOpen
  \bibfield  {author} {\bibinfo {author} {\bibfnamefont {A.}~\bibnamefont
  {Angelescu}}, \bibinfo {author} {\bibfnamefont {D.}~\bibnamefont
  {Be\v{c}irevi\'c}, \bibfnamefont {Damircirevi\'c}}, \bibinfo {author}
  {\bibfnamefont {D.}~\bibnamefont {Faroughy}}, \ and\ \bibinfo {author}
  {\bibfnamefont {O.}~\bibnamefont {Sumensari}},\ }\href {\doibase
  10.1007/JHEP10(2018)183} {\bibfield  {journal} {\bibinfo  {journal} {JHEP}\
  }\textbf {\bibinfo {volume} {10}},\ \bibinfo {pages} {183} (\bibinfo {year}
  {2018})},\ \Eprint {http://arxiv.org/abs/1808.08179} {arXiv:1808.08179
  [hep-ph]} \BibitemShut {NoStop}%
\bibitem [{\citenamefont {Cornella}\ \emph {et~al.}(2019)\citenamefont
  {Cornella}, \citenamefont {Fuentes-Martin},\ and\ \citenamefont
  {Isidori}}]{Cornella:2019hct}%
  \BibitemOpen
  \bibfield  {author} {\bibinfo {author} {\bibfnamefont {C.}~\bibnamefont
  {Cornella}}, \bibinfo {author} {\bibfnamefont {J.}~\bibnamefont
  {Fuentes-Martin}}, \ and\ \bibinfo {author} {\bibfnamefont {G.}~\bibnamefont
  {Isidori}},\ }\href {\doibase 10.1007/JHEP07(2019)168} {\bibfield  {journal}
  {\bibinfo  {journal} {JHEP}\ }\textbf {\bibinfo {volume} {07}},\ \bibinfo
  {pages} {168} (\bibinfo {year} {2019})},\ \Eprint
  {http://arxiv.org/abs/1903.11517} {arXiv:1903.11517 [hep-ph]} \BibitemShut
  {NoStop}%
\bibitem [{\citenamefont {Crivellin}\ \emph {et~al.}(2020)\citenamefont
  {Crivellin}, \citenamefont {M\"uller},\ and\ \citenamefont
  {Saturnino}}]{Crivellin:2019dwb}%
  \BibitemOpen
  \bibfield  {author} {\bibinfo {author} {\bibfnamefont {A.}~\bibnamefont
  {Crivellin}}, \bibinfo {author} {\bibfnamefont {D.}~\bibnamefont {M\"uller}},
  \ and\ \bibinfo {author} {\bibfnamefont {F.}~\bibnamefont {Saturnino}},\
  }\href {\doibase 10.1007/JHEP06(2020)020} {\bibfield  {journal} {\bibinfo
  {journal} {JHEP}\ }\textbf {\bibinfo {volume} {06}},\ \bibinfo {pages} {020}
  (\bibinfo {year} {2020})},\ \Eprint {http://arxiv.org/abs/1912.04224}
  {arXiv:1912.04224 [hep-ph]} \BibitemShut {NoStop}%
\bibitem [{\citenamefont {Altmannshofer}\ \emph {et~al.}(2020)\citenamefont
  {Altmannshofer}, \citenamefont {Dev}, \citenamefont {Soni},\ and\
  \citenamefont {Sui}}]{Altmannshofer:2020axr}%
  \BibitemOpen
  \bibfield  {author} {\bibinfo {author} {\bibfnamefont {W.}~\bibnamefont
  {Altmannshofer}}, \bibinfo {author} {\bibfnamefont {P.~S.~B.}\ \bibnamefont
  {Dev}}, \bibinfo {author} {\bibfnamefont {A.}~\bibnamefont {Soni}}, \ and\
  \bibinfo {author} {\bibfnamefont {Y.}~\bibnamefont {Sui}},\ }\href {\doibase
  10.1103/PhysRevD.102.015031} {\bibfield  {journal} {\bibinfo  {journal}
  {Phys. Rev. D}\ }\textbf {\bibinfo {volume} {102}},\ \bibinfo {pages}
  {015031} (\bibinfo {year} {2020})},\ \Eprint
  {http://arxiv.org/abs/2002.12910} {arXiv:2002.12910 [hep-ph]} \BibitemShut
  {NoStop}%
\bibitem [{\citenamefont {Saad}(2020)}]{Saad:2020ihm}%
  \BibitemOpen
  \bibfield  {author} {\bibinfo {author} {\bibfnamefont {S.}~\bibnamefont
  {Saad}},\ }\href {\doibase 10.1103/PhysRevD.102.015019} {\bibfield  {journal}
  {\bibinfo  {journal} {Phys. Rev. D}\ }\textbf {\bibinfo {volume} {102}},\
  \bibinfo {pages} {015019} (\bibinfo {year} {2020})},\ \Eprint
  {http://arxiv.org/abs/2005.04352} {arXiv:2005.04352 [hep-ph]} \BibitemShut
  {NoStop}%
\bibitem [{\citenamefont {Gherardi}\ \emph {et~al.}(2021)\citenamefont
  {Gherardi}, \citenamefont {Marzocca},\ and\ \citenamefont
  {Venturini}}]{Gherardi:2020qhc}%
  \BibitemOpen
  \bibfield  {author} {\bibinfo {author} {\bibfnamefont {V.}~\bibnamefont
  {Gherardi}}, \bibinfo {author} {\bibfnamefont {D.}~\bibnamefont {Marzocca}},
  \ and\ \bibinfo {author} {\bibfnamefont {E.}~\bibnamefont {Venturini}},\
  }\href {\doibase 10.1007/JHEP01(2021)138} {\bibfield  {journal} {\bibinfo
  {journal} {JHEP}\ }\textbf {\bibinfo {volume} {01}},\ \bibinfo {pages} {138}
  (\bibinfo {year} {2021})},\ \Eprint {http://arxiv.org/abs/2008.09548}
  {arXiv:2008.09548 [hep-ph]} \BibitemShut {NoStop}%
\bibitem [{\citenamefont {Babu}\ \emph {et~al.}(2021)\citenamefont {Babu},
  \citenamefont {Dev}, \citenamefont {Jana},\ and\ \citenamefont
  {Thapa}}]{Babu:2020hun}%
  \BibitemOpen
  \bibfield  {author} {\bibinfo {author} {\bibfnamefont {K.~S.}\ \bibnamefont
  {Babu}}, \bibinfo {author} {\bibfnamefont {P.~S.~B.}\ \bibnamefont {Dev}},
  \bibinfo {author} {\bibfnamefont {S.}~\bibnamefont {Jana}}, \ and\ \bibinfo
  {author} {\bibfnamefont {A.}~\bibnamefont {Thapa}},\ }\href {\doibase
  10.1007/JHEP03(2021)179} {\bibfield  {journal} {\bibinfo  {journal} {JHEP}\
  }\textbf {\bibinfo {volume} {03}},\ \bibinfo {pages} {179} (\bibinfo {year}
  {2021})},\ \Eprint {http://arxiv.org/abs/2009.01771} {arXiv:2009.01771
  [hep-ph]} \BibitemShut {NoStop}%
\bibitem [{\citenamefont {Angelescu}\ \emph {et~al.}(2021)\citenamefont
  {Angelescu}, \citenamefont {Be\v{c}irevi\'c}, \citenamefont {Faroughy},
  \citenamefont {Jaffredo},\ and\ \citenamefont
  {Sumensari}}]{Angelescu:2021lln}%
  \BibitemOpen
  \bibfield  {author} {\bibinfo {author} {\bibfnamefont {A.}~\bibnamefont
  {Angelescu}}, \bibinfo {author} {\bibfnamefont {D.}~\bibnamefont
  {Be\v{c}irevi\'c}}, \bibinfo {author} {\bibfnamefont {D.~A.}\ \bibnamefont
  {Faroughy}}, \bibinfo {author} {\bibfnamefont {F.}~\bibnamefont {Jaffredo}},
  \ and\ \bibinfo {author} {\bibfnamefont {O.}~\bibnamefont {Sumensari}},\
  }\href {\doibase 10.1103/PhysRevD.104.055017} {\bibfield  {journal} {\bibinfo
   {journal} {Phys. Rev. D}\ }\textbf {\bibinfo {volume} {104}},\ \bibinfo
  {pages} {055017} (\bibinfo {year} {2021})},\ \Eprint
  {http://arxiv.org/abs/2103.12504} {arXiv:2103.12504 [hep-ph]} \BibitemShut
  {NoStop}%
\bibitem [{\citenamefont {Cornella}\ \emph {et~al.}(2021)\citenamefont
  {Cornella}, \citenamefont {Faroughy}, \citenamefont {Fuentes-Martin},
  \citenamefont {Isidori},\ and\ \citenamefont {Neubert}}]{Cornella:2021sby}%
  \BibitemOpen
  \bibfield  {author} {\bibinfo {author} {\bibfnamefont {C.}~\bibnamefont
  {Cornella}}, \bibinfo {author} {\bibfnamefont {D.~A.}\ \bibnamefont
  {Faroughy}}, \bibinfo {author} {\bibfnamefont {J.}~\bibnamefont
  {Fuentes-Martin}}, \bibinfo {author} {\bibfnamefont {G.}~\bibnamefont
  {Isidori}}, \ and\ \bibinfo {author} {\bibfnamefont {M.}~\bibnamefont
  {Neubert}},\ }\href {\doibase 10.1007/JHEP08(2021)050} {\bibfield  {journal}
  {\bibinfo  {journal} {JHEP}\ }\textbf {\bibinfo {volume} {08}},\ \bibinfo
  {pages} {050} (\bibinfo {year} {2021})},\ \Eprint
  {http://arxiv.org/abs/2103.16558} {arXiv:2103.16558 [hep-ph]} \BibitemShut
  {NoStop}%
\bibitem [{\citenamefont {Marzocca}\ and\ \citenamefont
  {Trifinopoulos}(2021)}]{Marzocca:2021azj}%
  \BibitemOpen
  \bibfield  {author} {\bibinfo {author} {\bibfnamefont {D.}~\bibnamefont
  {Marzocca}}\ and\ \bibinfo {author} {\bibfnamefont {S.}~\bibnamefont
  {Trifinopoulos}},\ }\href {\doibase 10.1103/PhysRevLett.127.061803}
  {\bibfield  {journal} {\bibinfo  {journal} {Phys. Rev. Lett.}\ }\textbf
  {\bibinfo {volume} {127}},\ \bibinfo {pages} {061803} (\bibinfo {year}
  {2021})},\ \Eprint {http://arxiv.org/abs/2104.05730} {arXiv:2104.05730
  [hep-ph]} \BibitemShut {NoStop}%
\bibitem [{\citenamefont {Julio}\ \emph
  {et~al.}(2022{\natexlab{a}})\citenamefont {Julio}, \citenamefont {Saad},\
  and\ \citenamefont {Thapa}}]{Julio:2022ton}%
  \BibitemOpen
  \bibfield  {author} {\bibinfo {author} {\bibfnamefont {J.}~\bibnamefont
  {Julio}}, \bibinfo {author} {\bibfnamefont {S.}~\bibnamefont {Saad}}, \ and\
  \bibinfo {author} {\bibfnamefont {A.}~\bibnamefont {Thapa}},\ }\href@noop {}
  {\  (\bibinfo {year} {2022}{\natexlab{a}})},\ \Eprint
  {http://arxiv.org/abs/2202.10479} {arXiv:2202.10479 [hep-ph]} \BibitemShut
  {NoStop}%
\bibitem [{\citenamefont {Crivellin}\ \emph {et~al.}(2022)\citenamefont
  {Crivellin}, \citenamefont {Fuks},\ and\ \citenamefont
  {Schnell}}]{Crivellin:2022mff}%
  \BibitemOpen
  \bibfield  {author} {\bibinfo {author} {\bibfnamefont {A.}~\bibnamefont
  {Crivellin}}, \bibinfo {author} {\bibfnamefont {B.}~\bibnamefont {Fuks}}, \
  and\ \bibinfo {author} {\bibfnamefont {L.}~\bibnamefont {Schnell}},\
  }\href@noop {} {\  (\bibinfo {year} {2022})},\ \Eprint
  {http://arxiv.org/abs/2203.10111} {arXiv:2203.10111 [hep-ph]} \BibitemShut
  {NoStop}%
\bibitem [{\citenamefont {Julio}\ \emph
  {et~al.}(2022{\natexlab{b}})\citenamefont {Julio}, \citenamefont {Saad},\
  and\ \citenamefont {Thapa}}]{Julio:2022bue}%
  \BibitemOpen
  \bibfield  {author} {\bibinfo {author} {\bibfnamefont {J.}~\bibnamefont
  {Julio}}, \bibinfo {author} {\bibfnamefont {S.}~\bibnamefont {Saad}}, \ and\
  \bibinfo {author} {\bibfnamefont {A.}~\bibnamefont {Thapa}},\ }\href@noop {}
  {\  (\bibinfo {year} {2022}{\natexlab{b}})},\ \Eprint
  {http://arxiv.org/abs/2203.15499} {arXiv:2203.15499 [hep-ph]} \BibitemShut
  {NoStop}%
\bibitem [{\citenamefont {Shanker}(1982{\natexlab{a}})}]{Shanker:1981mj}%
  \BibitemOpen
  \bibfield  {author} {\bibinfo {author} {\bibfnamefont {O.~U.}\ \bibnamefont
  {Shanker}},\ }\href {\doibase 10.1016/0550-3213(82)90534-X} {\bibfield
  {journal} {\bibinfo  {journal} {Nucl. Phys. B}\ }\textbf {\bibinfo {volume}
  {206}},\ \bibinfo {pages} {253} (\bibinfo {year}
  {1982}{\natexlab{a}})}\BibitemShut {NoStop}%
\bibitem [{\citenamefont {Shanker}(1982{\natexlab{b}})}]{Shanker:1982nd}%
  \BibitemOpen
  \bibfield  {author} {\bibinfo {author} {\bibfnamefont {O.~U.}\ \bibnamefont
  {Shanker}},\ }\href {\doibase 10.1016/0550-3213(82)90196-1} {\bibfield
  {journal} {\bibinfo  {journal} {Nucl. Phys. B}\ }\textbf {\bibinfo {volume}
  {204}},\ \bibinfo {pages} {375} (\bibinfo {year}
  {1982}{\natexlab{b}})}\BibitemShut {NoStop}%
\bibitem [{\citenamefont {Leurer}(1994{\natexlab{a}})}]{Leurer:1993em}%
  \BibitemOpen
  \bibfield  {author} {\bibinfo {author} {\bibfnamefont {M.}~\bibnamefont
  {Leurer}},\ }\href {\doibase 10.1103/PhysRevD.49.333} {\bibfield  {journal}
  {\bibinfo  {journal} {Phys. Rev. D}\ }\textbf {\bibinfo {volume} {49}},\
  \bibinfo {pages} {333} (\bibinfo {year} {1994}{\natexlab{a}})},\ \Eprint
  {http://arxiv.org/abs/hep-ph/9309266} {arXiv:hep-ph/9309266} \BibitemShut
  {NoStop}%
\bibitem [{\citenamefont {Davidson}\ \emph {et~al.}(1994)\citenamefont
  {Davidson}, \citenamefont {Bailey},\ and\ \citenamefont
  {Campbell}}]{Davidson:1993qk}%
  \BibitemOpen
  \bibfield  {author} {\bibinfo {author} {\bibfnamefont {S.}~\bibnamefont
  {Davidson}}, \bibinfo {author} {\bibfnamefont {D.~C.}\ \bibnamefont
  {Bailey}}, \ and\ \bibinfo {author} {\bibfnamefont {B.~A.}\ \bibnamefont
  {Campbell}},\ }\href {\doibase 10.1007/BF01552629} {\bibfield  {journal}
  {\bibinfo  {journal} {Z. Phys. C}\ }\textbf {\bibinfo {volume} {61}},\
  \bibinfo {pages} {613} (\bibinfo {year} {1994})},\ \Eprint
  {http://arxiv.org/abs/hep-ph/9309310} {arXiv:hep-ph/9309310} \BibitemShut
  {NoStop}%
\bibitem [{\citenamefont {Leurer}(1994{\natexlab{b}})}]{Leurer:1993qx}%
  \BibitemOpen
  \bibfield  {author} {\bibinfo {author} {\bibfnamefont {M.}~\bibnamefont
  {Leurer}},\ }\href {\doibase 10.1103/PhysRevD.50.536} {\bibfield  {journal}
  {\bibinfo  {journal} {Phys. Rev. D}\ }\textbf {\bibinfo {volume} {50}},\
  \bibinfo {pages} {536} (\bibinfo {year} {1994}{\natexlab{b}})},\ \Eprint
  {http://arxiv.org/abs/hep-ph/9312341} {arXiv:hep-ph/9312341} \BibitemShut
  {NoStop}%
\bibitem [{\citenamefont {Carpentier}\ and\ \citenamefont
  {Davidson}(2010)}]{Carpentier:2010ue}%
  \BibitemOpen
  \bibfield  {author} {\bibinfo {author} {\bibfnamefont {M.}~\bibnamefont
  {Carpentier}}\ and\ \bibinfo {author} {\bibfnamefont {S.}~\bibnamefont
  {Davidson}},\ }\href {\doibase 10.1140/epjc/s10052-010-1482-4} {\bibfield
  {journal} {\bibinfo  {journal} {Eur. Phys. J. C}\ }\textbf {\bibinfo {volume}
  {70}},\ \bibinfo {pages} {1071} (\bibinfo {year} {2010})},\ \Eprint
  {http://arxiv.org/abs/1008.0280} {arXiv:1008.0280 [hep-ph]} \BibitemShut
  {NoStop}%
\bibitem [{\citenamefont {Bobeth}\ and\ \citenamefont
  {Buras}(2018)}]{Bobeth:2017ecx}%
  \BibitemOpen
  \bibfield  {author} {\bibinfo {author} {\bibfnamefont {C.}~\bibnamefont
  {Bobeth}}\ and\ \bibinfo {author} {\bibfnamefont {A.~J.}\ \bibnamefont
  {Buras}},\ }\href {\doibase 10.1007/JHEP02(2018)101} {\bibfield  {journal}
  {\bibinfo  {journal} {JHEP}\ }\textbf {\bibinfo {volume} {02}},\ \bibinfo
  {pages} {101} (\bibinfo {year} {2018})},\ \Eprint
  {http://arxiv.org/abs/1712.01295} {arXiv:1712.01295 [hep-ph]} \BibitemShut
  {NoStop}%
\bibitem [{\citenamefont {Dor\v{s}ner}\ \emph {et~al.}(2020)\citenamefont
  {Dor\v{s}ner}, \citenamefont {Fajfer},\ and\ \citenamefont
  {Patra}}]{Dorsner:2019vgp}%
  \BibitemOpen
  \bibfield  {author} {\bibinfo {author} {\bibfnamefont {I.}~\bibnamefont
  {Dor\v{s}ner}}, \bibinfo {author} {\bibfnamefont {S.}~\bibnamefont {Fajfer}},
  \ and\ \bibinfo {author} {\bibfnamefont {M.}~\bibnamefont {Patra}},\ }\href
  {\doibase 10.1140/epjc/s10052-020-7754-8} {\bibfield  {journal} {\bibinfo
  {journal} {Eur. Phys. J. C}\ }\textbf {\bibinfo {volume} {80}},\ \bibinfo
  {pages} {204} (\bibinfo {year} {2020})},\ \Eprint
  {http://arxiv.org/abs/1906.05660} {arXiv:1906.05660 [hep-ph]} \BibitemShut
  {NoStop}%
\bibitem [{\citenamefont {Mandal}\ and\ \citenamefont
  {Pich}(2019)}]{Mandal:2019gff}%
  \BibitemOpen
  \bibfield  {author} {\bibinfo {author} {\bibfnamefont {R.}~\bibnamefont
  {Mandal}}\ and\ \bibinfo {author} {\bibfnamefont {A.}~\bibnamefont {Pich}},\
  }\href {\doibase 10.1007/JHEP12(2019)089} {\bibfield  {journal} {\bibinfo
  {journal} {JHEP}\ }\textbf {\bibinfo {volume} {12}},\ \bibinfo {pages} {089}
  (\bibinfo {year} {2019})},\ \Eprint {http://arxiv.org/abs/1908.11155}
  {arXiv:1908.11155 [hep-ph]} \BibitemShut {NoStop}%
\bibitem [{\citenamefont {Su}\ and\ \citenamefont
  {Tandean}(2020)}]{Su:2019tjn}%
  \BibitemOpen
  \bibfield  {author} {\bibinfo {author} {\bibfnamefont {J.-Y.}\ \bibnamefont
  {Su}}\ and\ \bibinfo {author} {\bibfnamefont {J.}~\bibnamefont {Tandean}},\
  }\href {\doibase 10.1103/PhysRevD.102.075032} {\bibfield  {journal} {\bibinfo
   {journal} {Phys. Rev. D}\ }\textbf {\bibinfo {volume} {102}},\ \bibinfo
  {pages} {075032} (\bibinfo {year} {2020})},\ \Eprint
  {http://arxiv.org/abs/1912.13507} {arXiv:1912.13507 [hep-ph]} \BibitemShut
  {NoStop}%
\bibitem [{\citenamefont {Crivellin}\ and\ \citenamefont
  {Hoferichter}(2020)}]{Crivellin:2020lzu}%
  \BibitemOpen
  \bibfield  {author} {\bibinfo {author} {\bibfnamefont {A.}~\bibnamefont
  {Crivellin}}\ and\ \bibinfo {author} {\bibfnamefont {M.}~\bibnamefont
  {Hoferichter}},\ }\href {\doibase 10.1103/PhysRevLett.125.111801} {\bibfield
  {journal} {\bibinfo  {journal} {Phys. Rev. Lett.}\ }\textbf {\bibinfo
  {volume} {125}},\ \bibinfo {pages} {111801} (\bibinfo {year} {2020})},\
  \Eprint {http://arxiv.org/abs/2002.07184} {arXiv:2002.07184 [hep-ph]}
  \BibitemShut {NoStop}%
\bibitem [{\citenamefont {Crivellin}\ \emph {et~al.}(2021)\citenamefont
  {Crivellin}, \citenamefont {M\"uller},\ and\ \citenamefont
  {Schnell}}]{Crivellin:2021egp}%
  \BibitemOpen
  \bibfield  {author} {\bibinfo {author} {\bibfnamefont {A.}~\bibnamefont
  {Crivellin}}, \bibinfo {author} {\bibfnamefont {D.}~\bibnamefont {M\"uller}},
  \ and\ \bibinfo {author} {\bibfnamefont {L.}~\bibnamefont {Schnell}},\ }\href
  {\doibase 10.1103/PhysRevD.103.115023} {\bibfield  {journal} {\bibinfo
  {journal} {Phys. Rev. D}\ }\textbf {\bibinfo {volume} {103}},\ \bibinfo
  {pages} {115023} (\bibinfo {year} {2021})},\ \Eprint
  {http://arxiv.org/abs/2104.06417} {arXiv:2104.06417 [hep-ph]} \BibitemShut
  {NoStop}%
\bibitem [{\citenamefont {Marzocca}\ \emph {et~al.}(2021)\citenamefont
  {Marzocca}, \citenamefont {Trifinopoulos},\ and\ \citenamefont
  {Venturini}}]{Marzocca:2021miv}%
  \BibitemOpen
  \bibfield  {author} {\bibinfo {author} {\bibfnamefont {D.}~\bibnamefont
  {Marzocca}}, \bibinfo {author} {\bibfnamefont {S.}~\bibnamefont
  {Trifinopoulos}}, \ and\ \bibinfo {author} {\bibfnamefont {E.}~\bibnamefont
  {Venturini}},\ }\href@noop {} {\  (\bibinfo {year} {2021})},\ \Eprint
  {http://arxiv.org/abs/2106.15630} {arXiv:2106.15630 [hep-ph]} \BibitemShut
  {NoStop}%
\bibitem [{\citenamefont {Glashow}\ \emph {et~al.}(1970)\citenamefont
  {Glashow}, \citenamefont {Iliopoulos},\ and\ \citenamefont
  {Maiani}}]{Glashow:1970gm}%
  \BibitemOpen
  \bibfield  {author} {\bibinfo {author} {\bibfnamefont {S.~L.}\ \bibnamefont
  {Glashow}}, \bibinfo {author} {\bibfnamefont {J.}~\bibnamefont {Iliopoulos}},
  \ and\ \bibinfo {author} {\bibfnamefont {L.}~\bibnamefont {Maiani}},\ }\href
  {\doibase 10.1103/PhysRevD.2.1285} {\bibfield  {journal} {\bibinfo  {journal}
  {Phys. Rev. D}\ }\textbf {\bibinfo {volume} {2}},\ \bibinfo {pages} {1285}
  (\bibinfo {year} {1970})}\BibitemShut {NoStop}%
\bibitem [{\citenamefont {Burdman}\ \emph {et~al.}(2002)\citenamefont
  {Burdman}, \citenamefont {Golowich}, \citenamefont {Hewett},\ and\
  \citenamefont {Pakvasa}}]{Burdman:2001tf}%
  \BibitemOpen
  \bibfield  {author} {\bibinfo {author} {\bibfnamefont {G.}~\bibnamefont
  {Burdman}}, \bibinfo {author} {\bibfnamefont {E.}~\bibnamefont {Golowich}},
  \bibinfo {author} {\bibfnamefont {J.~L.}\ \bibnamefont {Hewett}}, \ and\
  \bibinfo {author} {\bibfnamefont {S.}~\bibnamefont {Pakvasa}},\ }\href
  {\doibase 10.1103/PhysRevD.66.014009} {\bibfield  {journal} {\bibinfo
  {journal} {Phys. Rev. D}\ }\textbf {\bibinfo {volume} {66}},\ \bibinfo
  {pages} {014009} (\bibinfo {year} {2002})},\ \Eprint
  {http://arxiv.org/abs/hep-ph/0112235} {arXiv:hep-ph/0112235} \BibitemShut
  {NoStop}%
\bibitem [{\citenamefont {Azizi}\ \emph {et~al.}(2010)\citenamefont {Azizi},
  \citenamefont {Bayar}, \citenamefont {Sarac},\ and\ \citenamefont
  {Sundu}}]{Azizi:2010zzb}%
  \BibitemOpen
  \bibfield  {author} {\bibinfo {author} {\bibfnamefont {K.}~\bibnamefont
  {Azizi}}, \bibinfo {author} {\bibfnamefont {M.}~\bibnamefont {Bayar}},
  \bibinfo {author} {\bibfnamefont {Y.}~\bibnamefont {Sarac}}, \ and\ \bibinfo
  {author} {\bibfnamefont {H.}~\bibnamefont {Sundu}},\ }\href {\doibase
  10.1088/0954-3899/37/11/115007} {\bibfield  {journal} {\bibinfo  {journal}
  {J. Phys. G}\ }\textbf {\bibinfo {volume} {37}},\ \bibinfo {pages} {115007}
  (\bibinfo {year} {2010})}\BibitemShut {NoStop}%
\bibitem [{\citenamefont {Paul}\ \emph {et~al.}(2011)\citenamefont {Paul},
  \citenamefont {Bigi},\ and\ \citenamefont {Recksiegel}}]{Paul:2011ar}%
  \BibitemOpen
  \bibfield  {author} {\bibinfo {author} {\bibfnamefont {A.}~\bibnamefont
  {Paul}}, \bibinfo {author} {\bibfnamefont {I.~I.}\ \bibnamefont {Bigi}}, \
  and\ \bibinfo {author} {\bibfnamefont {S.}~\bibnamefont {Recksiegel}},\
  }\href {\doibase 10.1103/PhysRevD.83.114006} {\bibfield  {journal} {\bibinfo
  {journal} {Phys. Rev. D}\ }\textbf {\bibinfo {volume} {83}},\ \bibinfo
  {pages} {114006} (\bibinfo {year} {2011})},\ \Eprint
  {http://arxiv.org/abs/1101.6053} {arXiv:1101.6053 [hep-ph]} \BibitemShut
  {NoStop}%
\bibitem [{\citenamefont {Cappiello}\ \emph {et~al.}(2013)\citenamefont
  {Cappiello}, \citenamefont {Cata},\ and\ \citenamefont
  {D'Ambrosio}}]{Cappiello:2012vg}%
  \BibitemOpen
  \bibfield  {author} {\bibinfo {author} {\bibfnamefont {L.}~\bibnamefont
  {Cappiello}}, \bibinfo {author} {\bibfnamefont {O.}~\bibnamefont {Cata}}, \
  and\ \bibinfo {author} {\bibfnamefont {G.}~\bibnamefont {D'Ambrosio}},\
  }\href {\doibase 10.1007/JHEP04(2013)135} {\bibfield  {journal} {\bibinfo
  {journal} {JHEP}\ }\textbf {\bibinfo {volume} {04}},\ \bibinfo {pages} {135}
  (\bibinfo {year} {2013})},\ \Eprint {http://arxiv.org/abs/1209.4235}
  {arXiv:1209.4235 [hep-ph]} \BibitemShut {NoStop}%
\bibitem [{\citenamefont {Fajfer}\ and\ \citenamefont
  {Ko\v{s}nik}(2015)}]{Fajfer:2015mia}%
  \BibitemOpen
  \bibfield  {author} {\bibinfo {author} {\bibfnamefont {S.}~\bibnamefont
  {Fajfer}}\ and\ \bibinfo {author} {\bibfnamefont {N.}~\bibnamefont
  {Ko\v{s}nik}},\ }\href {\doibase 10.1140/epjc/s10052-015-3801-2} {\bibfield
  {journal} {\bibinfo  {journal} {Eur. Phys. J. C}\ }\textbf {\bibinfo {volume}
  {75}},\ \bibinfo {pages} {567} (\bibinfo {year} {2015})},\ \Eprint
  {http://arxiv.org/abs/1510.00965} {arXiv:1510.00965 [hep-ph]} \BibitemShut
  {NoStop}%
\bibitem [{\citenamefont {\c{S}irvanli}(2016)}]{Sirvanli:2016wnr}%
  \BibitemOpen
  \bibfield  {author} {\bibinfo {author} {\bibfnamefont {B.~B.}\ \bibnamefont
  {\c{S}irvanli}},\ }\href {\doibase 10.1103/PhysRevD.93.034027} {\bibfield
  {journal} {\bibinfo  {journal} {Phys. Rev. D}\ }\textbf {\bibinfo {volume}
  {93}},\ \bibinfo {pages} {034027} (\bibinfo {year} {2016})}\BibitemShut
  {NoStop}%
\bibitem [{\citenamefont {Meinel}(2018)}]{Meinel:2017ggx}%
  \BibitemOpen
  \bibfield  {author} {\bibinfo {author} {\bibfnamefont {S.}~\bibnamefont
  {Meinel}},\ }\href {\doibase 10.1103/PhysRevD.97.034511} {\bibfield
  {journal} {\bibinfo  {journal} {Phys. Rev.}\ }\textbf {\bibinfo {volume}
  {D97}},\ \bibinfo {pages} {034511} (\bibinfo {year} {2018})},\ \Eprint
  {http://arxiv.org/abs/1712.05783} {arXiv:1712.05783 [hep-lat]} \BibitemShut
  {NoStop}%
%%CITATION = ARXIV:1712.05783;%%
\bibitem [{\citenamefont {Faustov}\ and\ \citenamefont
  {Galkin}(2018)}]{Faustov:2018dkn}%
  \BibitemOpen
  \bibfield  {author} {\bibinfo {author} {\bibfnamefont {R.~N.}\ \bibnamefont
  {Faustov}}\ and\ \bibinfo {author} {\bibfnamefont {V.~O.}\ \bibnamefont
  {Galkin}},\ }\href {\doibase 10.1140/epjc/s10052-018-6010-y} {\bibfield
  {journal} {\bibinfo  {journal} {Eur. Phys. J. C}\ }\textbf {\bibinfo {volume}
  {78}},\ \bibinfo {pages} {527} (\bibinfo {year} {2018})},\ \Eprint
  {http://arxiv.org/abs/1805.02516} {arXiv:1805.02516 [hep-ph]} \BibitemShut
  {NoStop}%
\bibitem [{\citenamefont {Gisbert}\ \emph {et~al.}(2021)\citenamefont
  {Gisbert}, \citenamefont {Golz},\ and\ \citenamefont
  {Mitzel}}]{Gisbert:2020vjx}%
  \BibitemOpen
  \bibfield  {author} {\bibinfo {author} {\bibfnamefont {H.}~\bibnamefont
  {Gisbert}}, \bibinfo {author} {\bibfnamefont {M.}~\bibnamefont {Golz}}, \
  and\ \bibinfo {author} {\bibfnamefont {D.~S.}\ \bibnamefont {Mitzel}},\
  }\href {\doibase 10.1142/S0217732321300020} {\bibfield  {journal} {\bibinfo
  {journal} {Mod. Phys. Lett. A}\ }\textbf {\bibinfo {volume} {36}},\ \bibinfo
  {pages} {2130002} (\bibinfo {year} {2021})},\ \Eprint
  {http://arxiv.org/abs/2011.09478} {arXiv:2011.09478 [hep-ph]} \BibitemShut
  {NoStop}%
\bibitem [{\citenamefont {Faisel}\ \emph {et~al.}(2021)\citenamefont {Faisel},
  \citenamefont {Su},\ and\ \citenamefont {Tandean}}]{Faisel:2020php}%
  \BibitemOpen
  \bibfield  {author} {\bibinfo {author} {\bibfnamefont {G.}~\bibnamefont
  {Faisel}}, \bibinfo {author} {\bibfnamefont {J.-Y.}\ \bibnamefont {Su}}, \
  and\ \bibinfo {author} {\bibfnamefont {J.}~\bibnamefont {Tandean}},\ }\href
  {\doibase 10.1007/JHEP04(2021)246} {\bibfield  {journal} {\bibinfo  {journal}
  {JHEP}\ }\textbf {\bibinfo {volume} {04}},\ \bibinfo {pages} {246} (\bibinfo
  {year} {2021})},\ \Eprint {http://arxiv.org/abs/2012.15847} {arXiv:2012.15847
  [hep-ph]} \BibitemShut {NoStop}%
\bibitem [{\citenamefont {Fajfer}\ and\ \citenamefont
  {Novosel}(2021)}]{Fajfer:2021woc}%
  \BibitemOpen
  \bibfield  {author} {\bibinfo {author} {\bibfnamefont {S.}~\bibnamefont
  {Fajfer}}\ and\ \bibinfo {author} {\bibfnamefont {A.}~\bibnamefont
  {Novosel}},\ }\href {\doibase 10.1103/PhysRevD.104.015014} {\bibfield
  {journal} {\bibinfo  {journal} {Phys. Rev. D}\ }\textbf {\bibinfo {volume}
  {104}},\ \bibinfo {pages} {015014} (\bibinfo {year} {2021})},\ \Eprint
  {http://arxiv.org/abs/2101.10712} {arXiv:2101.10712 [hep-ph]} \BibitemShut
  {NoStop}%
\bibitem [{\citenamefont {Bause}\ \emph {et~al.}(2021)\citenamefont {Bause},
  \citenamefont {Gisbert}, \citenamefont {Golz},\ and\ \citenamefont
  {Hiller}}]{Bause:2020xzj}%
  \BibitemOpen
  \bibfield  {author} {\bibinfo {author} {\bibfnamefont {R.}~\bibnamefont
  {Bause}}, \bibinfo {author} {\bibfnamefont {H.}~\bibnamefont {Gisbert}},
  \bibinfo {author} {\bibfnamefont {M.}~\bibnamefont {Golz}}, \ and\ \bibinfo
  {author} {\bibfnamefont {G.}~\bibnamefont {Hiller}},\ }\href {\doibase
  10.1103/PhysRevD.103.015033} {\bibfield  {journal} {\bibinfo  {journal}
  {Phys. Rev. D}\ }\textbf {\bibinfo {volume} {103}},\ \bibinfo {pages}
  {015033} (\bibinfo {year} {2021})},\ \Eprint
  {http://arxiv.org/abs/2010.02225} {arXiv:2010.02225 [hep-ph]} \BibitemShut
  {NoStop}%
\bibitem [{\citenamefont {de~Boer}\ and\ \citenamefont
  {Hiller}(2016)}]{deBoer:2015boa}%
  \BibitemOpen
  \bibfield  {author} {\bibinfo {author} {\bibfnamefont {S.}~\bibnamefont
  {de~Boer}}\ and\ \bibinfo {author} {\bibfnamefont {G.}~\bibnamefont
  {Hiller}},\ }\href {\doibase 10.1103/PhysRevD.93.074001} {\bibfield
  {journal} {\bibinfo  {journal} {Phys. Rev.}\ }\textbf {\bibinfo {volume}
  {D93}},\ \bibinfo {pages} {074001} (\bibinfo {year} {2016})},\ \Eprint
  {http://arxiv.org/abs/1510.00311} {arXiv:1510.00311 [hep-ph]} \BibitemShut
  {NoStop}%
%%CITATION = ARXIV:1510.00311;%%
\bibitem [{\citenamefont {De~Boer}\ and\ \citenamefont
  {Hiller}(2018)}]{DeBoer:2018pdx}%
  \BibitemOpen
  \bibfield  {author} {\bibinfo {author} {\bibfnamefont {S.}~\bibnamefont
  {De~Boer}}\ and\ \bibinfo {author} {\bibfnamefont {G.}~\bibnamefont
  {Hiller}},\ }\href {\doibase 10.1103/PhysRevD.98.035041} {\bibfield
  {journal} {\bibinfo  {journal} {Phys. Rev. D}\ }\textbf {\bibinfo {volume}
  {98}},\ \bibinfo {pages} {035041} (\bibinfo {year} {2018})},\ \Eprint
  {http://arxiv.org/abs/1805.08516} {arXiv:1805.08516 [hep-ph]} \BibitemShut
  {NoStop}%
\bibitem [{\citenamefont {Bause}\ \emph {et~al.}(2020)\citenamefont {Bause},
  \citenamefont {Golz}, \citenamefont {Hiller},\ and\ \citenamefont
  {Tayduganov}}]{Bause:2019vpr}%
  \BibitemOpen
  \bibfield  {author} {\bibinfo {author} {\bibfnamefont {R.}~\bibnamefont
  {Bause}}, \bibinfo {author} {\bibfnamefont {M.}~\bibnamefont {Golz}},
  \bibinfo {author} {\bibfnamefont {G.}~\bibnamefont {Hiller}}, \ and\ \bibinfo
  {author} {\bibfnamefont {A.}~\bibnamefont {Tayduganov}},\ }\href {\doibase
  10.1140/epjc/s10052-020-7621-7} {\bibfield  {journal} {\bibinfo  {journal}
  {Eur. Phys. J. C}\ }\textbf {\bibinfo {volume} {80}},\ \bibinfo {pages} {65}
  (\bibinfo {year} {2020})},\ \Eprint {http://arxiv.org/abs/1909.11108}
  {arXiv:1909.11108 [hep-ph]} \BibitemShut {NoStop}%
\bibitem [{\citenamefont {Golz}\ \emph {et~al.}(2021)\citenamefont {Golz},
  \citenamefont {Hiller},\ and\ \citenamefont {Magorsch}}]{Golz:2021imq}%
  \BibitemOpen
  \bibfield  {author} {\bibinfo {author} {\bibfnamefont {M.}~\bibnamefont
  {Golz}}, \bibinfo {author} {\bibfnamefont {G.}~\bibnamefont {Hiller}}, \ and\
  \bibinfo {author} {\bibfnamefont {T.}~\bibnamefont {Magorsch}},\ }\href
  {\doibase 10.1007/jhep09(2021)208} {\bibfield  {journal} {\bibinfo  {journal}
  {JHEP}\ }\textbf {\bibinfo {volume} {09}},\ \bibinfo {pages} {208} (\bibinfo
  {year} {2021})},\ \Eprint {http://arxiv.org/abs/2107.13010} {arXiv:2107.13010
  [hep-ph]} \BibitemShut {NoStop}%
\bibitem [{\citenamefont {Golz}\ \emph {et~al.}(2022)\citenamefont {Golz},
  \citenamefont {Hiller},\ and\ \citenamefont {Magorsch}}]{Golz:2022alh}%
  \BibitemOpen
  \bibfield  {author} {\bibinfo {author} {\bibfnamefont {M.}~\bibnamefont
  {Golz}}, \bibinfo {author} {\bibfnamefont {G.}~\bibnamefont {Hiller}}, \ and\
  \bibinfo {author} {\bibfnamefont {T.}~\bibnamefont {Magorsch}},\ }\href@noop
  {} {\  (\bibinfo {year} {2022})},\ \Eprint {http://arxiv.org/abs/2202.02331}
  {arXiv:2202.02331 [hep-ph]} \BibitemShut {NoStop}%
\bibitem [{\citenamefont {Lees}\ \emph {et~al.}(2011)\citenamefont {Lees} \emph
  {et~al.}}]{Lees:2011hb}%
  \BibitemOpen
  \bibfield  {author} {\bibinfo {author} {\bibfnamefont {J.}~\bibnamefont
  {Lees}} \emph {et~al.} (\bibinfo {collaboration} {BaBar}),\ }\href {\doibase
  10.1103/PhysRevD.84.072006} {\bibfield  {journal} {\bibinfo  {journal} {Phys.
  Rev. D}\ }\textbf {\bibinfo {volume} {84}},\ \bibinfo {pages} {072006}
  (\bibinfo {year} {2011})},\ \Eprint {http://arxiv.org/abs/1107.4465}
  {arXiv:1107.4465 [hep-ex]} \BibitemShut {NoStop}%
\bibitem [{\citenamefont {Aaij}\ \emph {et~al.}(2016)\citenamefont {Aaij} \emph
  {et~al.}}]{Aaij:2015qmj}%
  \BibitemOpen
  \bibfield  {author} {\bibinfo {author} {\bibfnamefont {R.}~\bibnamefont
  {Aaij}} \emph {et~al.} (\bibinfo {collaboration} {LHCb}),\ }\href {\doibase
  10.1016/j.physletb.2016.01.029} {\bibfield  {journal} {\bibinfo  {journal}
  {Phys. Lett. B}\ }\textbf {\bibinfo {volume} {754}},\ \bibinfo {pages} {167}
  (\bibinfo {year} {2016})},\ \Eprint {http://arxiv.org/abs/1512.00322}
  {arXiv:1512.00322 [hep-ex]} \BibitemShut {NoStop}%
\bibitem [{\citenamefont {Lees}\ \emph {et~al.}(2020)\citenamefont {Lees} \emph
  {et~al.}}]{BaBar:2020faa}%
  \BibitemOpen
  \bibfield  {author} {\bibinfo {author} {\bibfnamefont {J.~P.}\ \bibnamefont
  {Lees}} \emph {et~al.} (\bibinfo {collaboration} {BaBar}),\ }\href {\doibase
  10.1103/PhysRevD.101.112003} {\bibfield  {journal} {\bibinfo  {journal}
  {Phys. Rev. D}\ }\textbf {\bibinfo {volume} {101}},\ \bibinfo {pages}
  {112003} (\bibinfo {year} {2020})},\ \Eprint
  {http://arxiv.org/abs/2004.09457} {arXiv:2004.09457 [hep-ex]} \BibitemShut
  {NoStop}%
\bibitem [{\citenamefont {Aaij}\ \emph {et~al.}(2021)\citenamefont {Aaij} \emph
  {et~al.}}]{LHCb:2020car}%
  \BibitemOpen
  \bibfield  {author} {\bibinfo {author} {\bibfnamefont {R.}~\bibnamefont
  {Aaij}} \emph {et~al.} (\bibinfo {collaboration} {LHCb}),\ }\href {\doibase
  10.1007/JHEP06(2021)044} {\bibfield  {journal} {\bibinfo  {journal} {JHEP}\
  }\textbf {\bibinfo {volume} {06}},\ \bibinfo {pages} {044} (\bibinfo {year}
  {2021})},\ \Eprint {http://arxiv.org/abs/2011.00217} {arXiv:2011.00217
  [hep-ex]} \BibitemShut {NoStop}%
\bibitem [{\citenamefont {Angelescu}\ \emph {et~al.}(2020)\citenamefont
  {Angelescu}, \citenamefont {Faroughy},\ and\ \citenamefont
  {Sumensari}}]{Angelescu:2020uug}%
  \BibitemOpen
  \bibfield  {author} {\bibinfo {author} {\bibfnamefont {A.}~\bibnamefont
  {Angelescu}}, \bibinfo {author} {\bibfnamefont {D.~A.}\ \bibnamefont
  {Faroughy}}, \ and\ \bibinfo {author} {\bibfnamefont {O.}~\bibnamefont
  {Sumensari}},\ }\href {\doibase 10.1140/epjc/s10052-020-8210-5} {\bibfield
  {journal} {\bibinfo  {journal} {Eur. Phys. J. C}\ }\textbf {\bibinfo {volume}
  {80}},\ \bibinfo {pages} {641} (\bibinfo {year} {2020})},\ \Eprint
  {http://arxiv.org/abs/2002.05684} {arXiv:2002.05684 [hep-ph]} \BibitemShut
  {NoStop}%
\bibitem [{\citenamefont {Fuentes-Martin}\ \emph {et~al.}(2020)\citenamefont
  {Fuentes-Martin}, \citenamefont {Greljo}, \citenamefont {Martin~Camalich},\
  and\ \citenamefont {Ruiz-Alvarez}}]{Fuentes-Martin:2020lea}%
  \BibitemOpen
  \bibfield  {author} {\bibinfo {author} {\bibfnamefont {J.}~\bibnamefont
  {Fuentes-Martin}}, \bibinfo {author} {\bibfnamefont {A.}~\bibnamefont
  {Greljo}}, \bibinfo {author} {\bibfnamefont {J.}~\bibnamefont
  {Martin~Camalich}}, \ and\ \bibinfo {author} {\bibfnamefont {J.~D.}\
  \bibnamefont {Ruiz-Alvarez}},\ }\href {\doibase 10.1007/JHEP11(2020)080}
  {\bibfield  {journal} {\bibinfo  {journal} {JHEP}\ }\textbf {\bibinfo
  {volume} {11}},\ \bibinfo {pages} {080} (\bibinfo {year} {2020})},\ \Eprint
  {http://arxiv.org/abs/2003.12421} {arXiv:2003.12421 [hep-ph]} \BibitemShut
  {NoStop}%
\bibitem [{\citenamefont {Lai}\ \emph {et~al.}(2021)\citenamefont {Lai},
  \citenamefont {Li}, \citenamefont {Yan},\ and\ \citenamefont
  {Yang}}]{Lai:2021sww}%
  \BibitemOpen
  \bibfield  {author} {\bibinfo {author} {\bibfnamefont {L.-F.}\ \bibnamefont
  {Lai}}, \bibinfo {author} {\bibfnamefont {X.-Q.}\ \bibnamefont {Li}},
  \bibinfo {author} {\bibfnamefont {X.-S.}\ \bibnamefont {Yan}}, \ and\
  \bibinfo {author} {\bibfnamefont {Y.-D.}\ \bibnamefont {Yang}},\ }\href@noop
  {} {\  (\bibinfo {year} {2021})},\ \Eprint {http://arxiv.org/abs/2111.01463}
  {arXiv:2111.01463 [hep-ph]} \BibitemShut {NoStop}%
\bibitem [{\citenamefont {Abrahamyan}\ \emph {et~al.}(2011)\citenamefont
  {Abrahamyan} \emph {et~al.}}]{Abrahamyan:2011gv}%
  \BibitemOpen
  \bibfield  {author} {\bibinfo {author} {\bibfnamefont {S.}~\bibnamefont
  {Abrahamyan}} \emph {et~al.} (\bibinfo {collaboration} {APEX}),\ }\href
  {\doibase 10.1103/PhysRevLett.107.191804} {\bibfield  {journal} {\bibinfo
  {journal} {Phys. Rev. Lett.}\ }\textbf {\bibinfo {volume} {107}},\ \bibinfo
  {pages} {191804} (\bibinfo {year} {2011})},\ \Eprint
  {http://arxiv.org/abs/1108.2750} {arXiv:1108.2750 [hep-ex]} \BibitemShut
  {NoStop}%
\bibitem [{\citenamefont {Essig}\ \emph {et~al.}(2011)\citenamefont {Essig},
  \citenamefont {Schuster}, \citenamefont {Toro},\ and\ \citenamefont
  {Wojtsekhowski}}]{Essig:2010xa}%
  \BibitemOpen
  \bibfield  {author} {\bibinfo {author} {\bibfnamefont {R.}~\bibnamefont
  {Essig}}, \bibinfo {author} {\bibfnamefont {P.}~\bibnamefont {Schuster}},
  \bibinfo {author} {\bibfnamefont {N.}~\bibnamefont {Toro}}, \ and\ \bibinfo
  {author} {\bibfnamefont {B.}~\bibnamefont {Wojtsekhowski}},\ }\href {\doibase
  10.1007/JHEP02(2011)009} {\bibfield  {journal} {\bibinfo  {journal} {JHEP}\
  }\textbf {\bibinfo {volume} {02}},\ \bibinfo {pages} {009} (\bibinfo {year}
  {2011})},\ \Eprint {http://arxiv.org/abs/1001.2557} {arXiv:1001.2557
  [hep-ph]} \BibitemShut {NoStop}%
\bibitem [{\citenamefont {Essig}\ \emph {et~al.}(2013)\citenamefont {Essig}
  \emph {et~al.}}]{Essig:2013lka}%
  \BibitemOpen
  \bibfield  {author} {\bibinfo {author} {\bibfnamefont {R.}~\bibnamefont
  {Essig}} \emph {et~al.},\ }in\ \href@noop {} {\emph {\bibinfo {booktitle}
  {{Community Summer Study 2013}: {Snowmass on the Mississippi}}}}\ (\bibinfo
  {year} {2013})\ \Eprint {http://arxiv.org/abs/1311.0029} {arXiv:1311.0029
  [hep-ph]} \BibitemShut {NoStop}%
\bibitem [{\citenamefont {Allison}\ \emph {et~al.}(2015)\citenamefont {Allison}
  \emph {et~al.}}]{Allison:2014tpu}%
  \BibitemOpen
  \bibfield  {author} {\bibinfo {author} {\bibfnamefont {T.}~\bibnamefont
  {Allison}} \emph {et~al.} (\bibinfo {collaboration} {Qweak}),\ }\href
  {\doibase 10.1016/j.nima.2015.01.023} {\bibfield  {journal} {\bibinfo
  {journal} {Nucl. Instrum. Meth. A}\ }\textbf {\bibinfo {volume} {781}},\
  \bibinfo {pages} {105} (\bibinfo {year} {2015})},\ \Eprint
  {http://arxiv.org/abs/1409.7100} {arXiv:1409.7100 [physics.ins-det]}
  \BibitemShut {NoStop}%
\bibitem [{\citenamefont {Diaz}\ \emph {et~al.}(2017)\citenamefont {Diaz},
  \citenamefont {Schmaltz},\ and\ \citenamefont {Zhong}}]{Diaz:2017lit}%
  \BibitemOpen
  \bibfield  {author} {\bibinfo {author} {\bibfnamefont {B.}~\bibnamefont
  {Diaz}}, \bibinfo {author} {\bibfnamefont {M.}~\bibnamefont {Schmaltz}}, \
  and\ \bibinfo {author} {\bibfnamefont {Y.-M.}\ \bibnamefont {Zhong}},\ }\href
  {\doibase 10.1007/JHEP10(2017)097} {\bibfield  {journal} {\bibinfo  {journal}
  {JHEP}\ }\textbf {\bibinfo {volume} {10}},\ \bibinfo {pages} {097} (\bibinfo
  {year} {2017})},\ \Eprint {http://arxiv.org/abs/1706.05033} {arXiv:1706.05033
  [hep-ph]} \BibitemShut {NoStop}%
\bibitem [{\citenamefont {Schmaltz}\ and\ \citenamefont
  {Zhong}(2019)}]{Schmaltz:2018nls}%
  \BibitemOpen
  \bibfield  {author} {\bibinfo {author} {\bibfnamefont {M.}~\bibnamefont
  {Schmaltz}}\ and\ \bibinfo {author} {\bibfnamefont {Y.-M.}\ \bibnamefont
  {Zhong}},\ }\href {\doibase 10.1007/JHEP01(2019)132} {\bibfield  {journal}
  {\bibinfo  {journal} {JHEP}\ }\textbf {\bibinfo {volume} {01}},\ \bibinfo
  {pages} {132} (\bibinfo {year} {2019})},\ \Eprint
  {http://arxiv.org/abs/1810.10017} {arXiv:1810.10017 [hep-ph]} \BibitemShut
  {NoStop}%
\bibitem [{\citenamefont {Bandyopadhyay}\ \emph
  {et~al.}(2021{\natexlab{a}})\citenamefont {Bandyopadhyay}, \citenamefont
  {Dutta}, \citenamefont {Jakkapu},\ and\ \citenamefont
  {Karan}}]{Bandyopadhyay:2020wfv}%
  \BibitemOpen
  \bibfield  {author} {\bibinfo {author} {\bibfnamefont {P.}~\bibnamefont
  {Bandyopadhyay}}, \bibinfo {author} {\bibfnamefont {S.}~\bibnamefont
  {Dutta}}, \bibinfo {author} {\bibfnamefont {M.}~\bibnamefont {Jakkapu}}, \
  and\ \bibinfo {author} {\bibfnamefont {A.}~\bibnamefont {Karan}},\ }\href
  {\doibase 10.1016/j.nuclphysb.2021.115524} {\bibfield  {journal} {\bibinfo
  {journal} {Nucl. Phys. B}\ }\textbf {\bibinfo {volume} {971}},\ \bibinfo
  {pages} {115524} (\bibinfo {year} {2021}{\natexlab{a}})},\ \Eprint
  {http://arxiv.org/abs/2007.12997} {arXiv:2007.12997 [hep-ph]} \BibitemShut
  {NoStop}%
\bibitem [{\citenamefont {{CMS Collaboration}}(2012)}]{CMS:2012lqv}%
  \BibitemOpen
  \bibfield  {author} {\bibinfo {author} {\bibnamefont {{CMS Collaboration}}},\
  }\href@noop {} {\bibfield  {journal} {\bibinfo  {journal}
  {CMS-PAS-EXO-12-042}\ } (\bibinfo {year} {2012})}\BibitemShut {NoStop}%
\bibitem [{\citenamefont {Aad}\ \emph {et~al.}(2016)\citenamefont {Aad} \emph
  {et~al.}}]{Aad:2015caa}%
  \BibitemOpen
  \bibfield  {author} {\bibinfo {author} {\bibfnamefont {G.}~\bibnamefont
  {Aad}} \emph {et~al.} (\bibinfo {collaboration} {ATLAS}),\ }\href {\doibase
  10.1140/epjc/s10052-015-3823-9} {\bibfield  {journal} {\bibinfo  {journal}
  {Eur. Phys. J. C}\ }\textbf {\bibinfo {volume} {76}},\ \bibinfo {pages} {5}
  (\bibinfo {year} {2016})},\ \Eprint {http://arxiv.org/abs/1508.04735}
  {arXiv:1508.04735 [hep-ex]} \BibitemShut {NoStop}%
\bibitem [{\citenamefont {Sirunyan}\ \emph {et~al.}(2018)\citenamefont
  {Sirunyan} \emph {et~al.}}]{Sirunyan:2018ruf}%
  \BibitemOpen
  \bibfield  {author} {\bibinfo {author} {\bibfnamefont {A.~M.}\ \bibnamefont
  {Sirunyan}} \emph {et~al.} (\bibinfo {collaboration} {CMS}),\ }\href
  {\doibase 10.1103/PhysRevLett.121.241802} {\bibfield  {journal} {\bibinfo
  {journal} {Phys. Rev. Lett.}\ }\textbf {\bibinfo {volume} {121}},\ \bibinfo
  {pages} {241802} (\bibinfo {year} {2018})},\ \Eprint
  {http://arxiv.org/abs/1809.05558} {arXiv:1809.05558 [hep-ex]} \BibitemShut
  {NoStop}%
\bibitem [{\citenamefont {Sirunyan}\ \emph
  {et~al.}(2021{\natexlab{a}})\citenamefont {Sirunyan} \emph
  {et~al.}}]{CMS:2020wzx}%
  \BibitemOpen
  \bibfield  {author} {\bibinfo {author} {\bibfnamefont {A.~M.}\ \bibnamefont
  {Sirunyan}} \emph {et~al.} (\bibinfo {collaboration} {CMS}),\ }\href
  {\doibase 10.1016/j.physletb.2021.136446} {\bibfield  {journal} {\bibinfo
  {journal} {Phys. Lett. B}\ }\textbf {\bibinfo {volume} {819}},\ \bibinfo
  {pages} {136446} (\bibinfo {year} {2021}{\natexlab{a}})},\ \Eprint
  {http://arxiv.org/abs/2012.04178} {arXiv:2012.04178 [hep-ex]} \BibitemShut
  {NoStop}%
\bibitem [{\citenamefont {Zyla}\ \emph {et~al.}(2020)\citenamefont {Zyla} \emph
  {et~al.}}]{Zyla:2020zbs}%
  \BibitemOpen
  \bibfield  {author} {\bibinfo {author} {\bibfnamefont {P.}~\bibnamefont
  {Zyla}} \emph {et~al.} (\bibinfo {collaboration} {Particle Data Group}),\
  }\href {\doibase 10.1093/ptep/ptaa104} {\bibfield  {journal} {\bibinfo
  {journal} {PTEP}\ }\textbf {\bibinfo {volume} {2020}},\ \bibinfo {pages}
  {083C01} (\bibinfo {year} {2020})}\BibitemShut {NoStop}%
\bibitem [{\citenamefont {Bandyopadhyay}\ \emph
  {et~al.}(2021{\natexlab{b}})\citenamefont {Bandyopadhyay}, \citenamefont
  {Karan},\ and\ \citenamefont {Mandal}}]{Bandyopadhyay:2021pld}%
  \BibitemOpen
  \bibfield  {author} {\bibinfo {author} {\bibfnamefont {P.}~\bibnamefont
  {Bandyopadhyay}}, \bibinfo {author} {\bibfnamefont {A.}~\bibnamefont
  {Karan}}, \ and\ \bibinfo {author} {\bibfnamefont {R.}~\bibnamefont
  {Mandal}},\ }\href@noop {} {\  (\bibinfo {year} {2021}{\natexlab{b}})},\
  \Eprint {http://arxiv.org/abs/2108.06506} {arXiv:2108.06506 [hep-ph]}
  \BibitemShut {NoStop}%
\bibitem [{\citenamefont {Qian}\ \emph {et~al.}(2021)\citenamefont {Qian},
  \citenamefont {Li}, \citenamefont {Li}, \citenamefont {Meng}, \citenamefont
  {Xiao}, \citenamefont {Yang}, \citenamefont {Lu},\ and\ \citenamefont
  {You}}]{Qian:2021ihf}%
  \BibitemOpen
  \bibfield  {author} {\bibinfo {author} {\bibfnamefont {S.}~\bibnamefont
  {Qian}}, \bibinfo {author} {\bibfnamefont {C.}~\bibnamefont {Li}}, \bibinfo
  {author} {\bibfnamefont {Q.}~\bibnamefont {Li}}, \bibinfo {author}
  {\bibfnamefont {F.}~\bibnamefont {Meng}}, \bibinfo {author} {\bibfnamefont
  {J.}~\bibnamefont {Xiao}}, \bibinfo {author} {\bibfnamefont {T.}~\bibnamefont
  {Yang}}, \bibinfo {author} {\bibfnamefont {M.}~\bibnamefont {Lu}}, \ and\
  \bibinfo {author} {\bibfnamefont {Z.}~\bibnamefont {You}},\ }\href {\doibase
  10.1007/JHEP12(2021)047} {\bibfield  {journal} {\bibinfo  {journal} {JHEP}\
  }\textbf {\bibinfo {volume} {12}},\ \bibinfo {pages} {047} (\bibinfo {year}
  {2021})},\ \Eprint {http://arxiv.org/abs/2109.01265} {arXiv:2109.01265
  [hep-ph]} \BibitemShut {NoStop}%
\bibitem [{\citenamefont {Taxil}\ \emph {et~al.}(2000)\citenamefont {Taxil},
  \citenamefont {Tugcu},\ and\ \citenamefont {Virey}}]{Taxil:1999pf}%
  \BibitemOpen
  \bibfield  {author} {\bibinfo {author} {\bibfnamefont {P.}~\bibnamefont
  {Taxil}}, \bibinfo {author} {\bibfnamefont {E.}~\bibnamefont {Tugcu}}, \ and\
  \bibinfo {author} {\bibfnamefont {J.~M.}\ \bibnamefont {Virey}},\ }\href
  {\doibase 10.1007/s100520050743} {\bibfield  {journal} {\bibinfo  {journal}
  {Eur. Phys. J. C}\ }\textbf {\bibinfo {volume} {14}},\ \bibinfo {pages} {165}
  (\bibinfo {year} {2000})},\ \Eprint {http://arxiv.org/abs/hep-ph/9912272}
  {arXiv:hep-ph/9912272} \BibitemShut {NoStop}%
\bibitem [{\citenamefont {Abbon}\ \emph {et~al.}(2007)\citenamefont {Abbon}
  \emph {et~al.}}]{COMPASS:2007rjf}%
  \BibitemOpen
  \bibfield  {author} {\bibinfo {author} {\bibfnamefont {P.}~\bibnamefont
  {Abbon}} \emph {et~al.} (\bibinfo {collaboration} {COMPASS}),\ }\href
  {\doibase 10.1016/j.nima.2007.03.026} {\bibfield  {journal} {\bibinfo
  {journal} {Nucl. Instrum. Meth. A}\ }\textbf {\bibinfo {volume} {577}},\
  \bibinfo {pages} {455} (\bibinfo {year} {2007})},\ \Eprint
  {http://arxiv.org/abs/hep-ex/0703049} {arXiv:hep-ex/0703049} \BibitemShut
  {NoStop}%
\bibitem [{\citenamefont {Ahmed}\ \emph {et~al.}(2012)\citenamefont {Ahmed}
  \emph {et~al.}}]{HAPPEX:2011xlw}%
  \BibitemOpen
  \bibfield  {author} {\bibinfo {author} {\bibfnamefont {Z.}~\bibnamefont
  {Ahmed}} \emph {et~al.} (\bibinfo {collaboration} {HAPPEX}),\ }\href
  {\doibase 10.1103/PhysRevLett.108.102001} {\bibfield  {journal} {\bibinfo
  {journal} {Phys. Rev. Lett.}\ }\textbf {\bibinfo {volume} {108}},\ \bibinfo
  {pages} {102001} (\bibinfo {year} {2012})},\ \Eprint
  {http://arxiv.org/abs/1107.0913} {arXiv:1107.0913 [nucl-ex]} \BibitemShut
  {NoStop}%
\bibitem [{\citenamefont {Arrington}\ \emph {et~al.}(2021)\citenamefont
  {Arrington} \emph {et~al.}}]{Arrington:2021alx}%
  \BibitemOpen
  \bibfield  {author} {\bibinfo {author} {\bibfnamefont {J.}~\bibnamefont
  {Arrington}} \emph {et~al.},\ }\href@noop {} {\  (\bibinfo {year} {2021})},\
  \Eprint {http://arxiv.org/abs/2112.00060} {arXiv:2112.00060 [nucl-ex]}
  \BibitemShut {NoStop}%
\bibitem [{\citenamefont {Nishi}(2005)}]{Nishi:2004st}%
  \BibitemOpen
  \bibfield  {author} {\bibinfo {author} {\bibfnamefont {C.~C.}\ \bibnamefont
  {Nishi}},\ }\href {\doibase 10.1119/1.2074087} {\bibfield  {journal}
  {\bibinfo  {journal} {Am. J. Phys.}\ }\textbf {\bibinfo {volume} {73}},\
  \bibinfo {pages} {1160} (\bibinfo {year} {2005})},\ \Eprint
  {http://arxiv.org/abs/hep-ph/0412245} {arXiv:hep-ph/0412245 [hep-ph]}
  \BibitemShut {NoStop}%
%%CITATION = HEP-PH/0412245;%%
\bibitem [{\citenamefont {Gonz{\'a}lez-Alonso}\ \emph
  {et~al.}(2017)\citenamefont {Gonz{\'a}lez-Alonso}, \citenamefont
  {Martin~Camalich},\ and\ \citenamefont {Mimouni}}]{Gonzalez-Alonso:2017iyc}%
  \BibitemOpen
  \bibfield  {author} {\bibinfo {author} {\bibfnamefont {M.}~\bibnamefont
  {Gonz{\'a}lez-Alonso}}, \bibinfo {author} {\bibfnamefont {J.}~\bibnamefont
  {Martin~Camalich}}, \ and\ \bibinfo {author} {\bibfnamefont {K.}~\bibnamefont
  {Mimouni}},\ }\href {\doibase 10.1016/j.physletb.2017.07.003} {\bibfield
  {journal} {\bibinfo  {journal} {Phys. Lett. B}\ }\textbf {\bibinfo {volume}
  {772}},\ \bibinfo {pages} {777} (\bibinfo {year} {2017})},\ \Eprint
  {http://arxiv.org/abs/1706.00410} {arXiv:1706.00410 [hep-ph]} \BibitemShut
  {NoStop}%
\bibitem [{\citenamefont {Aebischer}\ \emph {et~al.}(2017)\citenamefont
  {Aebischer}, \citenamefont {Fael}, \citenamefont {Greub},\ and\ \citenamefont
  {Virto}}]{Aebischer:2017gaw}%
  \BibitemOpen
  \bibfield  {author} {\bibinfo {author} {\bibfnamefont {J.}~\bibnamefont
  {Aebischer}}, \bibinfo {author} {\bibfnamefont {M.}~\bibnamefont {Fael}},
  \bibinfo {author} {\bibfnamefont {C.}~\bibnamefont {Greub}}, \ and\ \bibinfo
  {author} {\bibfnamefont {J.}~\bibnamefont {Virto}},\ }\href {\doibase
  10.1007/JHEP09(2017)158} {\bibfield  {journal} {\bibinfo  {journal} {JHEP}\
  }\textbf {\bibinfo {volume} {09}},\ \bibinfo {pages} {158} (\bibinfo {year}
  {2017})},\ \Eprint {http://arxiv.org/abs/1704.06639} {arXiv:1704.06639
  [hep-ph]} \BibitemShut {NoStop}%
\bibitem [{\citenamefont {Feldmann}\ and\ \citenamefont
  {Yip}(2012)}]{Feldmann:2011xf}%
  \BibitemOpen
  \bibfield  {author} {\bibinfo {author} {\bibfnamefont {T.}~\bibnamefont
  {Feldmann}}\ and\ \bibinfo {author} {\bibfnamefont {M.~W.~Y.}\ \bibnamefont
  {Yip}},\ }\href {\doibase 10.1103/PhysRevD.85.014035,
  10.1103/physrevd.86.079901} {\bibfield  {journal} {\bibinfo  {journal} {Phys.
  Rev.}\ }\textbf {\bibinfo {volume} {D85}},\ \bibinfo {pages} {014035}
  (\bibinfo {year} {2012})},\ \bibinfo {note} {[Erratum: Phys.
  Rev.D86,079901(2012)]},\ \Eprint {http://arxiv.org/abs/1111.1844}
  {arXiv:1111.1844 [hep-ph]} \BibitemShut {NoStop}%
%%CITATION = ARXIV:1111.1844;%%
\bibitem [{\citenamefont {Das}(2018)}]{Das:2018sms}%
  \BibitemOpen
  \bibfield  {author} {\bibinfo {author} {\bibfnamefont {D.}~\bibnamefont
  {Das}},\ }\href {\doibase 10.1140/epjc/s10052-018-5731-2} {\bibfield
  {journal} {\bibinfo  {journal} {Eur. Phys. J. C}\ }\textbf {\bibinfo {volume}
  {78}},\ \bibinfo {pages} {230} (\bibinfo {year} {2018})},\ \Eprint
  {http://arxiv.org/abs/1802.09404} {arXiv:1802.09404 [hep-ph]} \BibitemShut
  {NoStop}%
\bibitem [{\citenamefont {Bourrely}\ \emph {et~al.}(2009)\citenamefont
  {Bourrely}, \citenamefont {Caprini},\ and\ \citenamefont
  {Lellouch}}]{Bourrely:2008za}%
  \BibitemOpen
  \bibfield  {author} {\bibinfo {author} {\bibfnamefont {C.}~\bibnamefont
  {Bourrely}}, \bibinfo {author} {\bibfnamefont {I.}~\bibnamefont {Caprini}}, \
  and\ \bibinfo {author} {\bibfnamefont {L.}~\bibnamefont {Lellouch}},\ }\href
  {\doibase 10.1103/PhysRevD.82.099902, 10.1103/PhysRevD.79.013008} {\bibfield
  {journal} {\bibinfo  {journal} {Phys. Rev.}\ }\textbf {\bibinfo {volume}
  {D79}},\ \bibinfo {pages} {013008} (\bibinfo {year} {2009})},\ \bibinfo
  {note} {[Erratum: Phys. Rev.D82,099902(2010)]},\ \Eprint
  {http://arxiv.org/abs/0807.2722} {arXiv:0807.2722 [hep-ph]} \BibitemShut
  {NoStop}%
%%CITATION = ARXIV:0807.2722;%%
\bibitem [{\citenamefont {Sobczyk}\ \emph {et~al.}(2019)\citenamefont
  {Sobczyk}, \citenamefont {Rocco}, \citenamefont {Lovato},\ and\ \citenamefont
  {Nieves}}]{Sobczyk:2019uej}%
  \BibitemOpen
  \bibfield  {author} {\bibinfo {author} {\bibfnamefont {J.~E.}\ \bibnamefont
  {Sobczyk}}, \bibinfo {author} {\bibfnamefont {N.}~\bibnamefont {Rocco}},
  \bibinfo {author} {\bibfnamefont {A.}~\bibnamefont {Lovato}}, \ and\ \bibinfo
  {author} {\bibfnamefont {J.}~\bibnamefont {Nieves}},\ }\href {\doibase
  10.1103/PhysRevC.99.065503} {\bibfield  {journal} {\bibinfo  {journal} {Phys.
  Rev. C}\ }\textbf {\bibinfo {volume} {99}},\ \bibinfo {pages} {065503}
  (\bibinfo {year} {2019})},\ \Eprint {http://arxiv.org/abs/1901.10192}
  {arXiv:1901.10192 [nucl-th]} \BibitemShut {NoStop}%
\bibitem [{\citenamefont {Barone}\ and\ \citenamefont
  {Ratcliffe}(2003)}]{Barone:2003fy}%
  \BibitemOpen
  \bibfield  {author} {\bibinfo {author} {\bibfnamefont {V.}~\bibnamefont
  {Barone}}\ and\ \bibinfo {author} {\bibfnamefont {P.~G.}\ \bibnamefont
  {Ratcliffe}},\ }\href@noop {} {\emph {\bibinfo {title} {{Transverse spin
  physics}}}}\ (\bibinfo {year} {2003})\BibitemShut {NoStop}%
\bibitem [{\citenamefont {Virey}(1999)}]{Virey:1998ny}%
  \BibitemOpen
  \bibfield  {author} {\bibinfo {author} {\bibfnamefont {J.~M.}\ \bibnamefont
  {Virey}},\ }\href {\doibase 10.1007/s100529901052} {\bibfield  {journal}
  {\bibinfo  {journal} {Eur. Phys. J. C}\ }\textbf {\bibinfo {volume} {8}},\
  \bibinfo {pages} {283} (\bibinfo {year} {1999})},\ \Eprint
  {http://arxiv.org/abs/hep-ph/9809439} {arXiv:hep-ph/9809439} \BibitemShut
  {NoStop}%
\bibitem [{\citenamefont {Petric}\ \emph {et~al.}(2010)\citenamefont {Petric}
  \emph {et~al.}}]{Petric:2010yt}%
  \BibitemOpen
  \bibfield  {author} {\bibinfo {author} {\bibfnamefont {M.}~\bibnamefont
  {Petric}} \emph {et~al.} (\bibinfo {collaboration} {Belle}),\ }\href
  {\doibase 10.1103/PhysRevD.81.091102} {\bibfield  {journal} {\bibinfo
  {journal} {Phys. Rev. D}\ }\textbf {\bibinfo {volume} {81}},\ \bibinfo
  {pages} {091102} (\bibinfo {year} {2010})},\ \Eprint
  {http://arxiv.org/abs/1003.2345} {arXiv:1003.2345 [hep-ex]} \BibitemShut
  {NoStop}%
\bibitem [{\citenamefont {Arnold}\ \emph {et~al.}(2013)\citenamefont {Arnold},
  \citenamefont {Fornal},\ and\ \citenamefont {Wise}}]{Arnold:2013cva}%
  \BibitemOpen
  \bibfield  {author} {\bibinfo {author} {\bibfnamefont {J.~M.}\ \bibnamefont
  {Arnold}}, \bibinfo {author} {\bibfnamefont {B.}~\bibnamefont {Fornal}}, \
  and\ \bibinfo {author} {\bibfnamefont {M.~B.}\ \bibnamefont {Wise}},\ }\href
  {\doibase 10.1103/PhysRevD.88.035009} {\bibfield  {journal} {\bibinfo
  {journal} {Phys. Rev. D}\ }\textbf {\bibinfo {volume} {88}},\ \bibinfo
  {pages} {035009} (\bibinfo {year} {2013})},\ \Eprint
  {http://arxiv.org/abs/1304.6119} {arXiv:1304.6119 [hep-ph]} \BibitemShut
  {NoStop}%
\bibitem [{\citenamefont {Gardner}\ and\ \citenamefont
  {Yan}(2019)}]{Gardner:2018azu}%
  \BibitemOpen
  \bibfield  {author} {\bibinfo {author} {\bibfnamefont {S.}~\bibnamefont
  {Gardner}}\ and\ \bibinfo {author} {\bibfnamefont {X.}~\bibnamefont {Yan}},\
  }\href {\doibase 10.1016/j.physletb.2019.01.054} {\bibfield  {journal}
  {\bibinfo  {journal} {Phys. Lett. B}\ }\textbf {\bibinfo {volume} {790}},\
  \bibinfo {pages} {421} (\bibinfo {year} {2019})},\ \Eprint
  {http://arxiv.org/abs/1808.05288} {arXiv:1808.05288 [hep-ph]} \BibitemShut
  {NoStop}%
\bibitem [{\citenamefont {Assad}\ \emph {et~al.}(2018)\citenamefont {Assad},
  \citenamefont {Fornal},\ and\ \citenamefont {Grinstein}}]{Assad:2017iib}%
  \BibitemOpen
  \bibfield  {author} {\bibinfo {author} {\bibfnamefont {N.}~\bibnamefont
  {Assad}}, \bibinfo {author} {\bibfnamefont {B.}~\bibnamefont {Fornal}}, \
  and\ \bibinfo {author} {\bibfnamefont {B.}~\bibnamefont {Grinstein}},\ }\href
  {\doibase 10.1016/j.physletb.2017.12.042} {\bibfield  {journal} {\bibinfo
  {journal} {Phys. Lett. B}\ }\textbf {\bibinfo {volume} {777}},\ \bibinfo
  {pages} {324} (\bibinfo {year} {2018})},\ \Eprint
  {http://arxiv.org/abs/1708.06350} {arXiv:1708.06350 [hep-ph]} \BibitemShut
  {NoStop}%
\bibitem [{\citenamefont {Ohlsen}(1972)}]{Ohlsen:1972zz}%
  \BibitemOpen
  \bibfield  {author} {\bibinfo {author} {\bibfnamefont {G.~G.}\ \bibnamefont
  {Ohlsen}},\ }\href {\doibase 10.1088/0034-4885/35/2/305} {\bibfield
  {journal} {\bibinfo  {journal} {Rept. Prog. Phys.}\ }\textbf {\bibinfo
  {volume} {35}},\ \bibinfo {pages} {717} (\bibinfo {year} {1972})}\BibitemShut
  {NoStop}%
\bibitem [{\citenamefont {Donnelly}\ and\ \citenamefont
  {Raskin}(1986)}]{Donnelly:1985ry}%
  \BibitemOpen
  \bibfield  {author} {\bibinfo {author} {\bibfnamefont {T.~W.}\ \bibnamefont
  {Donnelly}}\ and\ \bibinfo {author} {\bibfnamefont {A.~S.}\ \bibnamefont
  {Raskin}},\ }\href {\doibase 10.1016/0003-4916(86)90173-9} {\bibfield
  {journal} {\bibinfo  {journal} {Annals Phys.}\ }\textbf {\bibinfo {volume}
  {169}},\ \bibinfo {pages} {247} (\bibinfo {year} {1986})}\BibitemShut
  {NoStop}%
\bibitem [{\citenamefont {Adams}\ \emph {et~al.}(1997)\citenamefont {Adams}
  \emph {et~al.}}]{SpinMuonSMC:1997mkb}%
  \BibitemOpen
  \bibfield  {author} {\bibinfo {author} {\bibfnamefont {D.}~\bibnamefont
  {Adams}} \emph {et~al.} (\bibinfo {collaboration} {Spin Muon (SMC)}),\ }\href
  {\doibase 10.1103/PhysRevD.56.5330} {\bibfield  {journal} {\bibinfo
  {journal} {Phys. Rev. D}\ }\textbf {\bibinfo {volume} {56}},\ \bibinfo
  {pages} {5330} (\bibinfo {year} {1997})},\ \Eprint
  {http://arxiv.org/abs/hep-ex/9702005} {arXiv:hep-ex/9702005} \BibitemShut
  {NoStop}%
\bibitem [{\citenamefont {Arrington}\ \emph {et~al.}(2007)\citenamefont
  {Arrington} \emph {et~al.}}]{J.Arrington}%
  \BibitemOpen
  \bibfield  {author} {\bibinfo {author} {\bibfnamefont {J.}~\bibnamefont
  {Arrington}} \emph {et~al.} (\bibinfo {collaboration} {JLab E08-007}),\
  }\href
  {https://hallaweb.jlab.org/experiment/E08-007/home2_files/LowQ2Prop.pdf}
  {\enquote {\bibinfo {title} {{Measurement of the Proton Elastic Form Factor
  Ratio at Low $Q^2$}},}\ } (\bibinfo {year} {2007})\BibitemShut {NoStop}%
\bibitem [{\citenamefont {Camsonne}\ \emph {et~al.}(2012)\citenamefont
  {Camsonne} \emph {et~al.}}]{A.Camsonne}%
  \BibitemOpen
  \bibfield  {author} {\bibinfo {author} {\bibfnamefont {A.}~\bibnamefont
  {Camsonne}} \emph {et~al.} (\bibinfo {collaboration} {JLab E08-027}),\ }\href
  {https://hallaweb.jlab.org/experiment/g2p/docs/PAC33/dlt.pdf} {\enquote
  {\bibinfo {title} {{A Measurement of $g^2_p$ and the Longitudinal-Transverse
  Spin Polarizability}},}\ } (\bibinfo {year} {2012})\BibitemShut {NoStop}%
\bibitem [{\citenamefont {Gao}\ \emph {et~al.}(2011)\citenamefont {Gao} \emph
  {et~al.}}]{Allada2011}%
  \BibitemOpen
  \bibfield  {author} {\bibinfo {author} {\bibfnamefont {H.}~\bibnamefont
  {Gao}} \emph {et~al.} (\bibinfo {collaboration} {JLab E12-11-108}),\ }\href
  {https://www.jlab.org/exp_prog/proposals/11/PR12-11-108.pdf} {\enquote
  {\bibinfo {title} {{Target Single Spin Asymmetry in Semi-Inclusive
  Deep-Inelastic ($e$, $e^\prime\pi^{\pm}$) Reaction on a Transversely
  Polarized Proton Target}},}\ } (\bibinfo {year} {2011})\BibitemShut {NoStop}%
\bibitem [{\citenamefont {Zupanc}\ \emph {et~al.}(2014)\citenamefont {Zupanc}
  \emph {et~al.}}]{Belle:2013jfq}%
  \BibitemOpen
  \bibfield  {author} {\bibinfo {author} {\bibfnamefont {A.}~\bibnamefont
  {Zupanc}} \emph {et~al.} (\bibinfo {collaboration} {Belle}),\ }\href
  {\doibase 10.1103/PhysRevLett.113.042002} {\bibfield  {journal} {\bibinfo
  {journal} {Phys. Rev. Lett.}\ }\textbf {\bibinfo {volume} {113}},\ \bibinfo
  {pages} {042002} (\bibinfo {year} {2014})},\ \Eprint
  {http://arxiv.org/abs/1312.7826} {arXiv:1312.7826 [hep-ex]} \BibitemShut
  {NoStop}%
\bibitem [{\citenamefont {Ablikim}\ \emph {et~al.}(2016)\citenamefont {Ablikim}
  \emph {et~al.}}]{BESIII:2015bjk}%
  \BibitemOpen
  \bibfield  {author} {\bibinfo {author} {\bibfnamefont {M.}~\bibnamefont
  {Ablikim}} \emph {et~al.} (\bibinfo {collaboration} {BESIII}),\ }\href
  {\doibase 10.1103/PhysRevLett.116.052001} {\bibfield  {journal} {\bibinfo
  {journal} {Phys. Rev. Lett.}\ }\textbf {\bibinfo {volume} {116}},\ \bibinfo
  {pages} {052001} (\bibinfo {year} {2016})},\ \Eprint
  {http://arxiv.org/abs/1511.08380} {arXiv:1511.08380 [hep-ex]} \BibitemShut
  {NoStop}%
\bibitem [{\citenamefont {Sirunyan}\ \emph
  {et~al.}(2021{\natexlab{b}})\citenamefont {Sirunyan} \emph
  {et~al.}}]{CMS:2021ctt}%
  \BibitemOpen
  \bibfield  {author} {\bibinfo {author} {\bibfnamefont {A.~M.}\ \bibnamefont
  {Sirunyan}} \emph {et~al.} (\bibinfo {collaboration} {CMS}),\ }\href
  {\doibase 10.1007/JHEP07(2021)208} {\bibfield  {journal} {\bibinfo  {journal}
  {JHEP}\ }\textbf {\bibinfo {volume} {07}},\ \bibinfo {pages} {208} (\bibinfo
  {year} {2021}{\natexlab{b}})},\ \Eprint {http://arxiv.org/abs/2103.02708}
  {arXiv:2103.02708 [hep-ex]} \BibitemShut {NoStop}%
\bibitem [{\citenamefont {Anselmino}\ \emph {et~al.}(1995)\citenamefont
  {Anselmino}, \citenamefont {Efremov},\ and\ \citenamefont
  {Leader}}]{Anselmino:1994gn}%
  \BibitemOpen
  \bibfield  {author} {\bibinfo {author} {\bibfnamefont {M.}~\bibnamefont
  {Anselmino}}, \bibinfo {author} {\bibfnamefont {A.}~\bibnamefont {Efremov}},
  \ and\ \bibinfo {author} {\bibfnamefont {E.}~\bibnamefont {Leader}},\ }\href
  {\doibase 10.1016/0370-1573(95)00011-5} {\bibfield  {journal} {\bibinfo
  {journal} {Phys. Rept.}\ }\textbf {\bibinfo {volume} {261}},\ \bibinfo
  {pages} {1} (\bibinfo {year} {1995})},\ \bibinfo {note} {[Erratum: Phys.Rept.
  281, 399--400 (1997)]},\ \Eprint {http://arxiv.org/abs/hep-ph/9501369}
  {arXiv:hep-ph/9501369} \BibitemShut {NoStop}%
\end{thebibliography}%

\end{document}